\documentclass[pra,onecolumn,floatfix,a4paper,superscriptaddress]{revtex4}
\usepackage{bm,color,graphicx,amsmath,txfonts}
\usepackage{here}
\usepackage{array}
\usepackage{graphicx}
\usepackage[colorlinks, citecolor=blue,linkcolor=blue]{hyperref}
\usepackage{braket}
\usepackage{dsfont}
\newcommand{\e}{{e}}

\begin{document}

\makeatletter
\renewcommand{\@biblabel}[1]{\makebox[2em][l]{\textsuperscript{\textcolor{black}{\fontsize{10}{12}\selectfont[#1]}}}}
\makeatother

\let\oldbibliography\thebibliography
\renewcommand{\thebibliography}[1]{%
  \addcontentsline{toc}{section}{\refname}%
  \oldbibliography{#1}%
  \setlength\itemsep{0pt}%
}

\title{Characterizing quantum correlations and quantum teleportation in $gg \to t\bar{t}$ and $q\bar{q} \to t\bar{t}$ processes under noisy channels}

\author{Elhabib Jaloum}
\address{LPTHE-Department of Physics, Faculty of Sciences, Ibnou Zohr University, Agadir 80000, Morocco}

\author{Omar Bachain}
\address{LPHE-Modeling and Simulation, Faculty of Sciences, Mohammed V University in Rabat, Rabat, Morocco}

\author{Mohamed \surname{Amazioug} }
\email{m.amazioug@uiz.ac.ma}
\address{LPTHE-Department of Physics, Faculty of Sciences, Ibnou Zohr University, Agadir 80000, Morocco}

\author{Nazek Alessa}
\address{Department of Mathematical Sciences, College of Science, Princess Nourah bint Abdulrahman University, P.O. Box 84428, Riyadh 11671, Saudi Arabia}

\author{Rachid Ahl Laamara}
\address{LPHE-Modeling and Simulation, Faculty of Sciences, Mohammed V University in Rabat, Rabat, Morocco}
\address{Centre of Physics and Mathematics, CPM, Faculty of Sciences, Mohammed V University in Rabat, Rabat, Morocco}

\author{R. T. Matoog}
\address{Mathematics Department, Faculty of Sciences, Umm Al-Qura University, Makkah, Saudi Arabia}

\author{Abdel-Haleem Abdel-Aty}
\address{Department of Physics, College of Sciences, University of Bisha, Bisha, 61922, Saudi Arabia}

\date{\today}

\begin{abstract}
The measurement of top-quark spin correlations provides a key tool for probing its interactions with high precision. Owing to its extremely short lifetime ($\tau \sim 10^{-25}$ s), the top quark preserves its spin polarization information, making the $t\bar{t}$ system an ideal framework for investigating quantum correlations in high-energy physics. In this work, we analyze quantum correlations in $t\bar{t}$ pairs produced in QCD using several quantum information–theoretic measures, including Bell nonlocality, quantum steering, concurrence, and geometric quantum discord. Their dependence on kinematic variables is examined in both the $gg \to t\bar{t}$ and $q\bar{q} \to t\bar{t}$ channels, with convergence toward the $gg \to t\bar{t}$ dominated regime in the ultra-relativistic limit ($\beta = 1$). We also investigate the effect of three effective decoherence channels (AD, PD, and PF). The AD and PD channels lead to a monotonic degradation of correlations as the decoherence parameter $p$ increases, while the PF channel exhibits a symmetric behavior around $p=1/2$. The impact of these channels on quantum teleportation is analyzed, showing that it remains above the classical threshold of $2/3$ even in the presence of noise. These results indicate that certain quantum resources can persist despite decoherence, opening new perspectives at the interface of quantum information and particle physics.
\end{abstract}
\maketitle
\section{Introduction}    \label{sec:1}
Quantum superposition \citep{1,2} is a fundamental pillar of quantum information processing, allowing a qubit to exist in multiple states simultaneously. Quantum systems also display unique correlations with no classical equivalent, most notably quantum entanglement \citep{3,4}, along with Bell nonlocality \citep{5}, quantum steering \citep{8,9}, and quantum discord \citep{6,7}. Quantum entanglement \citep{10} is a fundamental phenomenon in quantum physics and quantum technologies \citep{11}. It occurs when two or more particles are correlated such that the quantum state of each cannot be described independently of the others, even at large distances. A change in one particle’s state thus instantly affects its partner. This correlation is a key resource for building powerful quantum computers \citep{12,13} and related quantum technologies \citep{14,15}. 

Quantum steering is a distinctive phenomenon in quantum mechanics \citep{16} that arises from the correlations shared by entangled particles \citep{17,18}. It refers to the ability of measurements performed on one particle to remotely influence the state of its partner, even across large spatial separations.
The concept was originally introduced by Erwin Schrödinger in response to the well-known Einstein–Podolsky–Rosen (EPR) paradox \cite{19,20}. Quantum steering can be viewed as an asymmetric form of quantum entanglement. In this framework, one party (commonly called Alice, the steerer) can steer or control the state of the distant particle held by the other party (usually Bob, the target) simply by choosing appropriate measurements on her own subsystem. Unlike Bell nonlocality \cite{21} or standard quantum entanglement, steering is inherently directional: Alice can influence Bob’s particle, but the reverse influence is not necessarily possible. This asymmetric property has generated considerable interest in both theoretical studies and experimental investigations in quantum physics.

The Geometric Quantum Discord (GQD) is a powerful quantifier of quantum correlations in bipartite systems. It is specifically designed to detect more general non-classical correlations that extend beyond quantum entanglement. Geometric approaches are widely used to define and quantify various quantum resources across different quantum systems \citep{30}. In particular, the Schatten 1-norm quantum discord \citep{30} serves as a robust geometric measure for evaluating the strength of quantum correlations in metal complexes \citep{75}. Unlike many conventional correlation measures, it explicitly incorporates the geometric structure of the quantum state space. This geometric perspective enables a richer and more nuanced characterization of quantum interactions. GQD has been extensively investigated both theoretically and experimentally, and it holds promising potential for the identification and practical exploitation of quantum resources in quantum information processing and quantum communication \citep{44,45}.

Over the past few years, the study of quantum entanglement has extended into high-energy physics, with particular interest devoted to spin-correlated top–antitop systems produced at the Large Hadron Collider. At the CERN Large Hadron Collider (LHC), a very large number of top quark pairs is produced~\cite{ref46,ref47,ref48}. During the low-luminosity phase, approximately eight million top quark pairs are expected to be produced per year~\cite{ref49}. This substantial statistics enables extremely precise measurements in the top quark sector. In particular, it allows for high-precision verification that the top quark indeed possesses the quantum numbers predicted by the Standard Model~\cite{ref50}. Furthermore, given the extremely high energy scale involved in processes featuring the top quark, top-quark physics also provides an ideal laboratory for searches for new physics beyond the Standard Model~\cite{ref51}. For instance, one can look for new resonances in the s-channel that could couple strongly to the top quark~\cite{ref52}. To probe the properties of such a hypothetical resonance, the spin correlations of the top quark represent a particularly powerful tool~\cite{ref53,ref54}. In particular, they can help determine the nature of the intermediate resonance.

A key feature of the top quark is that it decays before hadronizing, making it unique among all quarks~\cite{ref55}. This property ensures that its spin information remains undiluted by hadronization effects. In the Standard Model, the top quark decays primarily through the parity-violating weak interaction, thereby transferring its spin information to the angular distributions of the decay products~\cite{ref56,ref57}. Top-quark polarization is therefore an excellent observable, since it can be directly probed experimentally via a detailed analysis of the kinematics of its decay products~\cite{ref58}.

In this work, we study quantum correlations in $t\bar{t}$ pairs produced via QCD using key quantum information measures: Bell non-locality, quantum steering, concurrence, and geometric quantum discord. The analysis is performed in the $gg\to t\bar{t}$ and $q\bar{q}\to t\bar{t}$ channels, focusing on their dependence on kinematic variables. We confirm the hierarchy Bell non-locality $\subseteq$ Steering $\subseteq$ Entanglement $\subseteq$ Discord. The robustness of these correlations is tested through noisy decoherence channels. We show that the quantum teleportation fidelity remains above the classical threshold of $2/3$, even under significant decoherence.

The article is organized as follows. Section~\ref{sec:2} introduces the spin densities for top quark pair annihilation and gluon fusion processes, from which we derive the physical model of the $  t\bar{t}  $ system. Section~\ref{sec:3} presents the quantum correlation measures, analyzing these quantities as functions of various kinematic parameters. The dynamics of the quantum correlation measures under different dephasing noises are then discussed and compared in Section~\ref{sec:4}. In Section~\ref{sec:5}, we investigate the fidelity of quantum teleportation through various noisy channels. Finally, Section~\ref{sec:6} provides a concluding summary.
\section{Spin density matrix} \label{sec:2}
The state of a spin-$\frac{1}{2}$ particle pair (e.g., $t\bar{t}$) is generally given by the following density matrix
\begin{equation}
\rho_{{t\bar{\rm t}}} = \frac{1}{4}\sum_{\mu,\nu}^{3}C_{\mu\nu}\sigma^{t}_{\mu} \otimes \sigma^{\bar{t}}_{\nu},
\label{eq:1}
\end{equation}
The formalism employs a set of four Pauli matrices, $\sigma^{t}_{\mu}$ and $\sigma^{\bar{t}}_{\nu}$, defined in the rest frames of the top and antitop quark, with $C_{\mu,\nu}$ representing a $4\times 4$ real matrix encoding spin correlations and polarizations. The spin operators are expressed in coordinate frames with axes $\mathbf{\hat{k}},\mathbf{\hat{n}},\mathbf{\hat{r}}$, where $\mathbf{\hat{n}}$ is given by

\begin{equation}
\hat{\boldsymbol{\rm n}} = \frac{\hat{\boldsymbol{\rm P}} \times \hat{\boldsymbol{\rm k}}}{|\hat{\boldsymbol{\rm P}} \times \hat{\boldsymbol{\rm k}}|}, \quad \boldsymbol{\hat{\rm k}} = \boldsymbol{\hat{\rm n}} \times \boldsymbol{\hat{\rm r}},
\label{eq:haty}
\end{equation}
Here, $\mathbf{\hat{P}}$ denotes the direction of the initial incoming beam, while $\mathbf{\hat{k}}$ represents the momentum direction of the produced top quark.
In the case of top--antitop pair production at hadron colliders through QCD interactions, parity conservation implies that no net polarization appears at leading order, namely $\text{P}^{t}=\text{P}^{\bar{t}}=0$. Nevertheless, at the one-loop level, absorptive contributions generate a small polarization component perpendicular to the scattering plane \citep{SD1}. Furthermore, the structure of the correlation matrix $C_{\mu,\nu}$ is strongly restricted by the underlying symmetries of QCD. \citep{SD2}. The non-zero elements for quark–antiquark annihilation they are given by \cite{SD2,SD3}
\begin{equation}
\begin{aligned}
C_{k,k}&=\frac{(-1 + z^2)\left(-\beta^2 + 2 z^2 \left(-2 + \beta^2 + 2\sqrt{1-\beta^2}\right)\right)}{2-\beta^2\Big(1-z^2\Big)},\\
C_{k,r}&=\frac{-2 z \sqrt{1 - z^2}\left(1 - \beta^2 - \sqrt{1 - \beta^2} + z^2\left(-2 + \beta^2 + 2\sqrt{1 - \beta^2}\right)\right)}{2-\beta^2\Big(1-z^2\Big)},\\
C_{n,n}&=\frac{(-1 + z^2)\beta^2}{2-\beta^2\Big(1-z^2\Big)},\\
C_{r,r}&=\frac{2 - \beta^2 
- 2 z^4\left(-2 + \beta^2 + 2\sqrt{1-\beta^2}\right)
+ z^2\left(-4 + 3\beta^2 + 4\sqrt{1-\beta^2}\right)}{2-\beta^2\Big(1-z^2\Big)},
\end{aligned}
\end{equation}

For the gluon fusion process, the non-zero elements of matrix are given by
 
\begin{equation}
\begin{aligned}
C_{k,k}&=\frac{-2 + 4\beta^2 - 2(2 - 2z^2 + z^4)\beta^4 
+ 4(1 - z^2)\beta^2\left(-1 + z^4\beta^4 + 2z^2(-1 + z^2)\sqrt{1 - \beta^2} - 2(-1 + z^2)(z^2 + \beta^2)
\right)}{2 - 2 z^4 \beta^4 + 4(-1 + z^2)\beta^4(-1 + \beta^2)},\\
C_{k,r}&=\frac{4z\sqrt{1-z^2}\,\beta^2\left(
-2 + 2\beta^2 + \sqrt{1-\beta^2} + z^2\left(3 - 2\beta^2 - 3\sqrt{1-\beta^2}\right) + z^4\left(-2 + \beta^4 + 2\sqrt{1-\beta^2}\right)\right)}{2 - 2 z^4 \beta^4 + 4(-1 + z^2)\beta^4(-1 + \beta^2)},\\
C_{n,n}&=\frac{-2 + 4\beta^2 - 2(2 - 2z^2 + z^4)\beta^4}{2 - 2 z^4 \beta^4 + 4(-1 + z^2)\beta^4(-1 + \beta^2)},\\
C_{r,r}&=\frac{-2 - 2 (2 - 6 z^2 + 5 z^4)\beta^4 
+ 4 z^6 \beta^6  + 8 (-1 + z^2)\beta^2 \Big(-1 + z^2(-1 + z^2)(-1 + \sqrt{1 - \beta^2})\Big)}{2 - 2 z^4 \beta^4 + 4(-1 + z^2)\beta^4(-1 + \beta^2)},
\end{aligned}
\end{equation}
where $z$ is the cosine of the production angle $\theta$ ($z = \cos \theta$), and $\beta$ is the top-quark velocity given by 
\begin{equation}
\beta =\sqrt{1-\frac{4 m^{2}_{t}}{M^{2}_{t\bar{t}}}}
\end{equation} 
where $M_{t\bar{t}}$ denotes the invariant mass of the top--antitop system, while $m_t$ represents the top-quark mass.
It follows directly that the threshold production regime ($\beta=0$) corresponds to $M_{t\bar{t}} = 2m_t \approx 346~\text{GeV}$, which is the minimum invariant mass required for the production of a $t\bar{t}$ pair.
For convenience, the two-qubit state from Eq. (\ref{eq:1}) can be converted into an X-state by diagonalizing the matrix $C_{i,j}$. The resulting spin density matrix can be expressed as:

\begin{equation}
\rho^X = \frac{1}{4} \left( \mathbb{I}\otimes \mathbb{I} + \sum_{i=1}^{3} \mathbb{C}_{i} \sigma_i \otimes \sigma_i \right)
\label{varrhoX}
\end{equation}

The matrix $C^{\rm X}_{\mu\nu}$ for this state becomes
\begin{equation}
C_{\mu\nu}^{\rm X} =
\begin{pmatrix}
1 & 0 & 0 & 0 \\
0 & \mathbb{C}_1 & 0 & 0 \\
0 & 0 & \mathbb{C}_2 & 0 \\
0 & 0 & 0 & \mathbb{C}_3
\end{pmatrix}
\quad \text{where} \quad
\begin{aligned}
\mathbb{C}_{1,2} &= \frac{
C_{k,k} + C_{r,r}\pm\sqrt{\bigl(C_{k,k} - C_{r,r}\bigr)^2 + 4 C_{k,r}^2}}{2}, \\
\mathbb{C}_3 &= C_{n,n},
\end{aligned}
\label{eq:11}
\end{equation}

We note that $\mathbb{C}_3=C_{n,n}$, $\mathbb{C}_1$ and $\mathbb{C}_2$ come from diagonalizing the block matrix of $C_{ij}$, where $i,j = k,r$ in $C_{\mu\nu}$.

By applying Eq. (\ref{varrhoX}), the bipartite spin density operator can be explicitly reformulated in terms of the $\sigma_z$ eigenbasis. We rewrite the spin density operator for the quark-antiquark system in the $\sigma_{z}$ basis 
\begin{equation}
\rho_{t\bar{t}}^{\text{X}} =
\begin{pmatrix}
\rho_{1,1}& 0 & 0 & \rho_{1,4} \\
0 & \rho_{2,2} & \rho_{2,2} & 0 \\
0 & \rho_{2,2} & \rho_{2,2}& 0 \\
\rho_{1,4} & 0 & 0 & \rho_{4,4}
\end{pmatrix}, \quad \text{where} \quad
\begin{aligned}
 \rho_{1,1}&= \frac{1}{4}\bigg(1+\mathbb{C}_{3}\bigg), \quad
\rho_{1,4} = \rho_{4,1} = \frac{1}{4}\bigg(\mathbb{C}_{1}-\mathbb{C}_{2}\bigg), \\
\rho_{2,2} &= \rho_{3,3} = \frac{1}{4}\bigg(1-\mathbb{C}_{3}\bigg), \quad
\rho_{2,3} = \rho_{3,2} = \frac{1}{4}\bigg(\mathbb{C}_{1}+\mathbb{C}_{2}\bigg), \\
\rho_{4,4} &= \frac{1}{4}\bigg(1+\mathbb{C}_{3}\bigg).
\end{aligned}
\label{eq:DS}
\end{equation}

he nonzero elements of the top–antitop spin density matrix arising from quark–antiquark annihilation are given by  
\begin{equation}
\begin{aligned}
\rho_{1,1}&=\frac{1 + (-1 + z^2) \beta^2}{4 + 2 (-1 + z^2) \beta^2}, \quad
\rho_{1,4} = \rho_{4,1} = \frac{\sqrt{\bigl(1 + (-1 + z^2) \beta^2\bigr)^2}}{4 + 2 (-1 + z^2) \beta^2}, \\
\rho_{2,2} &= \rho_{3,3} = \frac{1}{4 + 2 (-1 + z^2) \beta^2}, \quad
\rho_{2,3} = \rho_{3,2} =\frac{1}{4 + 2 (-1 + z^2) \beta^2}, \\
\rho_{4,4} &= \frac{1 + (-1 + z^2) \beta^2}{4 + 2 (-1 + z^2) \beta^2}.
\end{aligned}
\label{eq:qq}
\end{equation}

Similarly, the nonzero elements of the top–antitop spin density matrix from gluon fusion are given by 

\begin{equation}
\begin{aligned}
\rho_{1,1}&=\frac{\beta^2 \left(-2 + z^2 + (2 - 2z^2 + z^4)\,\beta^2 \right)}{-2 + 4(-1 + z^2)\,\beta^2 + 2(2 - 2z^2 + z^4)\,\beta^4},\quad, \rho_{4,4} = \frac{\beta^2\left(-2 + z^2 + (2 - 2z^2 + z^4)\beta^2\right)}{-2 + 4(-1 + z^2)\beta^2 + 2(2 - 2z^2 + z^4)\beta^4},\\
\rho_{1,4} &= \rho_{4,1} = -\frac{\sqrt{\beta^4 \left(
4 - 8z^2 + 5z^4 + 2(-4 + 8z^2 - 6z^4 + z^6)\,\beta^2
+ (2 - 2z^2 + z^4)^2\,\beta^4\right)}}{-2 + 4(-1 + z^2)\,\beta^2 + 2(2 - 2z^2 + z^4)\,\beta^4},\\
\rho_{2,2} &= \rho_{3,3} = \frac{-1 + z^2 \beta^2}
{-2 + 4(-1 + z^2)\beta^2 + 2(2 - 2z^2 + z^4)\beta^4}, \quad
\rho_{2,3} = \rho_{3,2} \frac{1 + (-2 + z^2)\beta^2}
{-2 + 4(-1 + z^2)\beta^2 + 2(2 - 2z^2 + z^4)\beta^4},
\end{aligned}
\label{eq:gg}
\end{equation}

\section{Quantum resources} \label{sec:3}
\subsection{Bell non-locality}

The nonlocality of Bell represents one of the most distinctive and fundamental features of quantum mechanics. It demonstrates the existence of correlations between entangled subsystems that cannot be reproduced by any local hidden variable theory, thereby ruling out the classical notion of local realism.

For bipartite $2 \times 2$ quantum systems, the detection of nonlocality is equivalent to verifying whether a density matrix $\rho$ violates the Bell-CHSH inequality (Clauser--Horne--Shimony--Holt). According to the Horodecki criterion, the maximum value of the CHSH correlator for a given quantum state is \citep{Be1,Be2}
\begin{equation}
    \mathcal{B}_{\rm CHSH} = \sqrt{\max_{i<j}\big(\vartheta_{i}+\vartheta_{j}\big)},
\end{equation}
where $\vartheta_1 \geq \vartheta_2 \geq \vartheta_3$ are the eigenvalues of the matrix $U = C^T C$, and $C$ is the correlation matrix with elements $C_{ij} = \operatorname{Tr}[\rho (\sigma_i \otimes \sigma_j)]$ for $i,j = 1,2,3$. The three eigenvalues are given by
\begin{equation}
\begin{aligned}
\vartheta_1 &= 4\big(|\rho_{14}| + |\rho_{23}|\big)^2, \notag \\
\vartheta_2 &= 4\big(|\rho_{14}| - |\rho_{23}|\big)^2, \\
\vartheta_3 &= 4\big(\rho_{11} - \rho_{22} - \rho_{33} + \rho_{44}\big)^2.
\end{aligned}
\end{equation}

Since $\vartheta_1 \geq \vartheta_2$ always holds, the maximum CHSH violation for an X-state simplifies to \cite{Be3,Be4,Be5}
\begin{equation}
\mathcal{B}_{\rm CHSH} = 2 \max\bigl\{ \sqrt{\vartheta_1 + \vartheta_2},\ \sqrt{\vartheta_1 + \vartheta_3} \bigr\}.
\end{equation}

A convenient normalized measure of nonlocality can then be defined as 
\begin{equation}
\mathtt{B}(\rho) = \max\left[0,\ \frac{\mathcal{B}_{\rm CHSH}-2}{2\sqrt{2}-2}\right].
\end{equation}

This index equals zero for states compatible with local realism (no Bell violation) and reaches its maximum value of 1 at the Tsirel'son bound ($2\sqrt{2}$), corresponding to the strongest possible quantum violation. It thus provides a clear, continuous, and standardized quantification of Bell nonlocality for two-qubit X-states.

\begin{figure*}[!h]
\begin{center}
\includegraphics[scale=0.4]{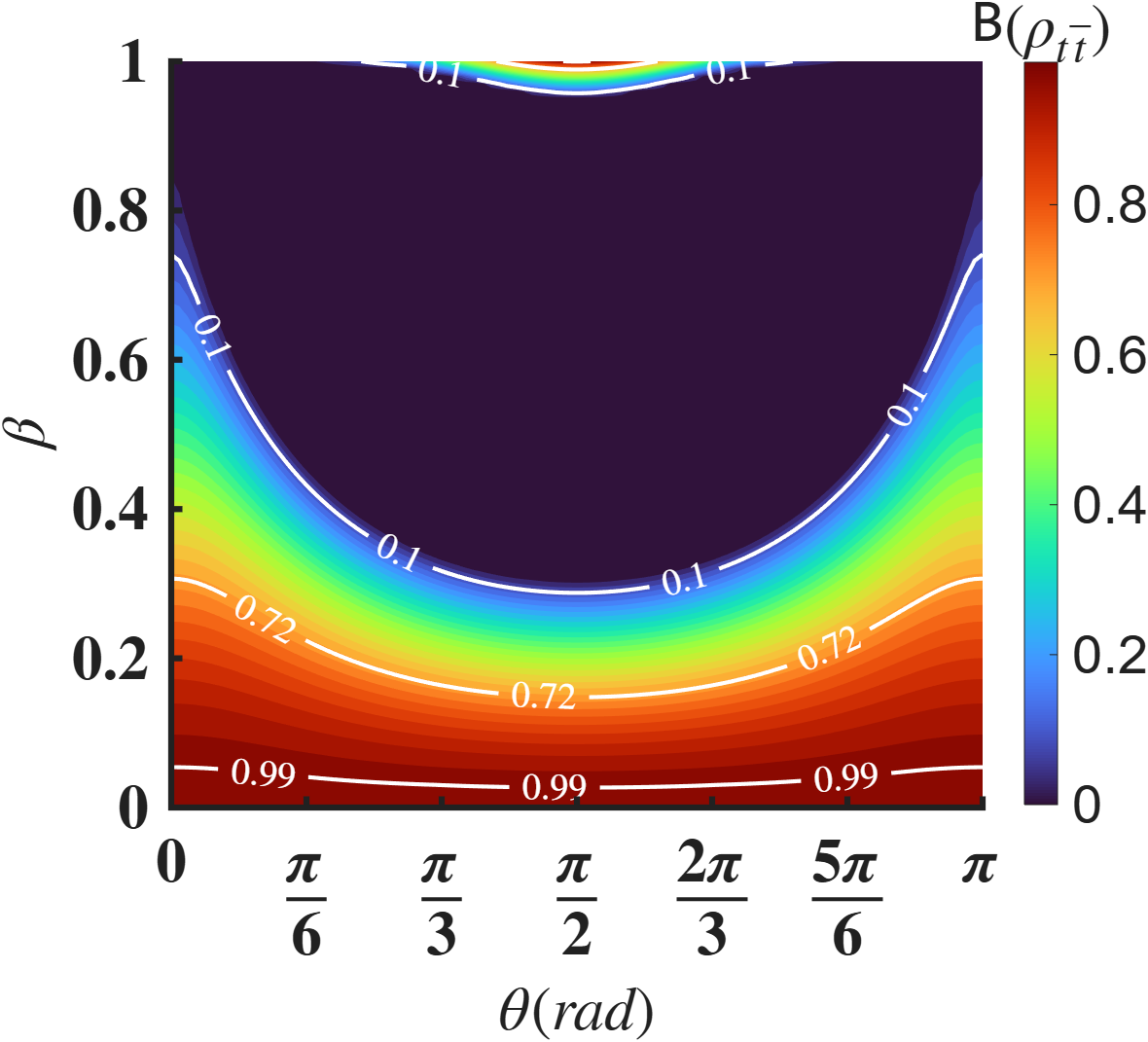}
\includegraphics[scale=0.4]{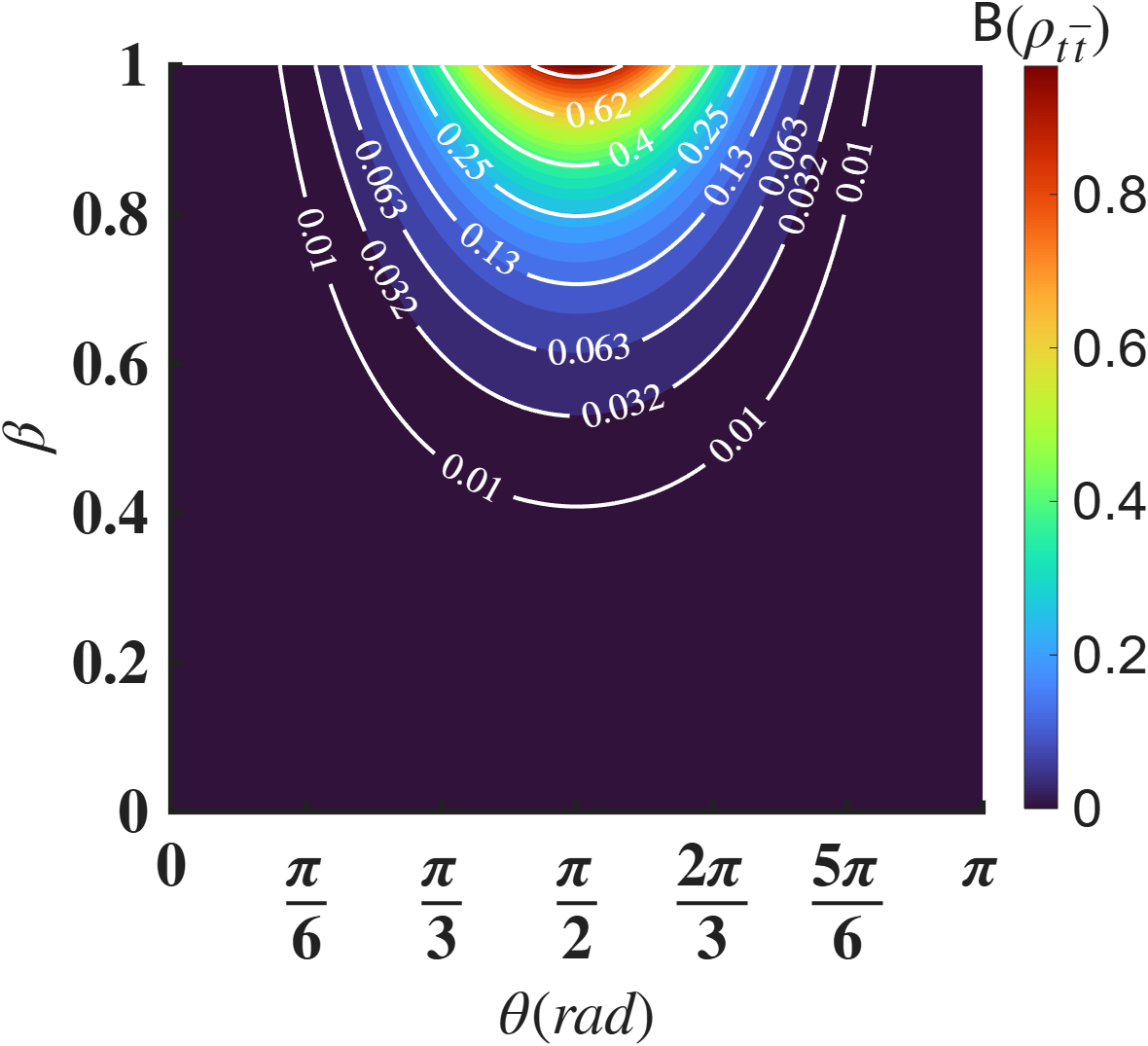}
\put(-420,200){{\bfseries (a)}}
\put(-190,200){{\bfseries (b)}}
\end{center}
\caption{Bell nonlocality as a function of the top-quark velocity $\beta$ and the production angle $\theta$. (a) Gluon-fusion channel $gg\to t\bar{t}$; (b) top-antitop quark annihilation channel $q\bar{q}\to t\bar{t}$, production at the LHC for $\text{p}\text{p}$ collisions at $\sqrt{s}=13\text{Tev}$}
\label{fig:e1}
\end{figure*}

Fig. \ref{fig:e1}(a) presents the results for the gluon fusion channel ($gg \to t\bar{t}$), where Bell non-locality depends on both the production angle $\theta$ and the top-quark velocity $\beta$. We observe that Bell non-locality decreases progressively as $\theta$ increases, reaching a maximum at $\theta = \pi/2$ and $\beta = 0$, which corresponds to the production threshold region $M_{t\bar{t}} = 2m_t$. In this limit, the top quarks are produced nearly at rest, which favors maximal spin correlations. In contrast, as the energy increases and $\beta \to 1$, Bell non-locality again reaches a maximal behavior for $\theta = \pi/2$, indicating a transition toward the ultra-relativistic regime.

Fig. \ref{fig:e1}(b) presents the quark–antiquark annihilation channel ($q\bar{q} \to t\bar{t}$). Bell non-locality is shown as a function of the production angle $\theta$ and the top-quark velocity $\beta$. We observe that the maximum of $\mathtt{B}(\rho_{q\bar{q}})$ is reached for $\beta = 1$ and $\theta = \pi/2$ (upper-right corner of the figure). This situation corresponds to the ultra-relativistic limit where $\frac{2m_t}{M_{t\bar{t}}} \to 0$, i.e., when $M_{t\bar{t}} \gg 2m_t$. In this kinematic regime, the helicity states of the top quark and the anti-top become strongly correlated, leading to a state close to a maximally entangled Bell state, and thus to a maximal violation of Bell’s inequality. We observe that the behavior of Bell non-locality converges to that of the pure $gg \to t\bar{t}$ case presented in Fig. 1(a), which confirms the robustness of the structure of quantum correlations over the entire kinematic domain.
\subsection{Quantum steering}

Quantum steering, a phenomenon that captures non-local correlations intermediate between entanglement and Bell nonlocality, can be characterized through the violation of specific steering inequalities. In this study, we employ the CJWR steering inequality, introduced by Cavalcanti, Jones, Wiseman, and Reid \citep{ST1, ST2}, to diagnose the steerability of the $\Lambda\bar{\Lambda}$ system. This criterion is particularly effective for two-qubit states subjected to three orthogonal measurement settings on each side ($N=3$). The steering function is defined as:

\begin{equation}
F_{3}^{\rm CJWR}(\rho)=\frac{1}{\sqrt{3}}\Bigg|\sum_{i=1}^{3}{\rm Tr}\left[ \rho\left( A_{i}\otimes B_{i}\right) \right] \Bigg|\leq 1,
\label{eq:CJWR}
\end{equation}

where $A_{i}=\mathbf{\hat{u}}_{i}\cdot\boldsymbol{\sigma}$ and $B_{i}=\mathbf{\hat{v}}_{i}\cdot\boldsymbol{\sigma}$ represent the projection operators for Alice and Bob, respectively, with $\boldsymbol{\sigma}$ being the vector of Pauli matrices. The unit vectors $\mathbf{\hat{u}}_{i}$ and orthonormal vectors $\mathbf{\hat{v}}_{i}$ define the local measurement directions.

The steerability of the state $\rho$ is fundamentally governed by its $3\times 3$ correlation matrix $C$, with elements $C_{ij} = {\rm Tr}[\rho(\sigma_i \otimes \sigma_j)]$. To extract the maximum steering violation, the measurement axes must be optimized relative to the principal axes of $C$. By setting $\parallel C\mathbf{\hat{v}}_{i}\parallel = \sqrt{{\rm Tr}(CC^{\rm T})/3}$ and choosing $\mathbf{\hat{u}}_{i}$ to align with the steered states \citep{ST3}, the function $F_{3}^{\rm CJWR}$ reaches its maximum:

\begin{equation}
\mathcal{F}_{3}(\rho)=\max_{\mathbf{\hat{u}}_{i},\mathbf{\hat{v}}_{i}}\left[ F_{3}^{\rm CJWR}(\rho)\right]  =\sqrt{{\rm Tr}\left( C C^{\rm T}\right)}
\end{equation}

For the specific case of $t\bar{t}$ pairs, which are described by an X-shaped density operator due to parity and CP conservation in the production process, the maximal violation simplifies to $\mathcal{F}_{3}(\rho_{\Lambda\bar{\Lambda}})=\sqrt{\mathbb{C}^{2}_{1}+\mathbb{C}^{2}_{2}+\mathbb{C}^{2}_{3}}$. Here, $\mathbb{C}_i$ correspond to the spin-correlation parameters along the principal axes.

Since a state is considered steerable only if $\mathcal{F}_{3} > 1$, and given that the theoretical maximum for a singlet state is $\mathcal{F}_{3}^{\max}=\sqrt{3}$, we introduce a modified measure $\mathtt{S}(\rho_{t\bar{t}})$ to quantify the degree of steering:

\begin{equation}
\mathtt{S}(\rho_{t\bar{t}})=\max\left\lbrace 0,\frac{\mathcal{F}_{3}(\rho_{\Lambda\bar{\Lambda}})-1}{\sqrt{3}-1}\right\rbrace 
\end{equation}

This normalized measure ensures that $\mathtt{S} = 0$ at the steering boundary and $\mathtt{S} = 1$ for maximally steerable states, providing a robust metric for comparing quantum correlations across different kinematic regions in hyperon decay experiments.

\begin{figure*}[t]
\includegraphics[scale=0.4]{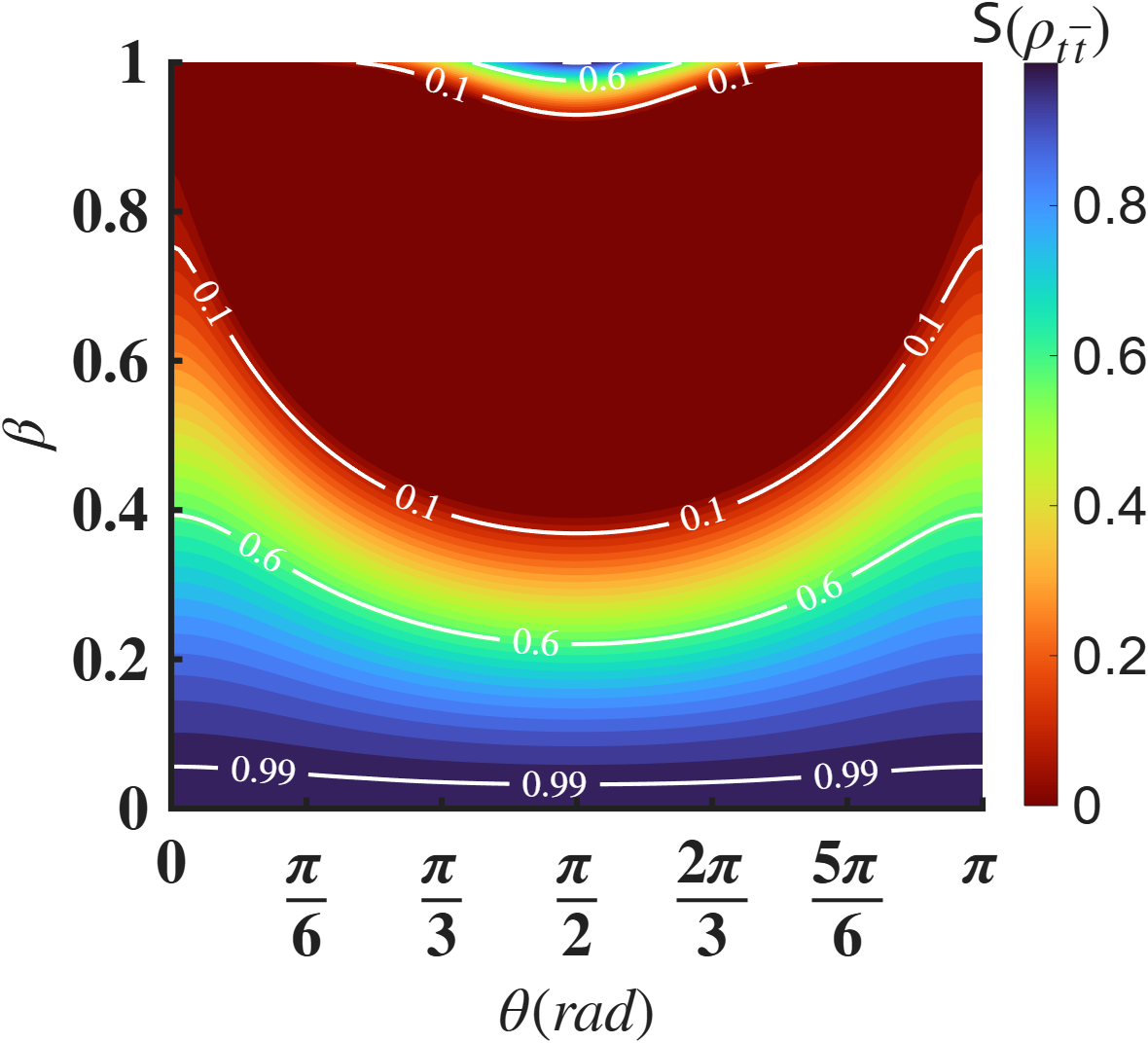}
\includegraphics[scale=0.4]{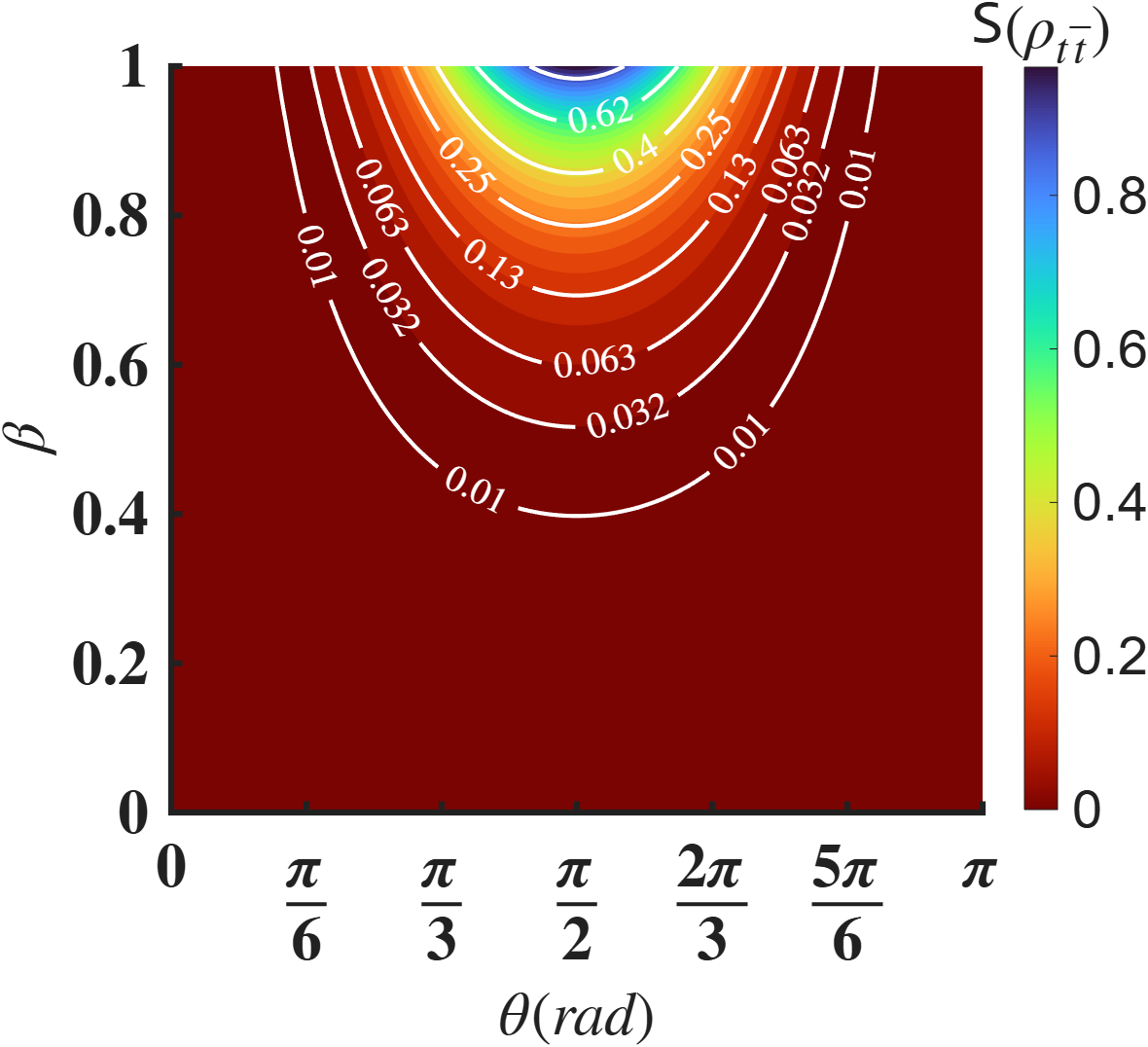}
\caption{Quantum steering as a function of the top-quark velocity $\beta$ and the production angle $\theta$. (a) Gluon-fusion channel $gg\to t\bar{t}$; (b) top-antitop annihilation channel $q\bar{q}\to t\bar{t}$.}
\label{fig:e2}
\end{figure*}

We now focus on the behavior of quantum steering in the two production channels as a function of the production angle $\theta$ and the top-quark velocity $\beta$.

In Fig.~\ref{fig:e2}(a), corresponding to the gluon fusion channel ($gg \to t\bar{t}$), quantum steering exhibits a smooth dependence on both $\theta$ and $\beta$. Strong steering is observed around $\theta = \pi/2$, where spin correlations are maximized due to the symmetry of the production process. In the threshold region ($\beta \to 0$), the top quarks are produced nearly at rest, which enhances spin correlations and leads to a significant steering signal. As $\beta$ increases, steering remains non-zero over a wide range of angles, indicating that quantum correlations are preserved even away from the threshold regime.

In Fig.~\ref{fig:e2}(b), for the quark--antiquark annihilation channel ($q\bar{q} \to t\bar{t}$), steering also reaches its highest values in the ultra-relativistic regime ($\beta \to 1$), particularly around $\theta = \pi/2$. This regime is characterized by strong helicity correlations between the top quark and the anti-top, which favor the emergence of steerable quantum states. Unlike the threshold region, where correlations arise from near-static configurations, the ultra-relativistic limit reflects the dominance of dynamical spin alignment driven by the kinematics of the process.

Overall, quantum steering remains significant across a broad kinematic domain in both channels, with pronounced regions around $\theta = \pi/2$ and in the limiting regimes $\beta \to 0$ and $\beta \to 1$. This behavior highlights the persistence of directional quantum correlations in top-quark pair production, governed by both spin dynamics and relativistic effects.
\subsection{Concurrence}

It is well established that concurrence is a practical and widely used measure for quantifying entanglement in bipartite quantum states. Accordingly, we adopt concurrence as the entanglement measure throughout this work. It is defined as \citep{A1,A2}
\begin{equation}
C=\max\left\lbrace 0,\sqrt{\mu_{1}}-\sqrt{\mu_{2}}-\sqrt{\mu_{3}}-\sqrt{\mu_{4}}\right\rbrace ,\mu_{1}>\mu_{2}>\mu_{3}>\mu_{4}
\end{equation}

where $\mu_{i}(i=1,2,3,4)$ are the eigenvalues of the matrix $\mathcal{R}=\rho(\sigma_{y}\otimes\sigma_{y})\rho^{*}(\sigma_{y}\otimes\sigma_{y})$. In the present case, the density matrix has an X-structure. For such states, the concurrence admits a convenient closed-form expression given by \citep{A3}

\begin{equation}\label{crx}
\mathtt{C}(\rho_{t\bar{t}}^{\rm X})=2\max \left\lbrace |\rho_{2,3}|-\sqrt{\rho_{1,1}\rho_{4,4}},\; |\rho_{1,4}|-\sqrt{\rho_{2,2}\rho_{3,3}},\;0\right\rbrace .
\end{equation}

The concurrence ranges from zero to one by construction. A vanishing concurrence indicates that the state is separable, meaning it exhibits no quantum entanglement. Conversely, a strictly positive concurrence signals the presence of entanglement, with higher values reflecting stronger quantum correlations between the two subsystems.

\begin{figure*}[t]
\includegraphics[scale=0.4]{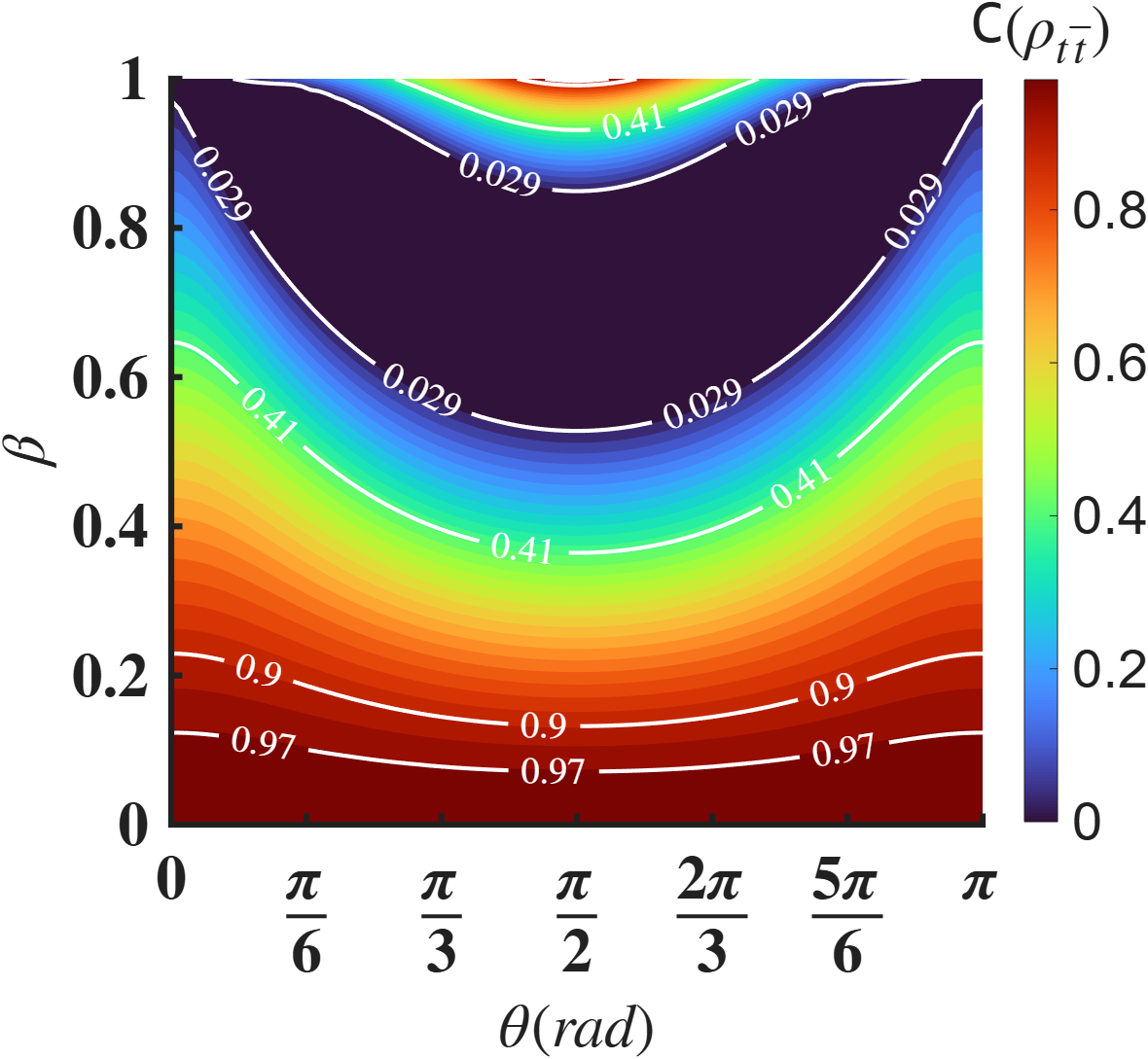}
\includegraphics[scale=0.4]{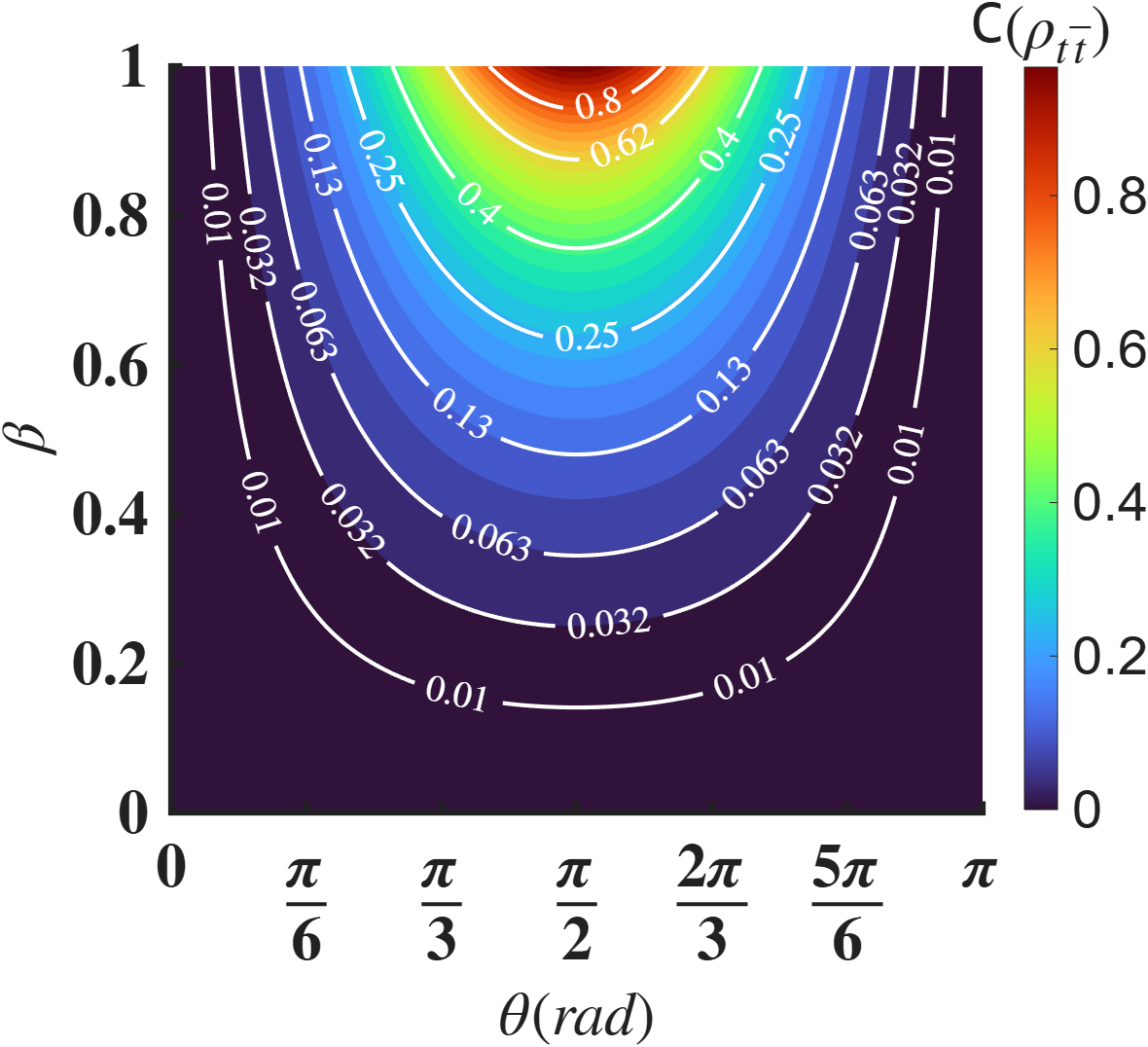}
\caption{Concurrence as a function of the top-quark velocity $\beta$ and the production angle $\theta$. (a) Gluon-fusion channel $gg\to t\bar{t}$; (b) top-antitop quark annihilation channel $q\bar{q}\to t\bar{t}$.}
\label{fig:e3}
\end{figure*}

We analyze the quantum concurrence in the two dominant production channels as a function of the top-quark scattering angle $\theta$ and velocity $\beta$. The concurrence serves as a measure of the overall spin entanglement between the top quark and antiquark.

In the gluon-fusion channel $gg\to t\bar{t}$ [Fig.~\ref{fig:e3}(a)], the concurrence exhibits a strong dependence on both $\theta$ and $\beta$. Near the production threshold ($\beta\to 0$), the $t\bar{t}$ system is nearly pure and displays strong spin correlations, resulting in high entanglement. As $\beta$ increases, the concurrence decreases; however, it remains maximal near $\theta=\pi/2$, particularly in the ultrarelativistic regime ($\beta\to 1$), where the states stay strongly entangled.

For the quark-antiquark annihilation channel $q\bar{q}\to t\bar{t}$ [Fig.~\ref{fig:e3}(b)], the concurrence becomes more pronounced in the ultrarelativistic limit ($\beta\to 1$), with a maximum also occurring around $\theta=\pi/2$. This behavior arises from helicity conservation at high energies, which enhances the spin correlations.

Overall, significant entanglement is present over a broad kinematic range in both channels. The regions of enhanced concurrence depend on $\theta$ and $\beta$. Notably, the behavior in the full $\text{p}\text{p}$ collision environment converges toward that of the pure $gg\to t\bar{t}$ channel, underscoring the robustness of the quantum correlation structure across the explored kinematic domain.
\subsection{Geometric quantum discord}
The quantum discord provides a complete measure of quantum correlations, capturing all non-classical correlations beyond entanglement alone \citep{d1}. For the top-antitop quark pair state \(\rho_{t\bar{t}}\), it is defined as
\begin{equation}
{\rm Q}(\rho_{t\bar{t}}) = {\rm I}(\rho_t:\rho_{\bar{t}}) - \mathcal{C}(\rho_{t\bar{t}}),
\end{equation}
where \({\rm I}(\rho_t:\rho_{\bar{t}}) = S(\rho_t) + S(\rho_{\bar{t}}) - S(\rho_{t\bar{t}})\) is the mutual information and \(\mathcal{C}(\rho_{t\bar{t}})\) is the maximum classical correlation, obtained by maximizing over all local positive-operator-valued measures (POVMs) \(\{\Pi_k\}\) on one subsystem:
\begin{equation}
\mathcal{C}(\rho_{t\bar{t}}) = \max_{\{\Pi_k\}} \Bigl[S(\rho_{t}) - \sum_k p_k S(\rho_{k|\bar{t}}) \Bigr].
\end{equation}
Since this optimization is computationally demanding even for two-qubit systems, geometric approaches have been developed to quantify quantum correlations more efficiently \citep{d2,d0,d5,d7}.

In this work, we employ the geometric quantum discord (GQD) based on the Schatten 1-norm (trace norm), which quantifies quantum correlations via the minimum distance to the closest classical-quantum state \citep{d10}:
\begin{equation}
\mathcal{D}_G(\rho_{t\bar{t}}) = \min_{\rho_c \in \Omega} \|\rho_{t\bar{t}} - \rho_c\|_1,
\end{equation}
where \(\|\cdot\|_1 = \operatorname{tr}\sqrt{(\cdot)^\dagger(\cdot)}\) is the Schatten 1-norm and \(\Omega\) denotes the set of classical-quantum states of the form
\begin{equation}
\hat{\varrho}_c = \sum_k p_k \, \Pi_{k,t} \otimes \rho_{k,\bar{t}},
\end{equation}
corresponding to local measurements on the top quark. For generic two-qubit X states, which naturally arise in \(t\bar{t}\) spin correlations, the minimization admits an analytical solution. The GQD can then be expressed directly in terms of the Fano-Bloch correlation matrix elements \(\mathtt{S}_{\alpha\beta}\) \cite{d10}:
\begin{equation}
\mathcal{D}_{G}(\hat{\varrho}_{\Lambda\bar{\Lambda}})=\frac{1}{2}\sqrt{\frac{\mathtt{S}_{1,1}^{2}\mathtt{S}_{\max}^2 -\mathtt{S}_{2,2}^{2}\mathtt{S}_{\min}^2}{\mathtt{S}_{\max}^2-\mathtt{S}_{\min}^2 +\mathtt{S}_{1,1}^{2}-\mathtt{S}_{2,2}^{2}}},
\label{eq:Q}
\end{equation}
with \(\mathtt{S}_{\max}^2=\max\left\lbrace \mathtt{S}_{2,2}^{2}+\mathtt{S}_{3,0}^{2},\mathtt{S}_{3,3}^{2}\right\rbrace\) and \(\mathtt{S}_{\min}^2=\min\left\lbrace \mathtt{S}_{1,1}^{2},\mathtt{S}_{3,3}^{2}\right\rbrace\). Here the relevant correlation matrix elements are obtained from the Fano-Bloch decomposition
\begin{equation}
R=\frac{1}{4}\sum_{\alpha,\beta=0}^{3}\mathtt{S}_{\alpha,\beta}\sigma_{\alpha}\otimes\sigma_{\beta},
\end{equation}
where the non-vanishing components are given by 
\begin{align}
\mathtt{S}_{1,1} &= 2(\rho_{2,3} + \rho_{1,4}), &
\mathtt{S}_{2,2} &= 2(\rho_{2,3} - \rho_{1,4}), \nonumber \\
\mathtt{S}_{3,3} &= 1 - 2(\rho_{2,2} + \rho_{3,3}), &
\mathtt{S}_{3,0} &= 2(\rho_{1,1} + \rho_{2,2}) - 1.
\end{align}

\begin{figure*}[!h]
\includegraphics[scale=0.4]{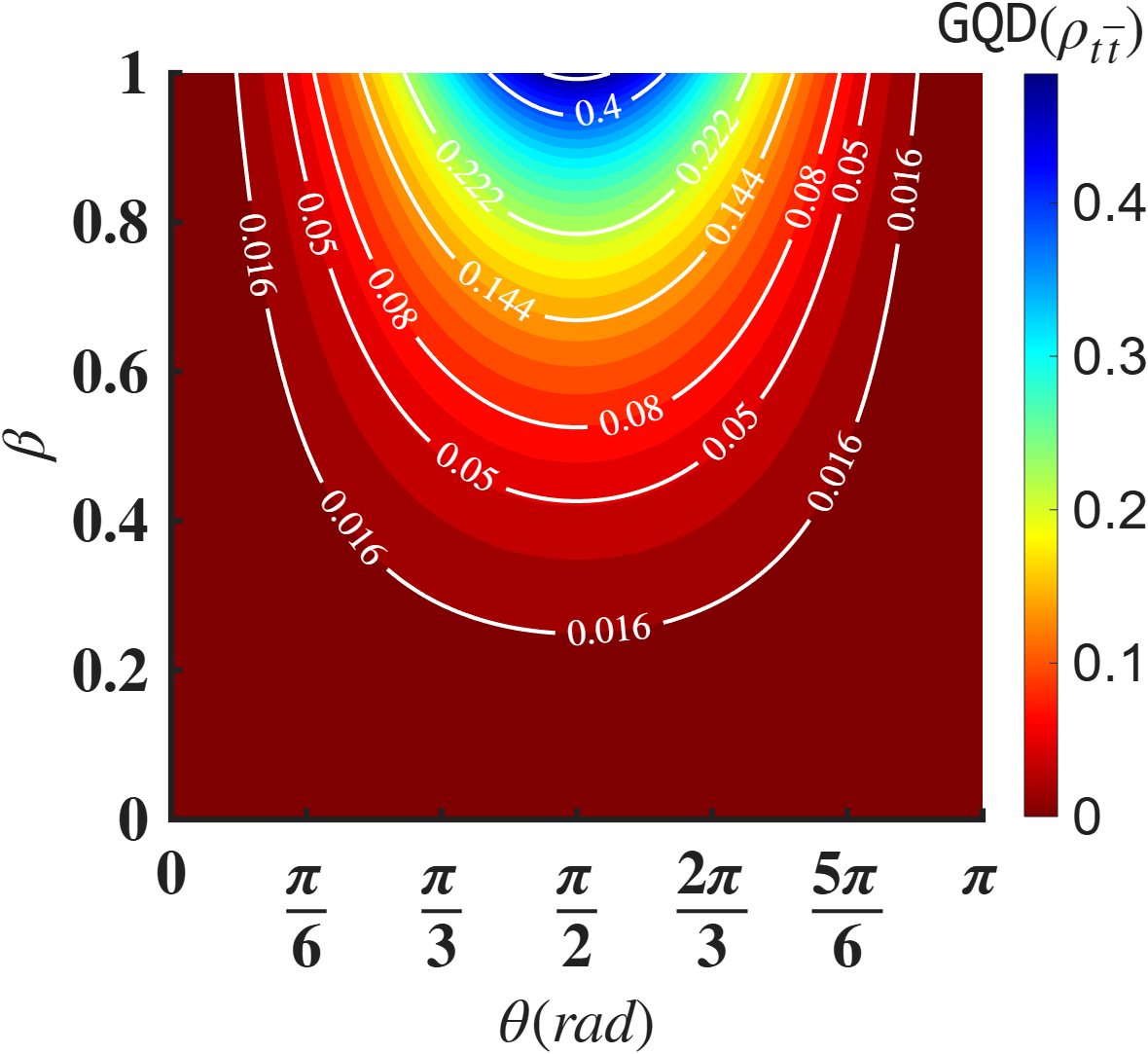}
\includegraphics[scale=0.4]{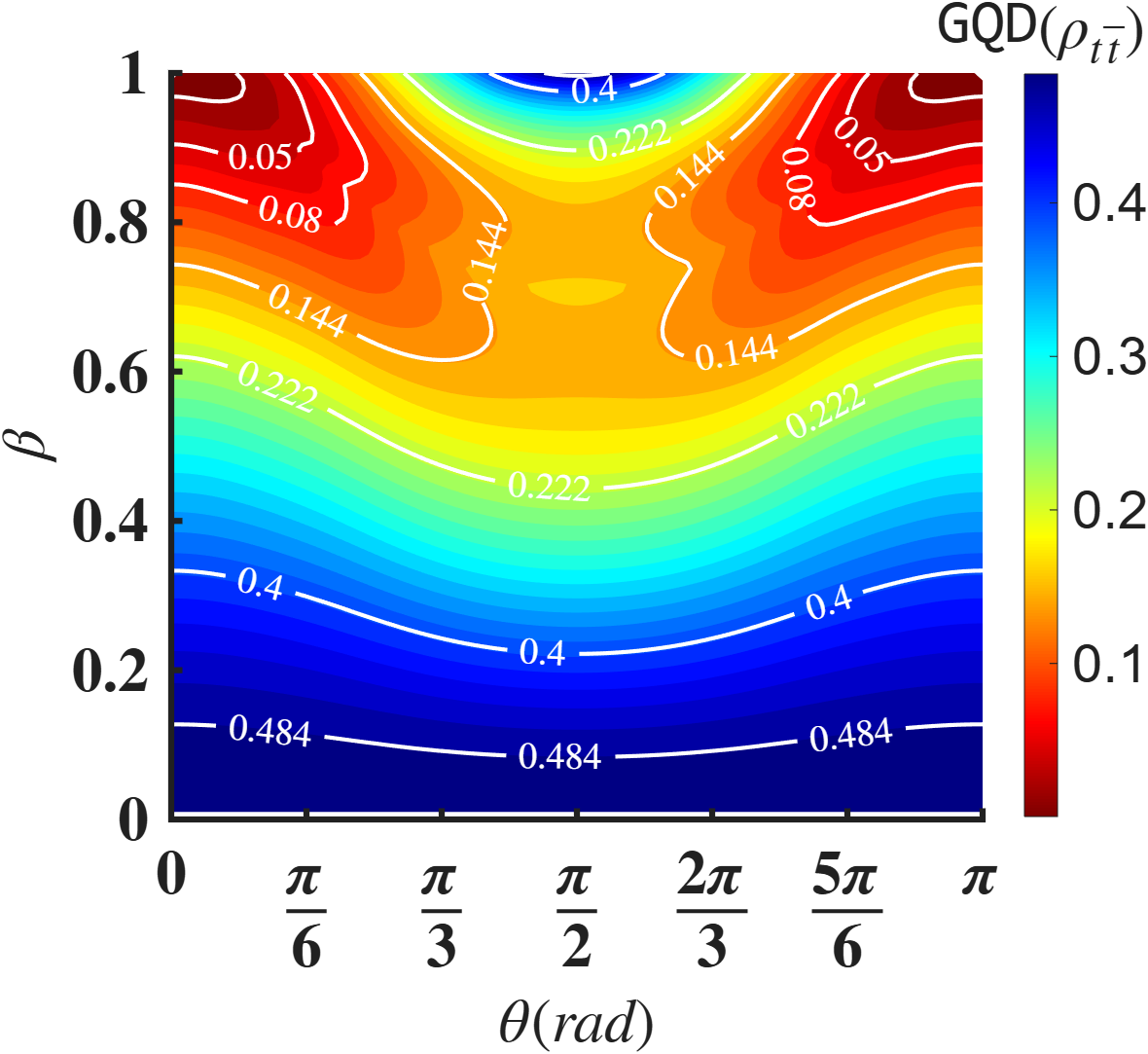}
\caption{Geometric quantum discord as a function of the top-quark velocity $\beta$ and the production angle $\theta$. (a) Gluon-fusion channel $gg\to t\bar{t}$; (b) top-antitop quark annihilation channel $q\bar{q}\to t\bar{t}$.}
\label{fig:e4}
\end{figure*}

We now investigate the behavior of geometric quantum discord (GQD) in the two production channels as a function of the production angle $\theta$ and the top-quark velocity $\beta$. Unlike concurrence, which strictly quantifies entanglement, GQD captures more general quantum correlations, including those present in separable states.

In Fig.~\ref{fig:e4}(a), corresponding to the gluon fusion channel ($gg \to t\bar{t}$), the geometric quantum discord exhibits a pronounced dependence on both $\theta$ and $\beta$. It reaches its maximum around $\theta = \pi/2$ and $\beta \to 1$, where the spin configuration and relativistic effects combine to enhance quantum correlations. Interestingly, GQD remains relatively large even in the threshold region ($\beta \to 0$), indicating that non-classical correlations persist even when entanglement may be reduced. This highlights the robustness of discord-type correlations against variations in the kinematic regime.

In Fig.~\ref{fig:e4}(b), for the quark--antiquark annihilation channel ($q\bar{q} \to t\bar{t}$), a similar behavior is observed. The GQD attains its maximum at $\theta = \pi/2$ and $\beta \to 1$, reflecting the strong spin correlations induced by helicity conservation in the ultra-relativistic limit. The geometric discord remains non-zero over a broad range of parameters, confirming the persistence of quantum correlations beyond entanglement.

Overall, the geometric quantum discord shows that quantum correlations are widely distributed across the kinematic space in both production channels. Its non-vanishing values even in regions where other measures may decrease emphasize that discord provides a more general and robust characterization of quantum correlations in top-quark pair production.

We observe directly from the spin density matrices that, for $\beta=1$ (ultra-relativistic limit) and production angle $\theta=\pi/2$ in the center-of-mass frame, the spin density matrix arising from gluon fusion \(gg\to q\bar{q}\) becomes identical to that from quark-antiquark annihilation $q\bar{q}\to q\bar{q}$

\begin{equation}
\rho_{t\bar{t}}=\frac{1}{2}
\begin{pmatrix}
0 & 0 & 0 & 0\\
0 & 1 & 1 & 0 \\
0 & 1 & 1 & 0 \\
0 & 0 & 0 & 0
\end{pmatrix}
\end{equation}
In this specific kinematic regime, the quark annihilation channel converges to the gluon fusion channel. This equivalence occurs because, at high energies and $\theta=\pi/2$, the transverse helicity contributions dominate strongly, the mass-suppressed terms vanish, and the spin structures projected onto the final-state quarks by the initial gluons and by the virtual quark in the $s$-channel become equivalent, rendering the two production mechanisms indistinguishable in their spin correlations.

\section{Description of the model and noisy channels}\label{sec:4}
The interaction between a quantum system and its surrounding environment generally induces decoherence, leading to the degradation of quantum coherence and the weakening of quantum correlations. To describe these environmental effects, several standard noisy quantum channels are commonly considered, including amplitude damping (AD), phase flip (PF), and phase damping (PD) channels \citep{Ns1}. These channels model different types of noise that alter the quantum state, progressively transforming a pure state into a mixed one. For an initial bipartite quantum state $\rho_{\text{Y}\bar{\text{Y}}}$, the evolved state in the presence of decoherence can be obtained using the Kraus operator formalism as
\begin{equation}
\hat{\varepsilon}(\rho^{\text{X}}_{\text{Y}\bar{\text{Y}}})=\sum_{kl}\hat{K}_{kl} \rho^{\text{X}}_{\text{Y}\bar{\text{Y}}} \hat{K}_{kl}^{\dagger}.
\label{eq:epsilon}
\end{equation}

Here, the Kraus operators are defined as $\hat{K}_{kl} = \hat{K}_k \otimes \hat{K}_l$, where $\hat{K}_k$ and $\hat{K}_l$ describe the local one-qubit quantum channels acting on each subsystem independently. These operators must fulfill the completeness relation
\[
\sum_{kl} \hat{K}_{kl}^\dagger \hat{K}_{kl} = \mathbb{I},
\]
which guarantees the trace-preserving character of the quantum evolution and ensures the physical consistency of the quantum operation.

\begin{table}[H]
\centering
\caption{Kraus operator representations of the amplitude damping (AD), phase flip (PF), and phase damping (PD) quantum channels, where $p$ denotes the decoherence probability.}
\label{tab1}

\begin{tabular}{c c}
\hline
\hline
Channel description & Kraus operators \\
\hline
\hline
AD channel &
$
\hat{K}_1 =
\begin{pmatrix}
1 & 0 \\
0 & \sqrt{1-p}
\end{pmatrix}
\quad , \quad
\hat{K}_2 =
\begin{pmatrix}
0 & \sqrt{p} \\
0 & 0
\end{pmatrix}
$
\\[20pt]

PF channel &
$
\hat{K}_1 =
\begin{pmatrix}
\sqrt{p} & 0 \\
0 & \sqrt{p}
\end{pmatrix}
\quad , \quad
\hat{K}_2 =
\begin{pmatrix}
\sqrt{1-p} & 0 \\
0 & \sqrt{1-p}
\end{pmatrix}
$
\\[20pt]

PD channel &
$
\hat{K}_1 =
\begin{pmatrix}
1 & 0 \\
0 & \sqrt{1-p}
\end{pmatrix}
\quad , \quad
\hat{K}_2 =
\begin{pmatrix}
0 & 0 \\
0 & \sqrt{p}
\end{pmatrix}
$
\\[10pt]
\hline
\end{tabular}
\label{tab1}
\end{table}
The Kraus operators corresponding to the considered noisy quantum channels are summarized in Table~\ref{tab1}. Here, $0 \leq p \leq 1$ denotes the decoherence parameter, which characterizes the strength of the interaction between the quantum system and its surrounding environment, and thus represents the probability of losing system excitation. In the following, we briefly discuss the physical interpretation of each type of noise together with its associated Kraus operators. Furthermore, we investigate the influence of these noisy channels on the quantum teleportation protocol by employing the entangled states introduced in Eq.~(\ref{eq:DS}).

In the preceding section, we examined the behavior of quantum correlations with respect to the production angle $\theta$ and the top-quark velocity $\beta$.

Here, we study the effect of the parameter $p$ on these quantum correlations. For simplicity, we focus on the ultrarelativistic limit $\beta = 1$, in which the $q\bar{q}\to t\bar{t}$ channel approaches the gluon-initiated process ($q\bar{q}\to gg$).
\subsection{Amplitude damping channel}

We now study the effect of the amplitude damping (AD) channel on the system. This type of noise describes the energy exchange between the quantum system and its surrounding environment; for example, an excited atom may emit a photon and decay to a lower energy state. Using Eqs. (\ref{eq:DS}) and Eq. (\ref{eq:epsilon}) together with the Kraus operators given in Table~\ref{tab1}, the elements of the resulting density matrix in the presence of noise can be readily obtained as follows

\begin{equation}
\hat{\rho}^{\text{AD}}_{\text{Y}\bar{\text{Y}}} =
\begin{pmatrix}
\eta_{1,1}& 0 & 0 & \eta_{1,4}  \\
0 & \eta_{2,2} & \eta_{2,3} & 0 \\
0 & \eta_{2,3} & \eta_{3,3}& 0 \\
\eta_{1,4}  & 0 & 0 & \eta_{4,4}
\end{pmatrix}, \quad \text{where} \quad
\begin{aligned}
\eta_{1,1} &= \rho_{1,1} + p\left(2\rho_{2,2} + p\rho_{4,4}\right), \\
\eta_{1,4} &= (1 - p)\rho_{1,4}, \\
\eta_{2,2} &=\eta_{3,3}= -(p - 1)\left(\rho_{2,2} + p\rho_{4,4}\right), \\
\eta_{2,3} &= (1 - p)\rho_{2,3}, \\
\eta_{4,4} &= (p - 1)^2 \rho_{4,4}.
\end{aligned}
\label{eq:AD}
\end{equation}

\subsection{Phase flip channel}

The phase-flip (PF) channel represents an important class of decoherence processes that affect only the phase coherence of the quantum state, while leaving the energy populations unchanged. In this model, the Pauli operator $\sigma_z$ is applied to the first qubit with probability $1-p$, whereas the quantum state remains unchanged with probability $p$.
Under the influence of the PF channel, the density matrix of the system takes the following form

\begin{equation}
\hat{\rho}^{\text{PF}}_{\text{Y}\bar{\text{Y}}}=
\begin{pmatrix}
\rho_{1,1} & 0 & 0 & \rho_{1,4}\Gamma \\
0 & \rho_{2,2} & \rho_{2,3}\Gamma & 0 \\
0 & \rho_{2,3}\Gamma & \rho_{3,3} & 0 \\
\rho_{1,4}\Gamma & 0 & 0 & \rho_{4,4}
\end{pmatrix},
\quad \text{with} \quad \Gamma = (1-2p)^2.
\label{eq:PF}
\end{equation}
\subsection{Phase Damping channel}

The phase-damping (PD) channel is a common model of environmental noise that originates from the interaction between a quantum system and its surroundings. This type of decoherence leads to the loss of relative phase information between the energy eigenstates, while leaving the population terms of the density matrix unchanged. Consequently, quantum coherence gradually decreases without any associated energy exchange with the environment.

By employing Eqs.~(\ref{eq:DS}) and (\ref{eq:epsilon}) together with the Kraus representation given in Table~\ref{tab1}, the noisy density matrix can be explicitly derived as follows:

\begin{equation}
\hat{\rho}^{\text{PD}}_{\text{Y}\bar{\text{Y}}} =
\begin{pmatrix}
\rho_{1,1} & 0 & 0 & \Gamma\rho_{1,4} \\
0 & \rho_{2,2} & \Gamma\rho_{2,3} & 0 \\
0 & \Gamma\rho_{2,3} & \rho_{3,3} & 0 \\
\Gamma\rho_{1,4} & 0 & 0 & \rho_{4,4}
\end{pmatrix},
\quad \text{with} \quad \Gamma = (1-p).
\label{eq:PD}
\end{equation}

\begin{figure*}[t]
\includegraphics[scale=0.28]{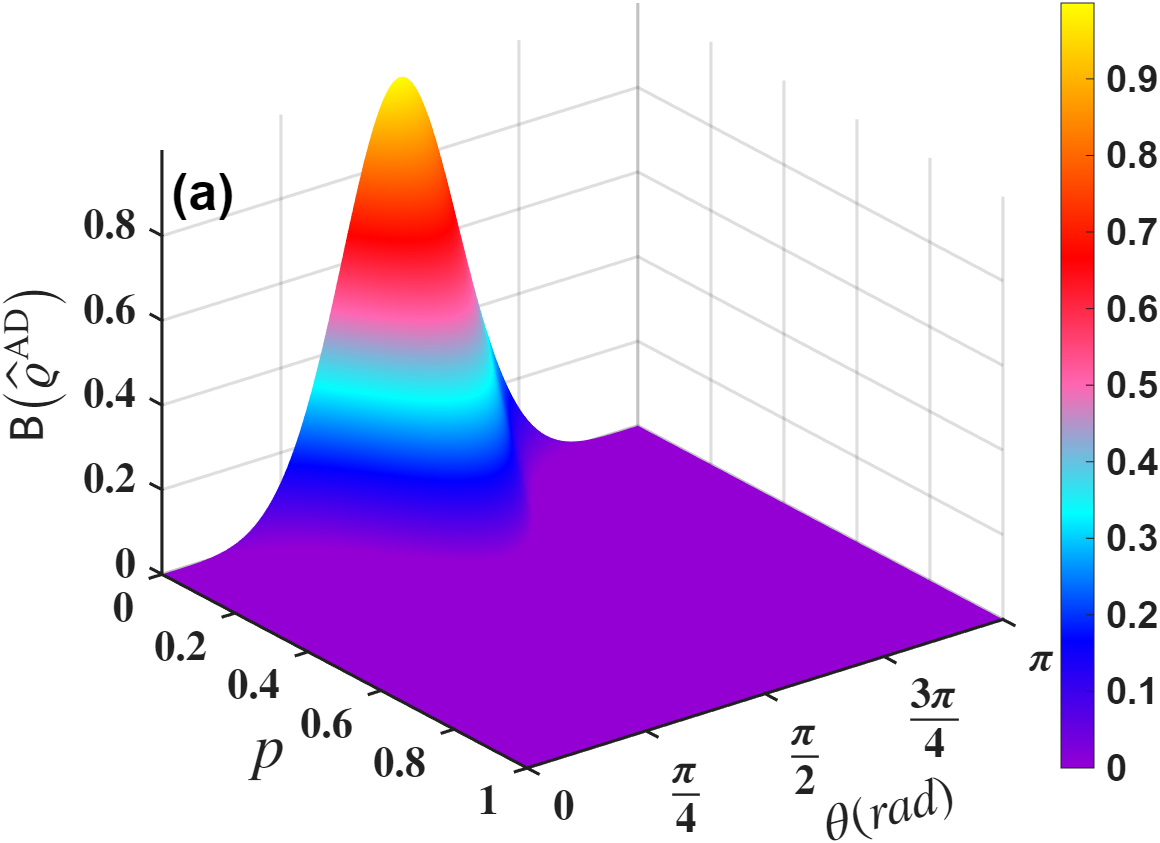}
\includegraphics[scale=0.28]{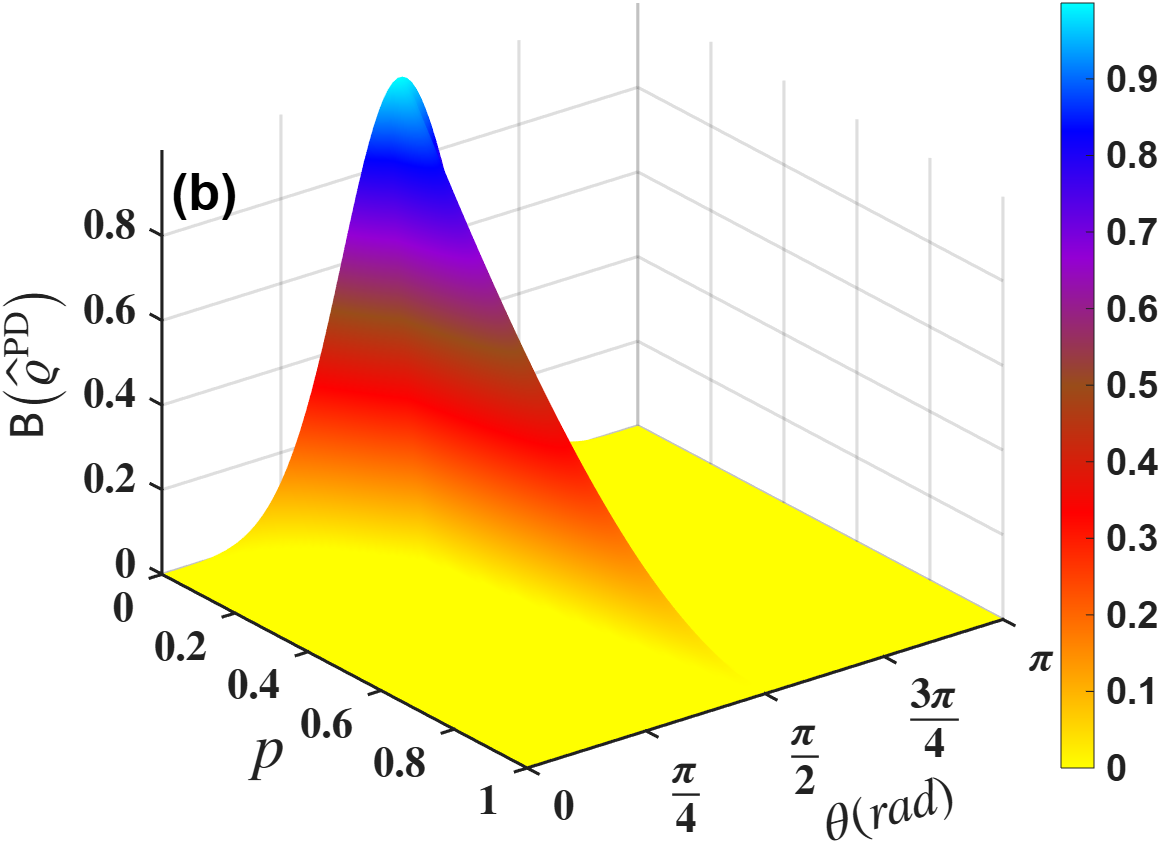}
\includegraphics[scale=0.28]{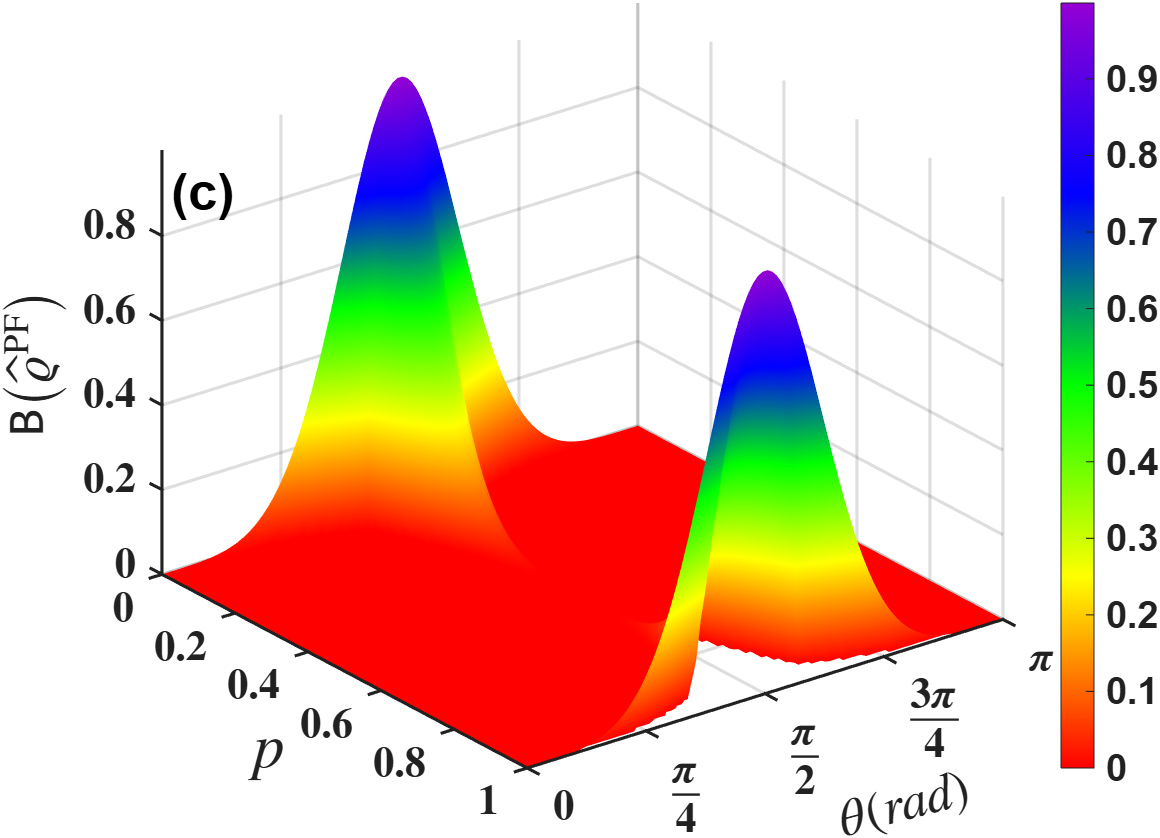}\\
\includegraphics[scale=0.28]{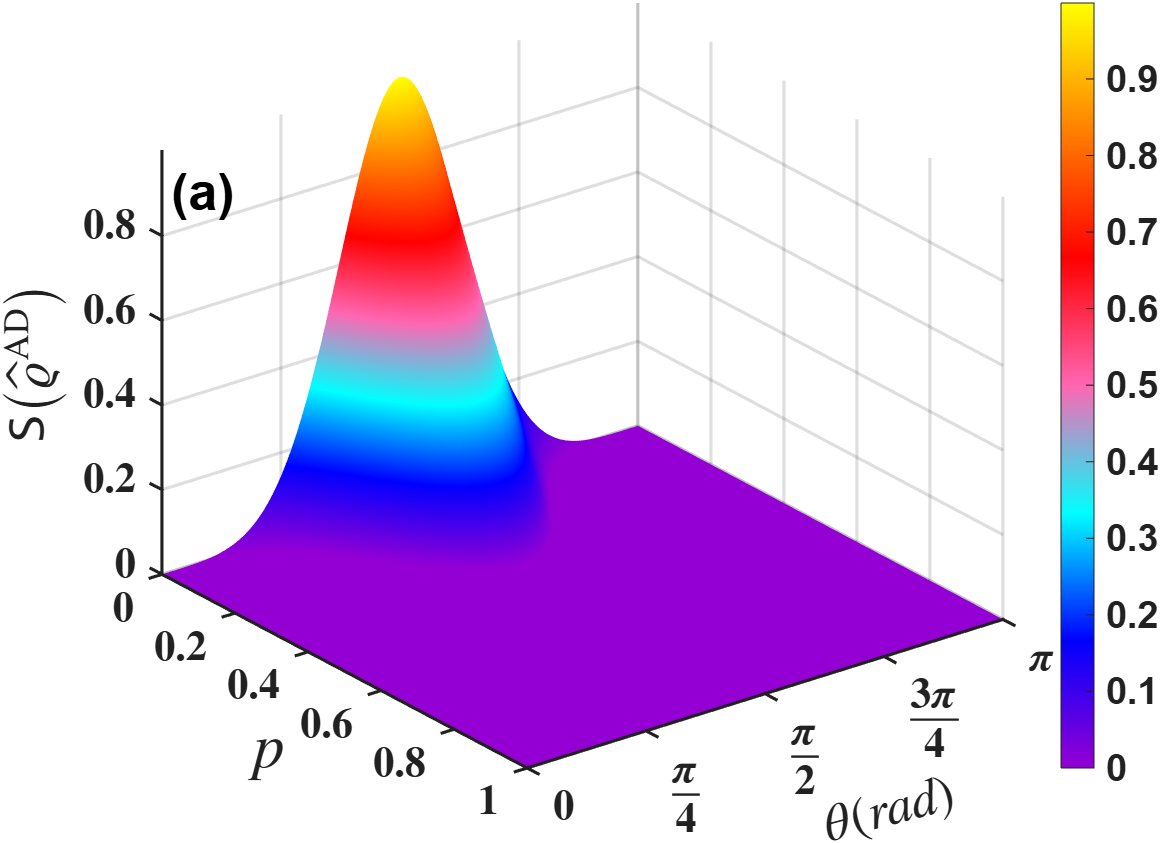}
\includegraphics[scale=0.28]{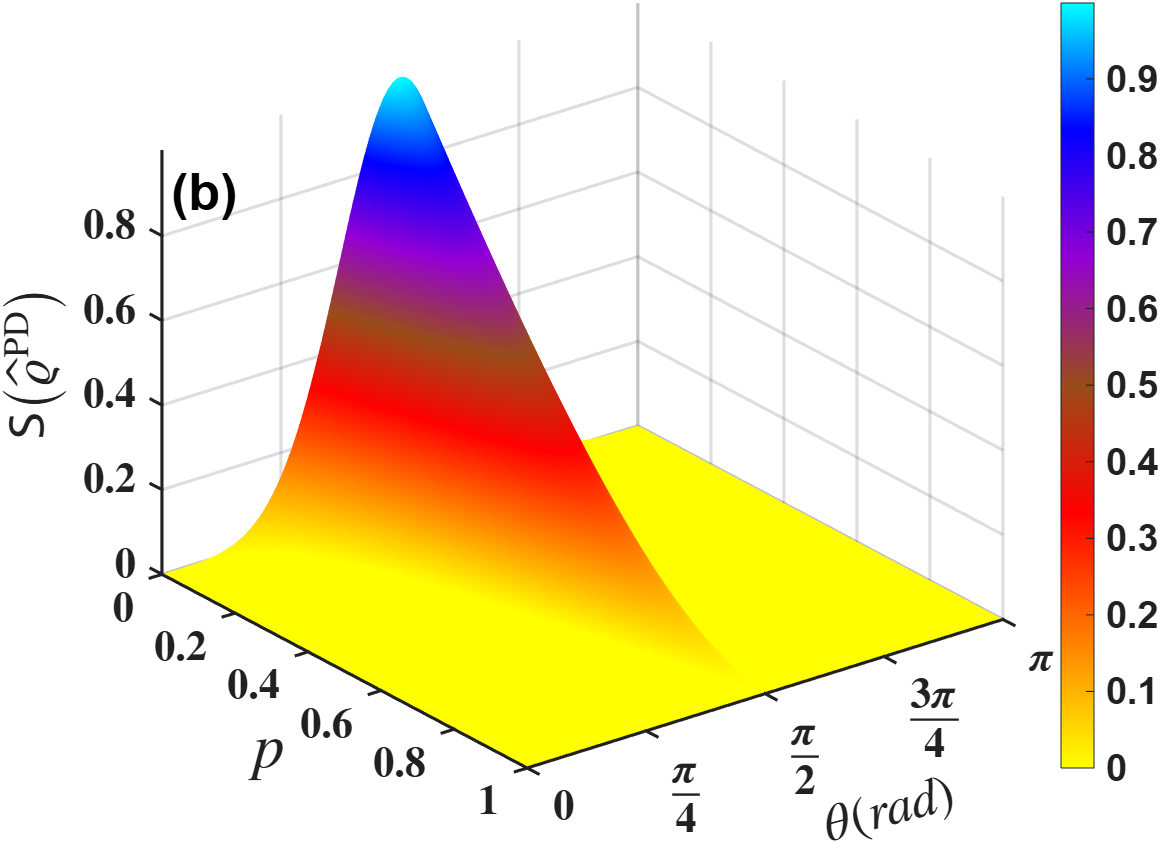}
\includegraphics[scale=0.28]{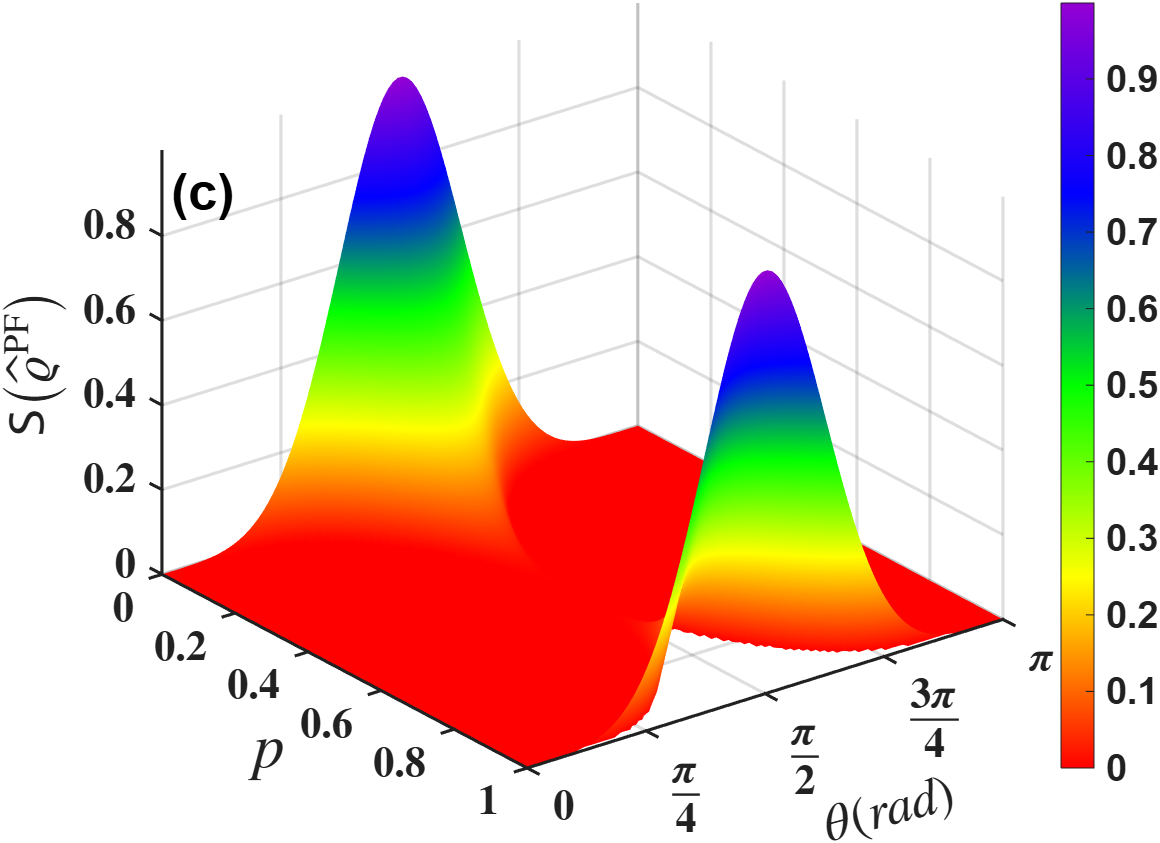}\\
\includegraphics[scale=0.28]{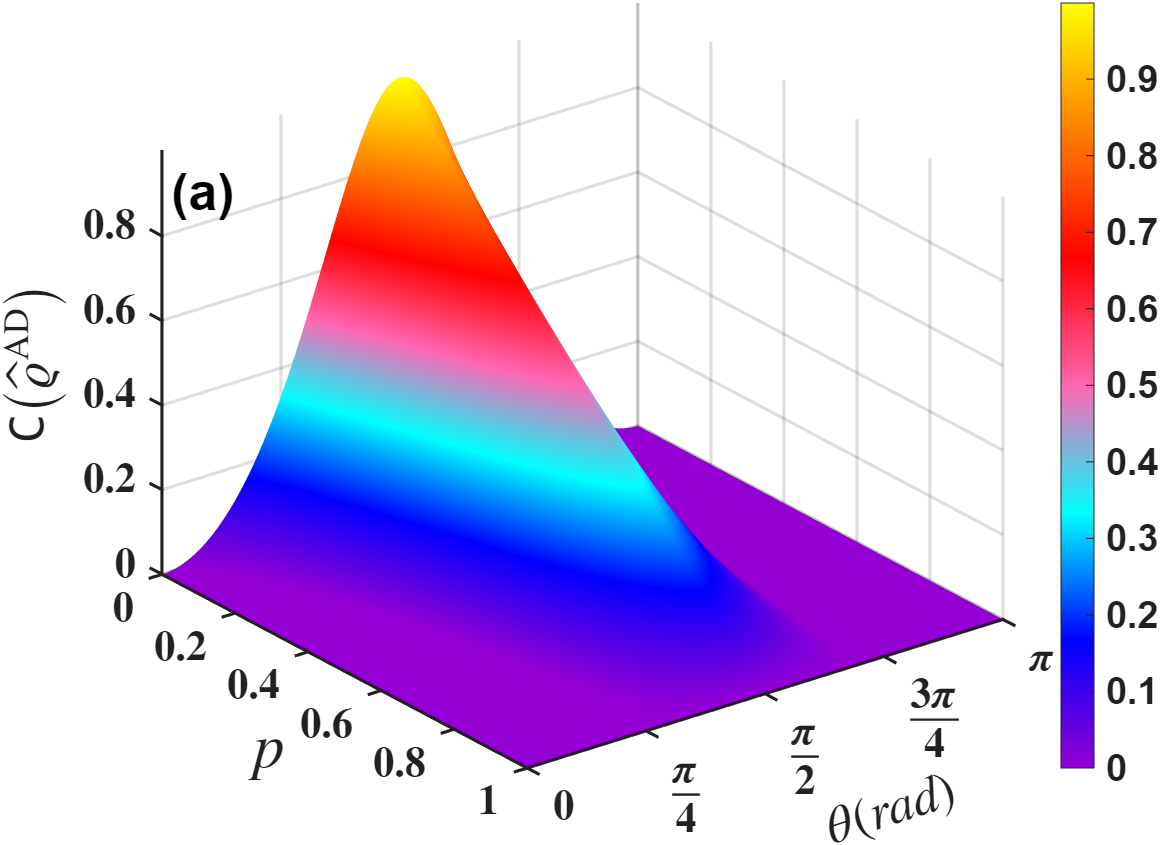}
\includegraphics[scale=0.28]{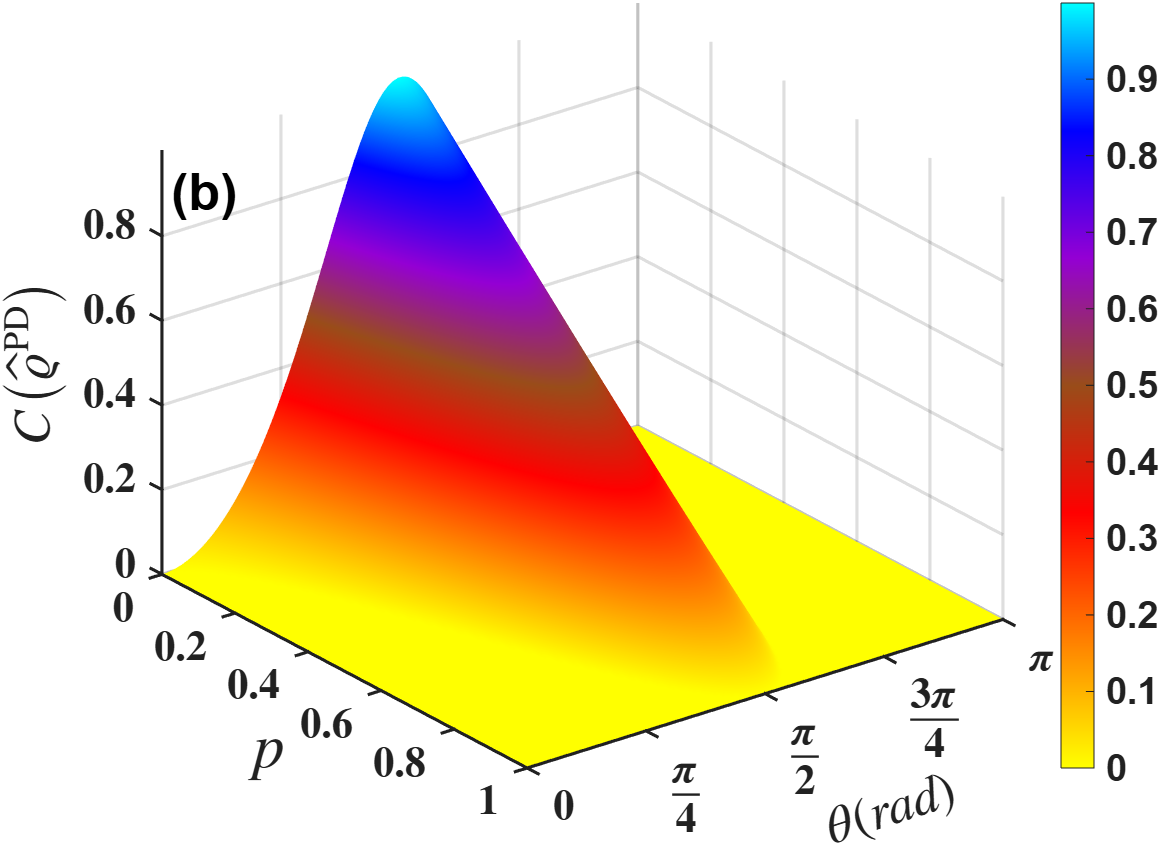}
\includegraphics[scale=0.28]{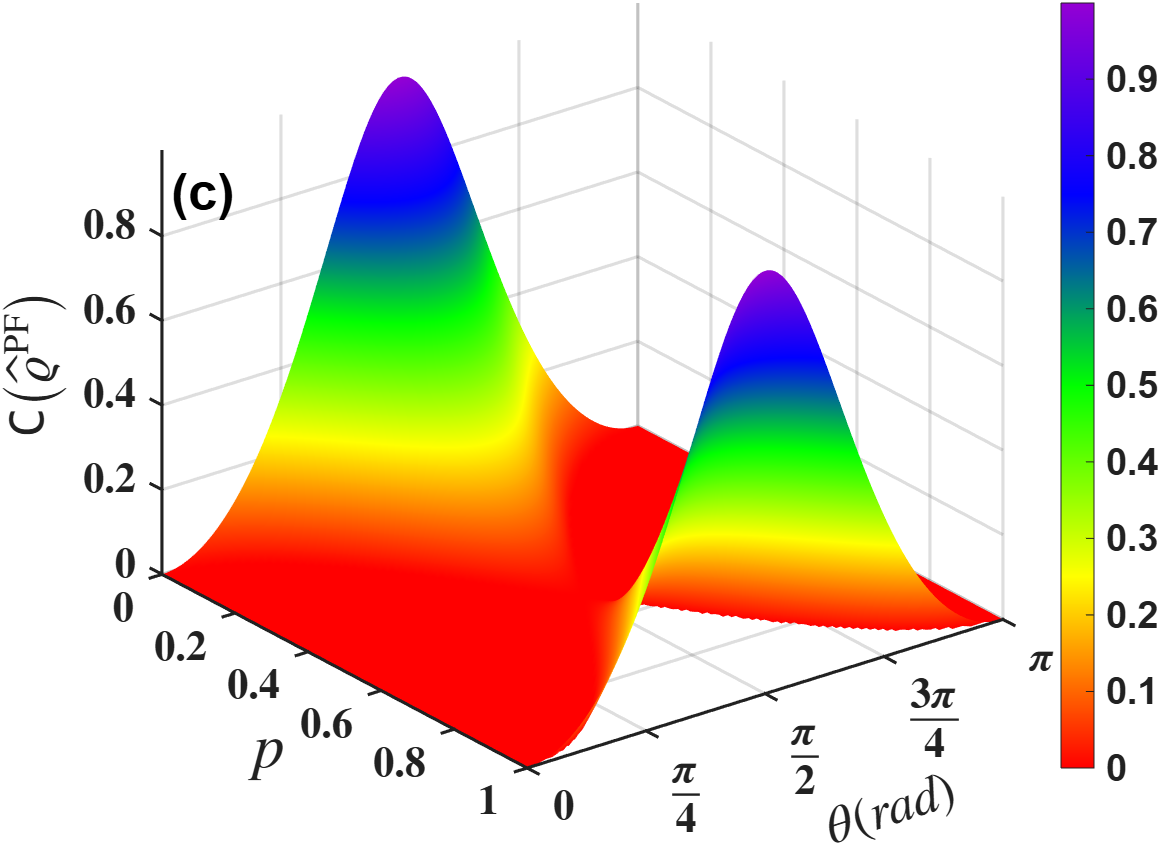}\\
\includegraphics[scale=0.28]{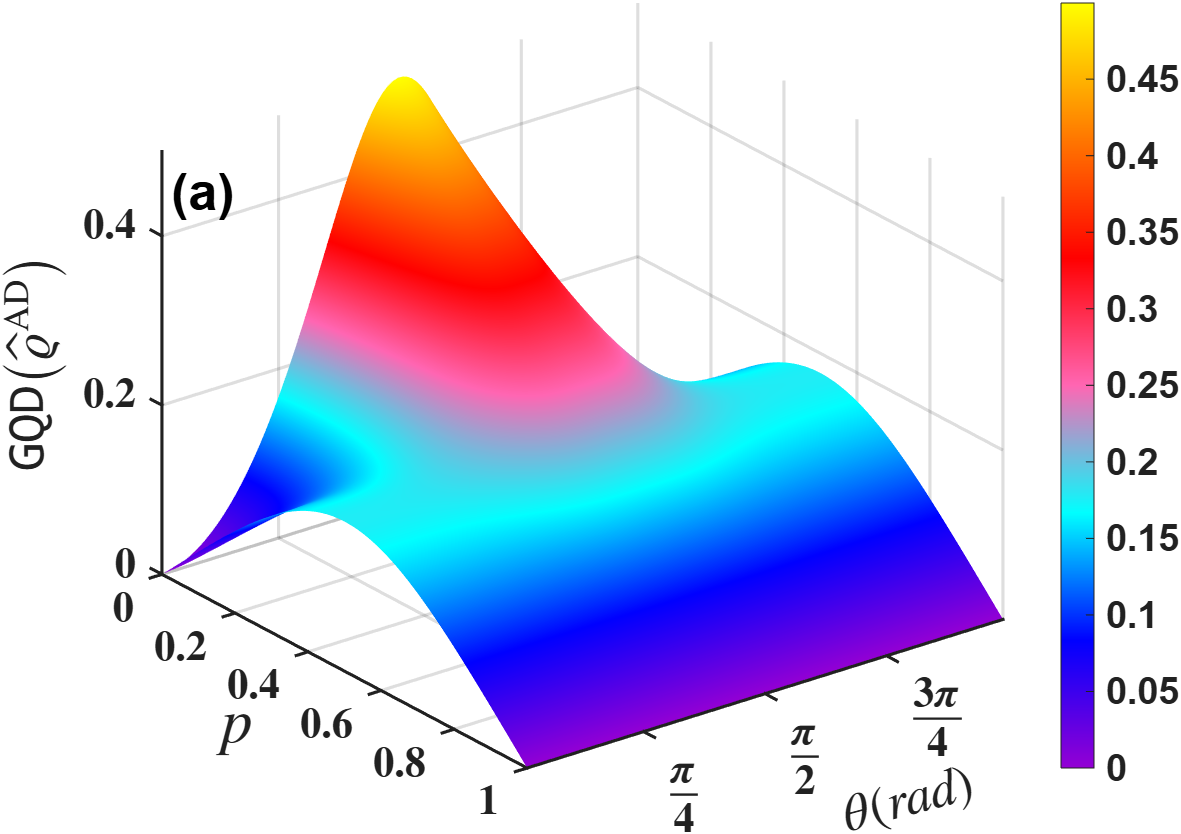}
\includegraphics[scale=0.28]{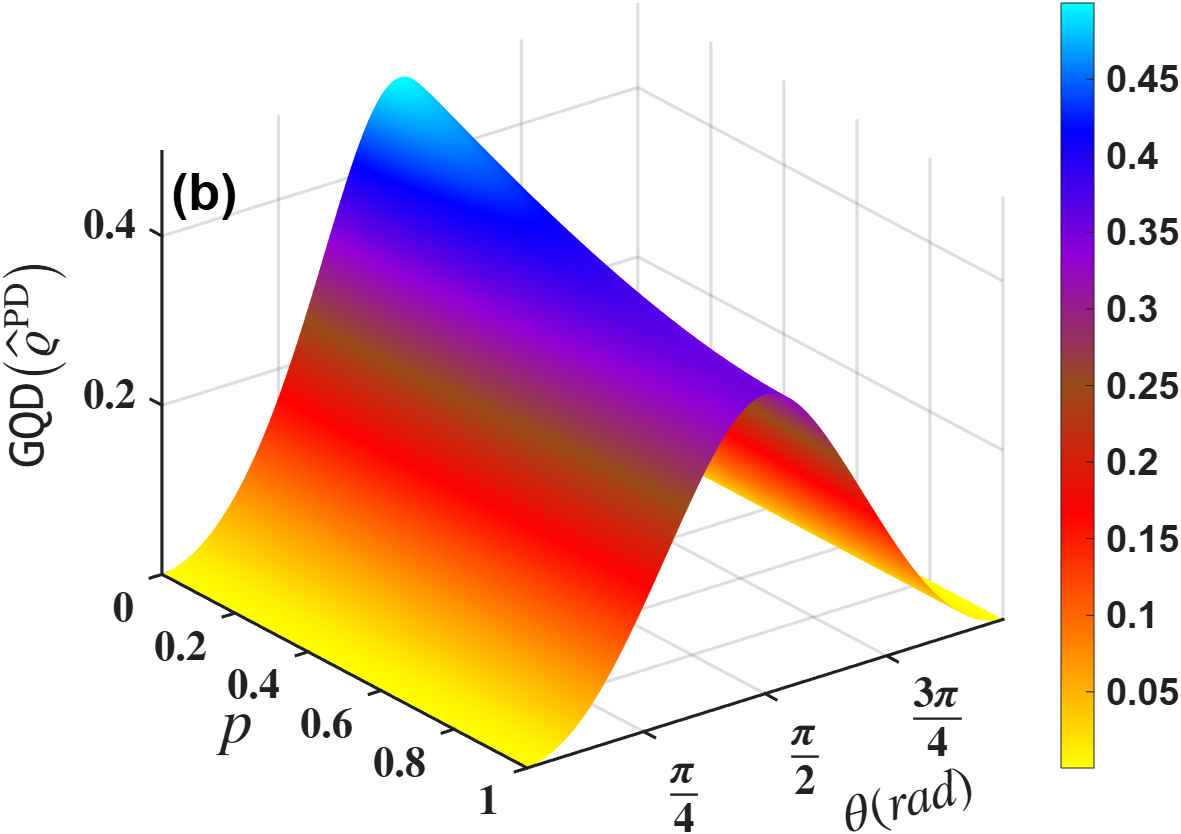}
\includegraphics[scale=0.28]{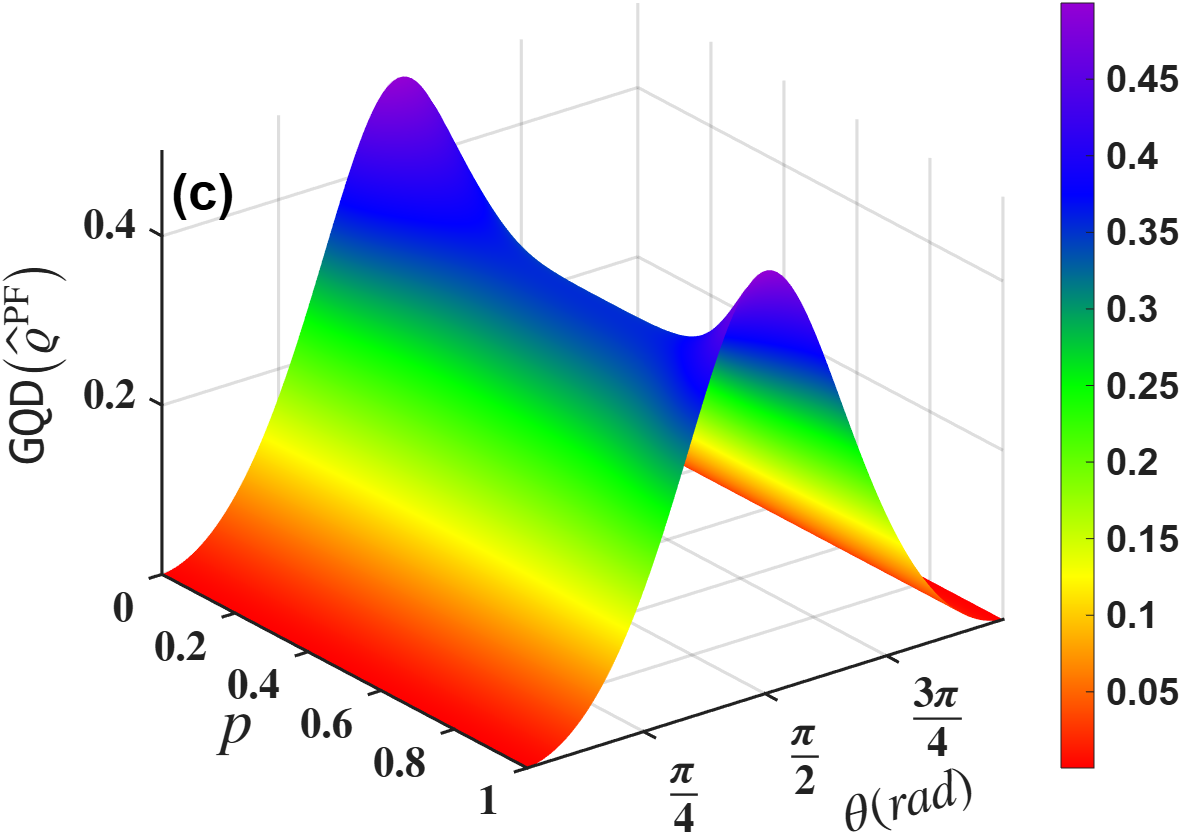}
\caption{Quantum correlations in the presence of decoherence: Bell nonlocality, quantum steering, concurrence, and geometric quantum discord as functions of the production angle $\theta$ and the decoherence parameter $p$. Results are shown for $\beta=1$ in the (a) amplitude-damping (AD), (b) phase-damping (PD) and (c) phase-flip (PF) channels. The limit $\beta=1$ corresponds to the regime where the $q\bar{q}\to t\bar{t}$ process approaches the gluon-initiated channel $gg\to t\bar{t}$.}
\label{fig:De}
\end{figure*}

This subsection presents a three-dimensional analysis of Bell nonlocality, quantum steering, concurrence, and geometric quantum discord for the reconstructed spin density matrix of the $t\bar{t}$ system. We examine how these quantum-correlation measures depend on the effective noise parameter $p$ and the production angle $\theta$ when the state is subjected to standard quantum channels—Amplitude Damping (AD), Phase Damping (PD), and Phase Flip (PF)—introduced here as phenomenological transformations to probe the robustness and structural features of the correlations, rather than as a description of a physical dynamical evolution. The results are shown for $\beta=1$, corresponding to the ultra-relativistic limit in which the $q\bar{q}\to t\bar{t}$ contribution approaches the regime dominated by the $gg\to t\bar{t}$ channel.

Figure~\ref{fig:De}(a) presents the three-dimensional evolution of these four quantum correlation quantifiers under the Amplitude Damping channel. All four measures reach their maximum values at $p=0$ and $\theta=\pi/2$. As $p$ increases, the AD channel induces a monotonic decay of the correlations, eventually leading to their complete disappearance. This behavior highlights the detrimental impact of decoherence on quantum resources and the challenges it poses for quantum information processing.

The Phase Damping (PD) channel exhibits a behavior similar to that of the Amplitude Damping (AD) channel for most quantum correlation measures. As shown in Fig.~\ref{fig:De}(b), the Bell non-locality, quantum steering, and quantum concurrence all reach their maximum values at $p=0$ and $\theta=\pi/2$. These quantifiers then undergo a monotonic decay as the decoherence parameter $p$ increases, eventually vanishing completely for sufficiently large $p$. However, the geometric quantum discord (QGD) displays greater robustness compared to the AD channel, decreasing more slowly with increasing $p$. This slower degradation highlights the relative resistance of geometric quantum discord to pure dephasing noise. Overall, the results emphasize the destructive role of phase damping decoherence on quantum correlations and underline the importance of developing efficient strategies to protect quantum resources in realistic noisy environments.

Fig.~\ref{fig:De}(c) displays the evolution of the quantum correlations under the Phase Flip (PF) channel. The correlations first decrease with rising $p$ until they vanish completely, before being restored to their maximum values as $p \to 1$. The quantum concurrence, in particular, reaches its highest value at $\theta = \pi/2$ for both $p=0$ and $p=1$. As detailed in Figure~\ref{fig:De}(c), the concurrence exhibits three distinct regimes: a monotonic decay at low $p$, a complete vanishing over an intermediate range, and a revival that brings it back to its maximum as $p$ approaches unity. This non-monotonic behavior underscores the complex dynamics of entanglement under phase flip decoherence and the possibility of partial recovery of quantum correlations despite environmental noise.

\section{Quantum teleportation}\label{sec:5}

Quantum teleportation allows the transfer of an arbitrary unknown quantum state by utilizing shared entanglement as a resource. In this work, we investigate quantum teleportation protocols employing mixed entangled states originating from top-antitop ($t\bar{t}$) pairs, which effectively behave as a generalized depolarizing channel \cite{Tel1,Tel2,Tel3}. We particularly analyze the impact of the amplitude angle $\varphi$, which plays a pivotal role in optimizing the fidelity of the teleportation process in such systems.

Let the input state be an arbitrary unknown two-qubit pure state $\ket{\psi_{\rm in}}$ parameterized as
\begin{equation}
\ket{\psi_{\rm in}} = \cos\left(\frac{\varphi}{2}\right)\ket{0 1} + \e^{i \phi} \sin\left(\frac{\varphi}{2}\right)\ket{1 0},
\end{equation}

where, $\varphi \in [0, \pi]$ describes all possible states with varying amplitudes, while $\phi \in [0, 2\pi]$ represents their phase. The corresponding density matrix writes as

\begin{align}
\hat{\varrho}_{\text{in}} &= \ket{\psi_{\text{in}}}\bra{\psi_{\text{in}}} \notag \\
&=
\begin{pmatrix}
0 & 0 & 0 & 0 \\
0 & \sin^2\!\left(\tfrac{\varphi}{2}\right) & \tfrac{1}{2} e^{i\phi} \sin\varphi & 0 \\
0 & \tfrac{1}{2} e^{-i\phi} \sin\varphi & \cos^2\!\left(\tfrac{\varphi}{2}\right) & 0 \\
0 & 0 & 0 & 0
\end{pmatrix}.
\label{eq:in}
\end{align}

A quantum channel is formally described by a completely positive and trace-preserving (CPTP) map, which governs the transformation of an input density matrix into an output density matrix. In the framework of top--antitop ($t\bar{t}$) pair production at hadron colliders such as the LHC, this approach provides a useful tool for analyzing the transfer and survival of quantum information, including spin correlations and entanglement, between the top quark and the antitop quark without requiring any physical transfer of the quantum state itself.
In the presence of noisy or mixed quantum channels, where decoherence effects may arise from QCD interactions, gluon emission, or the extremely short lifetime of the top quark, the resulting state is described by a mixed density operator. The teleportation protocol is then implemented through Bell-state measurements together with local unitary operations acting on the input state $\hat{\varrho}_{\rm in}$, leading to the reconstructed output state $\hat{\varrho}_{\rm out}$ given by
\begin{equation}
\hat{\varrho}_{\rm out} = \sum_{i,j} \mathtt{p}_{i,j} \, (\tau_i \otimes \tau_j) \, \hat{\varrho}_{\rm in} \, (\tau_i \otimes \tau_j),
\label{eq:out}
\end{equation}
where $\tau_i$ and $\tau_j$ are local unitary operators acting on the degrees of freedom of the top and antitop, respectively, and $\mathtt{p}_{i,j}$ denote the probabilities associated with different measurement outcomes or decoherence paths in the $t\bar{t}$ quantum channel.

This framework allows one to study how the initial entanglement of the top-antitop pair is affected by realistic quantum channels and opens the door to exploring whether concepts inspired by quantum teleportation or entanglement distillation can be theoretically adapted to this heavy-quark system.

The probabilities are given by 
\begin{equation}
\mathtt{p}_{ij}=\text{Tr}[\mathcal{E}_i \rho_{t\bar{t}}] \, \text{Tr}[\mathcal{E}_j \rho_{t\bar{t}}],
\end{equation}

with $\sum \mathtt{p}_{i,j}=1$. The measurement operators correspond to the Bell states
\[
\mathcal{E}_0 = \ket{\Phi^-}\bra{\Phi^-}, \quad
\mathcal{E}_1 = \ket{\Psi^-}\bra{\Psi^-}, 
\]
\[
\mathcal{E}_2 = \ket{\Psi^+}\bra{\Psi^+},\quad
\mathcal{E}_3 = \ket{\Phi^+}\bra{\Phi^+},
\]
with
\[
\ket{\Phi^\pm} = \frac{1}{\sqrt{2}}(\ket{01} \pm \ket{10}), \quad
\ket{\Psi^\pm} = \frac{1}{\sqrt{2}}(\ket{00} \pm \ket{11}).
\]
Thereafter, by applying the local unitary transformation to the input state (Eq.~(\ref{eq:out})), the teleported output state at Bob’s lab remains an X-shaped matrix

\begin{equation}
\hat{\varrho}_{out}=
\begin{pmatrix}
\hat{\varrho}_{\text{out}}^{1,1} & 0 & 0 & \hat{\varrho}_{\text{out}}^{1,4} \\
0 & \hat{\varrho}_{\text{out}}^{2,2} & \hat{\varrho}_{\text{out}}^{2,3} & 0 \\
0 & \hat{\varrho}_{\text{out}}^{2,3} & \hat{\varrho}_{\text{out}}^{3,3}& 0 \\
\hat{\varrho}_{\text{out}}^{1,4} & 0 & 0 & \hat{\varrho}_{\text{out}}^{1,1}
\end{pmatrix}, 
\end{equation}
the elements of this matrix are

\begin{equation}
\begin{aligned}
\hat{\varrho}_{\text{out}}^{1,1} &= 2 \rho_{2,2} (\rho_{1,1} + \rho_{4,4}), \\
\hat{\varrho}_{\text{out}}^{2,2} &= (\rho_{1,1} + \rho_{4,4})^2 \sin^2\left(\frac{\varphi}{2}\right) + 4 \rho_{2,2}^2 \cos^2\left(\frac{\varphi}{2}\right), \\
\hat{\varrho}_{\text{out}}^{3,3} &= (\rho_{1,1} + \rho_{4,4})^2 \cos^2\left(\frac{\varphi}{2}\right) + 4 \rho_{2,2}^2 \sin^2\left(\frac{\varphi}{2}\right), \\
\hat{\varrho}_{\text{out}}^{2,3} &= \hat{\varrho}_{3,2} = 2\left(\rho_{2,3}^2 e^{-i\phi} + \rho_{1,4}^2 e^{i\phi} \right) \sin\varphi, \\
\hat{\varrho}_{\text{out}}^{1,4} &= 4 \rho_{2,3} \rho_{1,4} \sin\varphi \cos\phi.
\end{aligned}
\label{eq:OF}
\end{equation}
\subsection{Fidelity}

In order to further investigate the structure and robustness of quantum correlations in top–antitop systems, we consider the reconstructed spin density matrix obtained from collider observables as an effective bipartite quantum state. Within this framework, we analyze how different classes of quantum correlations behave under the action of standard quantum noise models, including amplitude damping (AD), phase damping (PD), and phase-flip (PF) channels. These channels are introduced here as phenomenological tools to probe the stability and structural properties of the correlations encoded in the state, rather than as a description of a physical dynamical evolution of the system.

In addition, we examine the quantum teleportation fidelity associated with the reconstructed state under these effective transformations. This allows us to quantify the degree to which the underlying spin correlations can support quantum-information tasks in a formal sense, providing further insight into the non-classical features of the system.

We emphasize that the use of such quantum channels does not imply the existence of a controllable system–environment interaction in the top–antitop system, but rather serves as a convenient and widely used framework to characterize the response of quantum correlations to generic perturbations.

The performance of the teleportation protocol is commonly evaluated through the fidelity between the input state $\hat{\varrho}_{\rm in}$ and the reconstructed output state $\hat{\varrho}_{\rm out}$. For pure input states, the fidelity provides a reliable measure of the efficiency of the quantum channel in preserving and transmitting quantum information \citep{Fe1,Tel3}. It is defined as
\begin{equation}
F(\hat{\varrho}_{in},\hat{\varrho}^{\text{X}}_{out})=\bra{\psi_{in}}\hat{\varrho}^{\text{X}}_{out}\ket{\psi_{in}}=\Big\lbrace\text{Tr}\Big( \sqrt{\sqrt{\hat{\varrho}_{in}}\hat{\varrho}^{\text{X}}_{out}\sqrt{\hat{\varrho}_{in}}}\Big)\Big\rbrace^{2},
\label{eq:50}
\end{equation}
a straightforward calculation, we find that

\begin{equation}
\begin{aligned}
F (\hat{\varrho}_{in},\hat{\varrho}^{\text{X}}_{out})&= \hat{\varrho}^{\text{X}}_{2,2}\sin^{2}\Big(\frac{\varphi}{2}\Big)+\hat{\varrho}^{\text{X}}_{3,3}\cos^{2}\Big(\frac{\varphi}{2}\Big)+\Re\text{e}\Big(\hat{\varrho}^{\text{X}}_{2,3}\e^{-i\phi}\Big)
\end{aligned}
\label{eq:F1}
\end{equation}

To assess the performance of quantum state transfer in our system, we study the evolution of the fidelity between the input state $\hat{\varrho}_{\text{in}}$ and the output state $\hat{\varrho}^{\text{X}}_{\text{out}}$. This quantity provides a fundamental measure of the closeness between the teleported state and the initial state.

The analysis focuses on the effect of the system's geometric parameters, the production angle $\theta$, the amplitude angle $\varphi$, and the phase $\phi$, which control the characteristics of the states involved in the teleportation. A clear understanding of this dependence allows the optimization of quantum teleportation protocols according to the relevant physical and experimental constraints associated with angular interactions.

The quantum teleportation fidelity evaluated in this work should be interpreted as an information-theoretic indicator of the non-classical correlations encoded in the reconstructed spin density matrix, rather than as evidence of a physically realizable teleportation protocol. In particular, values exceeding the classical threshold $2/3$ signal that the underlying correlations are sufficiently strong to outperform any classical strategy in a formal sense, thereby providing a quantitative measure of the usefulness of the state as an effective quantum resource.

\begin{figure*}[!h]
\includegraphics[scale=0.28]{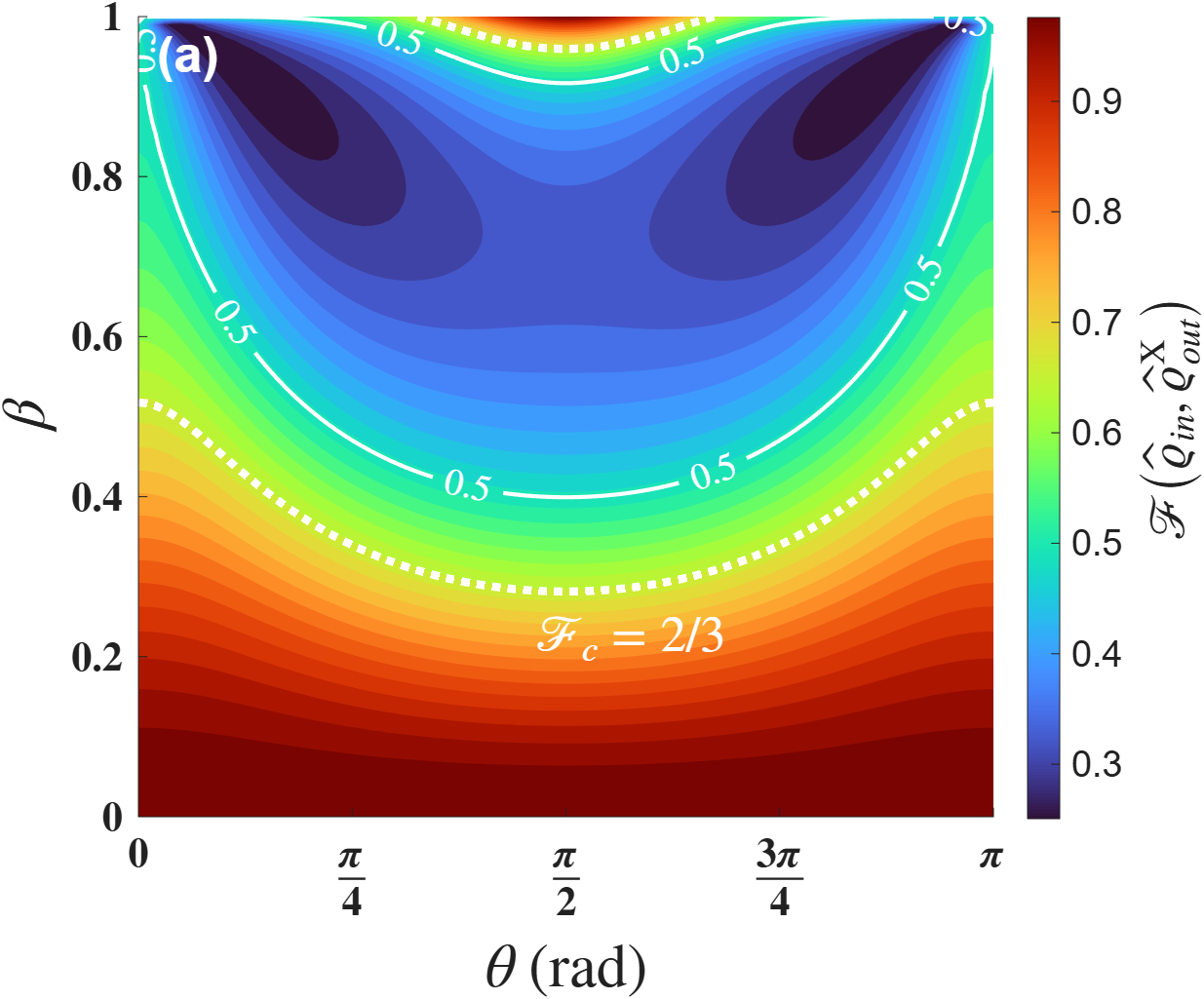}
\includegraphics[scale=0.28]{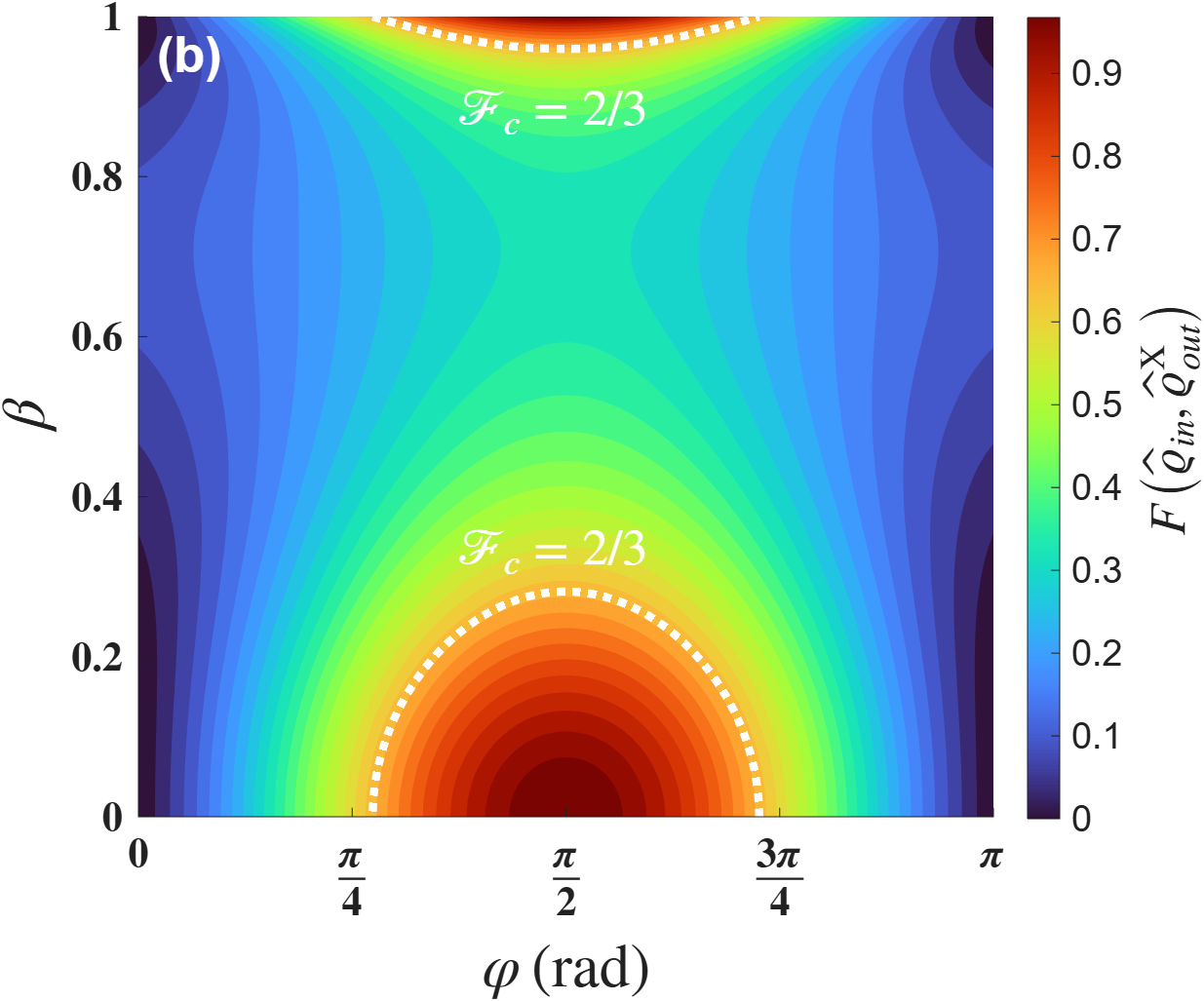}
\includegraphics[scale=0.28]{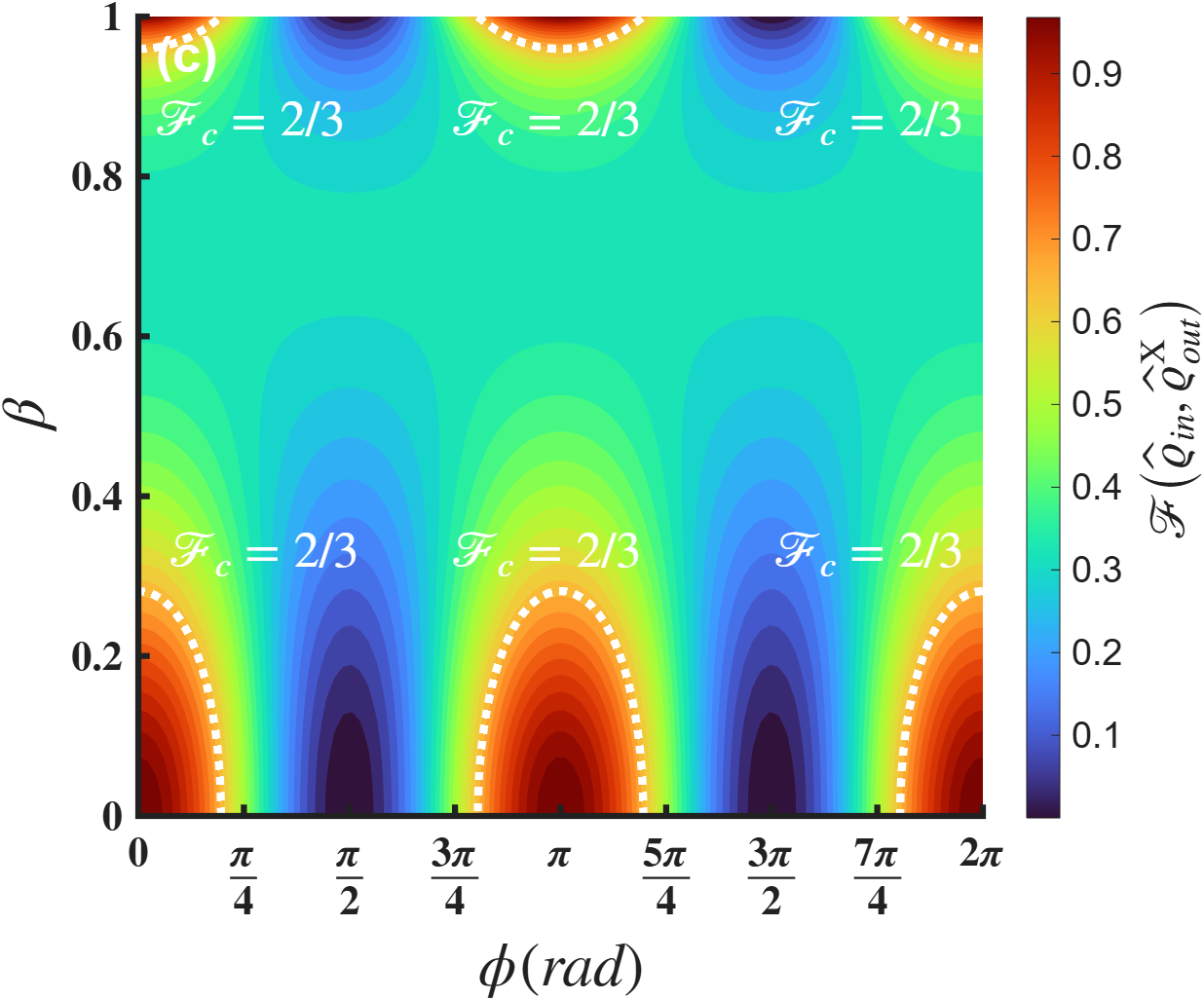}
\caption{Plots of the fidelity $F$ versus (a) production angle $\theta$ and top quark velocity $\beta$ ($\varphi=\pi/2$, $\phi=0$), 
(b) amplitude $\varphi$ and $\beta$ ($\theta=\pi/2$, $\phi=0$), and (c) phase $\phi$ and $\beta$ ($\theta=\varphi=\pi/2$), for gluon fusion $gg\to t\bar{t}$.}
\label{fig:F1}
\end{figure*}

The analysis of the results presented in Fig.~\ref{fig:F1}(a) reveals a strong dependence of the quantum teleportation fidelity $\mathcal{F}$ on both the production angle $\theta$ and the top-quark velocity $\beta$, for fixed amplitude angle $\varphi = \pi/2$ and phase $\phi = 0$. In the low-velocity regime, $\mathcal{F}$ attains its maximum at $\beta = 0$ and exhibits a nontrivial angular dependence: it decreases with increasing $\theta$ up to $\theta = \pi/2$, then increases symmetrically to recover its maximal value at $\theta = \pi$. The range $\beta \in [0,0.5]$ for which the quantum regime ($\mathcal{F} > 2/3$) holds is widest at $\theta = 0$ and $\theta = \pi$, and narrows progressively as $\theta$ approaches $\pi/2$. In the ultrarelativistic limit $\beta \to 1$, the fidelity reaches its peak at $\theta = \pi/2$ while remaining above the classical threshold of $2/3$, thereby confirming the persistence of quantum behavior. Furthermore, a clear symmetry about $\theta = \pi/2$ is observed, indicating equivalent fidelity values for angles symmetrically distributed around this point.

The analysis of Fig.~\ref{fig:F1}(b) highlights the significant influence of the amplitude angle $\varphi$ and the top-quark velocity $\beta$ on the teleportation fidelity, with the production angle fixed at $\theta = \pi/2$ and the phase $\phi = 0$. The fidelity vanishes at the extreme values of both $\varphi$ and $\beta$. It reaches its maximum at $\varphi = \pi/2$, then decreases with increasing $\beta$, before recovering its maximum value in the ultrarelativistic limit $\beta \to 1$, again at $\varphi = \pi/2$. A fidelity threshold of $\mathcal{F} = 2/3$ is adopted to distinguish between the classical and quantum regimes in the gluon-fusion process $gg \to t\bar{t}$. The maximum fidelity is achieved at $\varphi = \pi/2$ for both $\beta = 0$ and $\beta = 1$, indicating that the quantum regime emerges at the extrema of the velocity. Furthermore, the system exhibits a clear symmetry with respect to $\varphi = \pi/2$, with amplitude angles symmetrically distributed around this point leading to equivalent behavior.

In Fig.~\ref{fig:F1}(c), we present the evolution of the fidelity as a function of the phase $\phi$ and the top quark velocity $\beta$ for the process $gg \to t\bar{t}$. The figure reveals a clear sinusoidal dependence of the fidelity on $\phi$, highlighting the significant influence of this phase on the entanglement and quantum superposition properties of the $t\bar{t}$ pair. The fidelity reaches its maximum values for $\phi = k\pi$ ($k = 0, 1, 2$), as well as at the extreme values of $\beta$. These configurations correspond to optimal states of the top--antitop system. In these regimes, the fully quantum character of the system is achieved, as the fidelity of the teleported state clearly exceeds the classical limit of $2/3$.

\begin{figure*}[!h]
\includegraphics[scale=0.28]{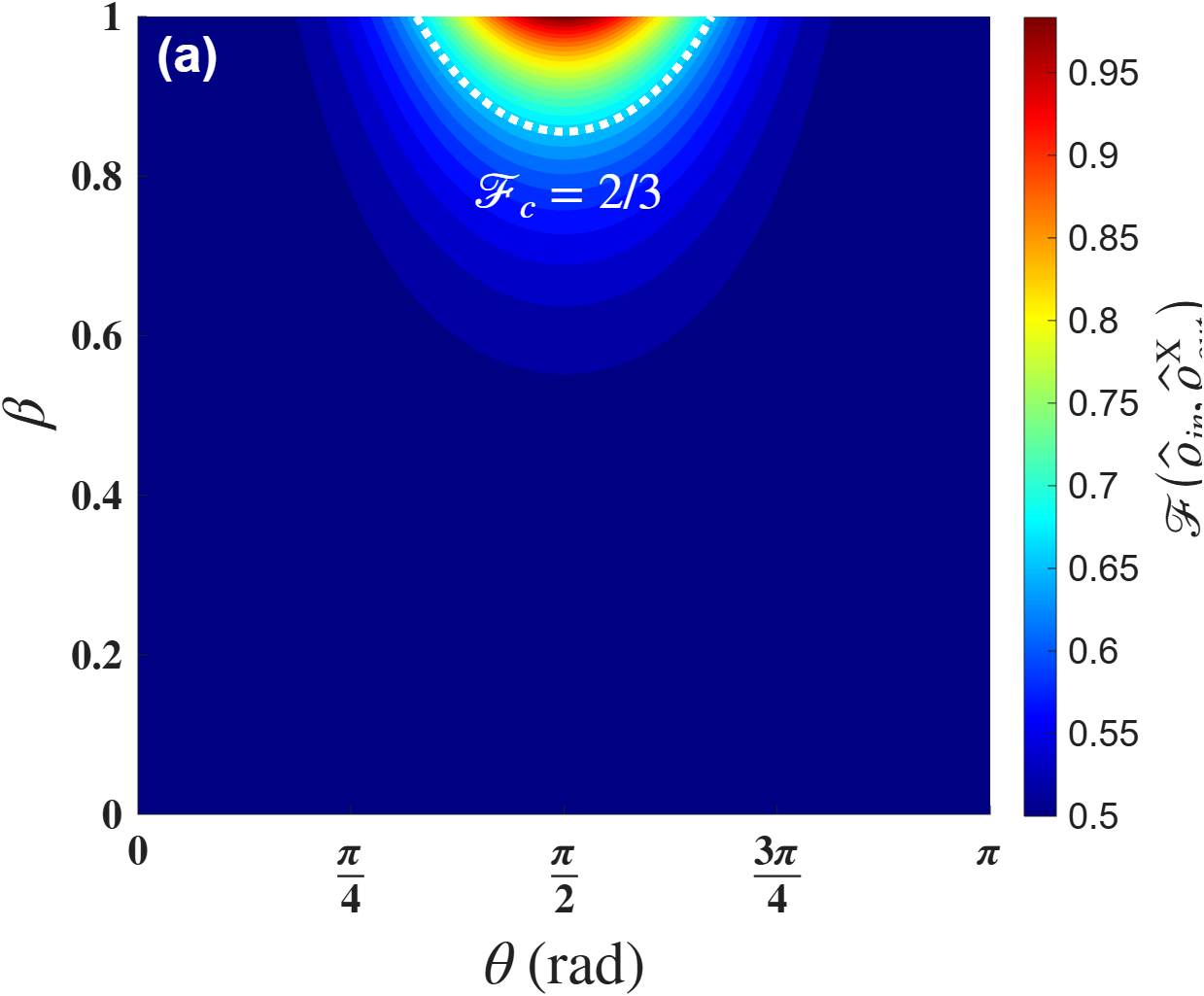}
\includegraphics[scale=0.28]{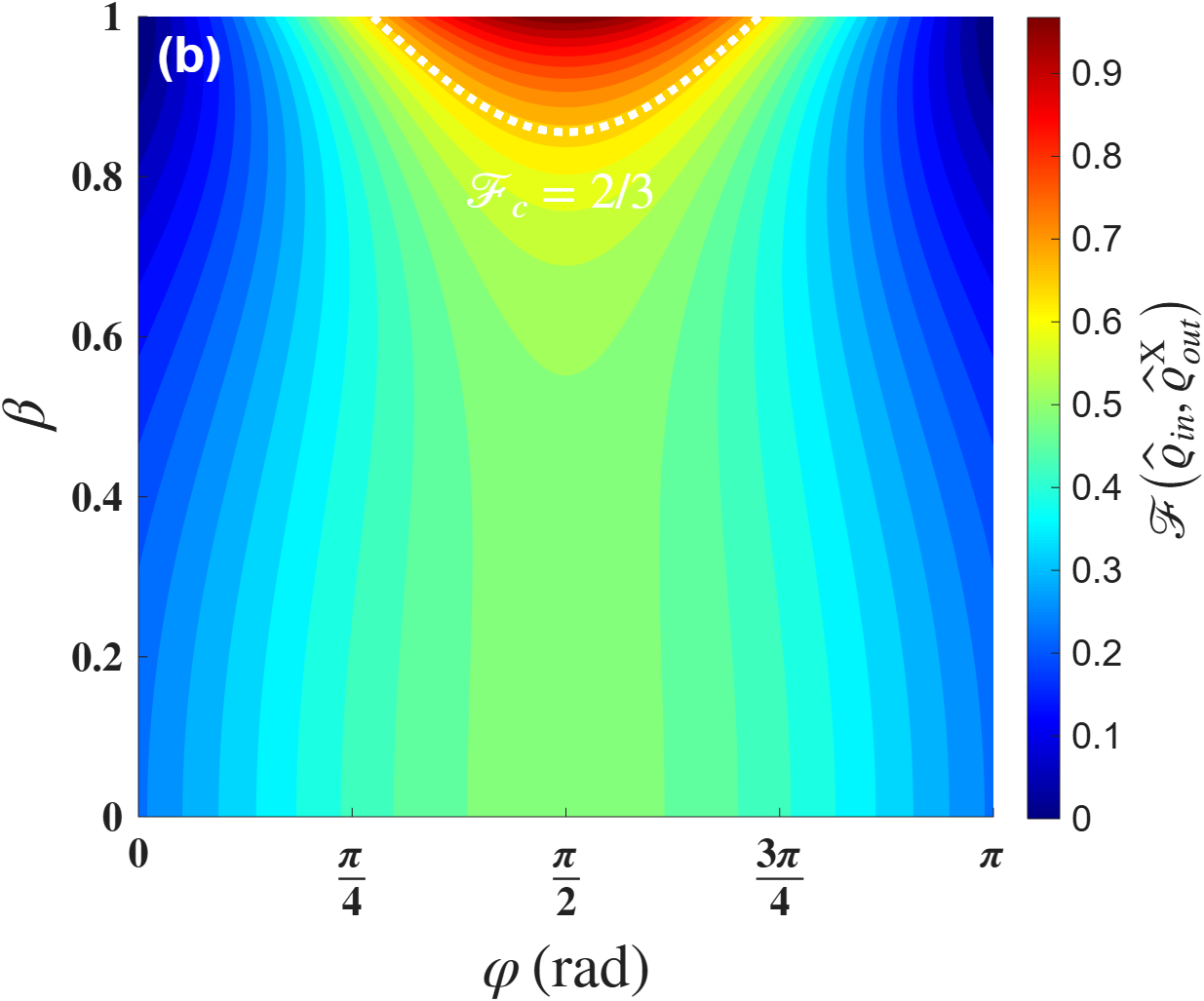}
\includegraphics[scale=0.28]{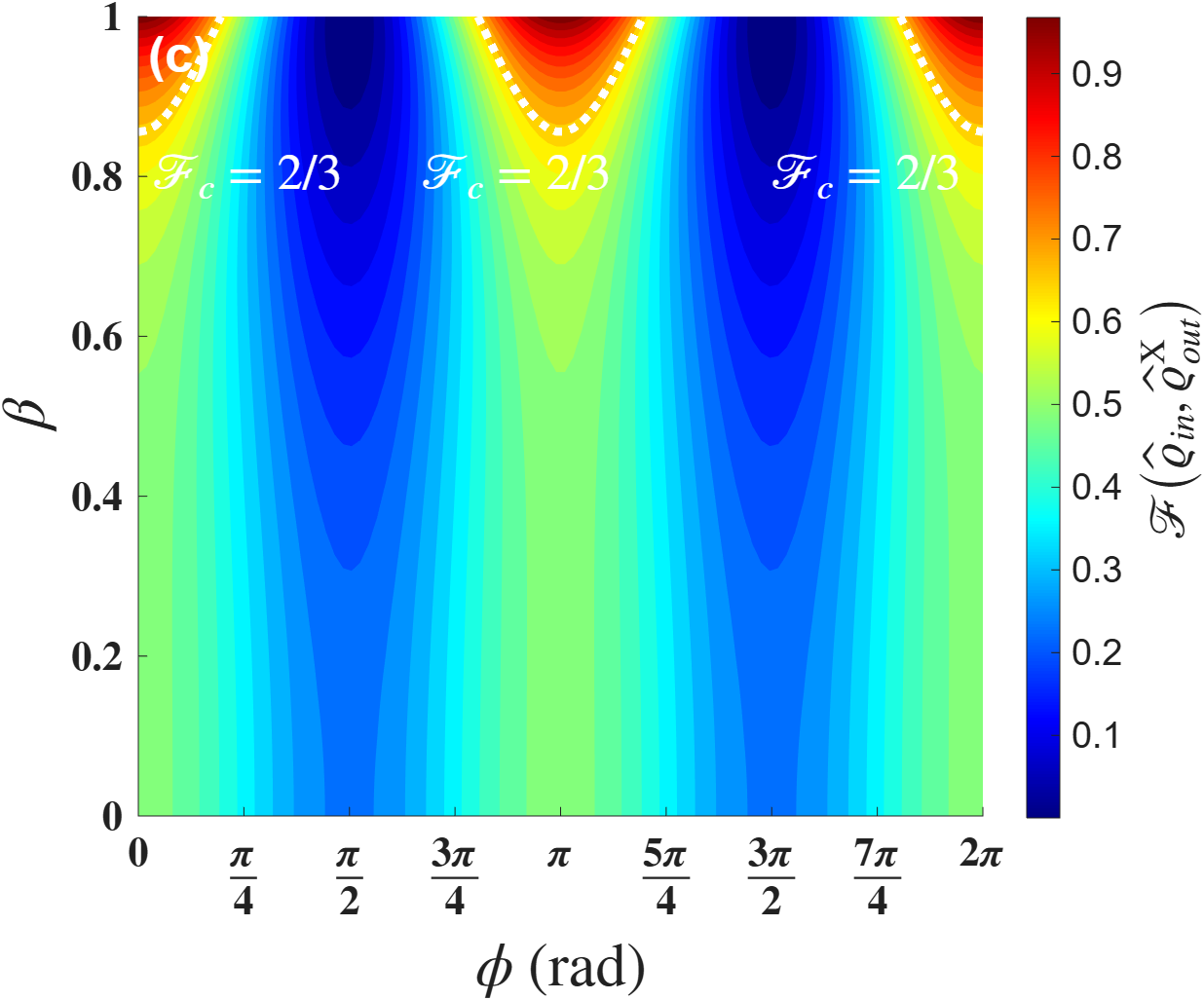}
\caption{Plots of the fidelity $\mathcal{F}$ as a function of (a) production angle $\theta$ and top quark velocity $\beta$ ($\varphi = \pi/2$, $\phi = 0$), (b) amplitude $\varphi$ and $\beta$ ($\theta = \pi/2$, $\phi = 0$), and (c) phase $\phi$ and $\beta$ ($\theta = \varphi = \pi/2$), for quark-antiquark annihilation $q\bar{q} \to t\bar{t}$.}
\label{fig:F2}
\end{figure*}

In Fig.~\ref{fig:F2}(a), we present the fidelity as a function of the production angle $  \theta  $ and the top quark velocity $  \beta  $ for the quark-antiquark annihilation process $q\bar{q} \to t\bar{t}$. The fidelity remains constant over the entire range of $  \theta$ and $\beta$, except in the vicinity of $  \theta=\pi/2$ when $\beta$ approaches 1. Indeed, in this ultra-relativistic limit, the quantum state of the $t\bar{t}$ pair converges to the same state as that obtained in the gluon-gluon ($  gg  $) production channel. In the ultra-relativistic limit ($\beta\to 1  $), the fidelity attains its maximum at $\theta=\pi/2$ while staying above the classical bound of $2/3$, thus indicating the survival of quantum features. The figure also displays a symmetric behavior with respect to $\theta=\pi/2$.

In Fig.~\ref{fig:F2}(b) and Fig.~\ref{fig:F2}(c), we present the behavior of the fidelity for the quark--antiquark annihilation process $q\bar{q} \to t\bar{t}$ as a function of the top-quark velocity $\beta$ and different parameters. In Fig.~\ref{fig:F2}(b), the fidelity is shown as a function of the production angle $\theta$ and $\beta$, while Fig.~\ref{fig:F2}(c) illustrates its dependence on the phase $\phi$ and $\beta$. In both cases, the fidelity reaches its maximum in the ultra-relativistic limit ($\beta \to 1$). More precisely, the maximum occurs at $\theta = \pi/2$ in Fig.~\ref{fig:F2}(b), and at $\phi = k\pi$ ($k = 0,1,2$) in Fig.~\ref{fig:F2}(c). Furthermore, the quantum regime consistently emerges in this ultra-relativistic limit.

\begin{figure*}[!h]
\includegraphics[scale=0.28]{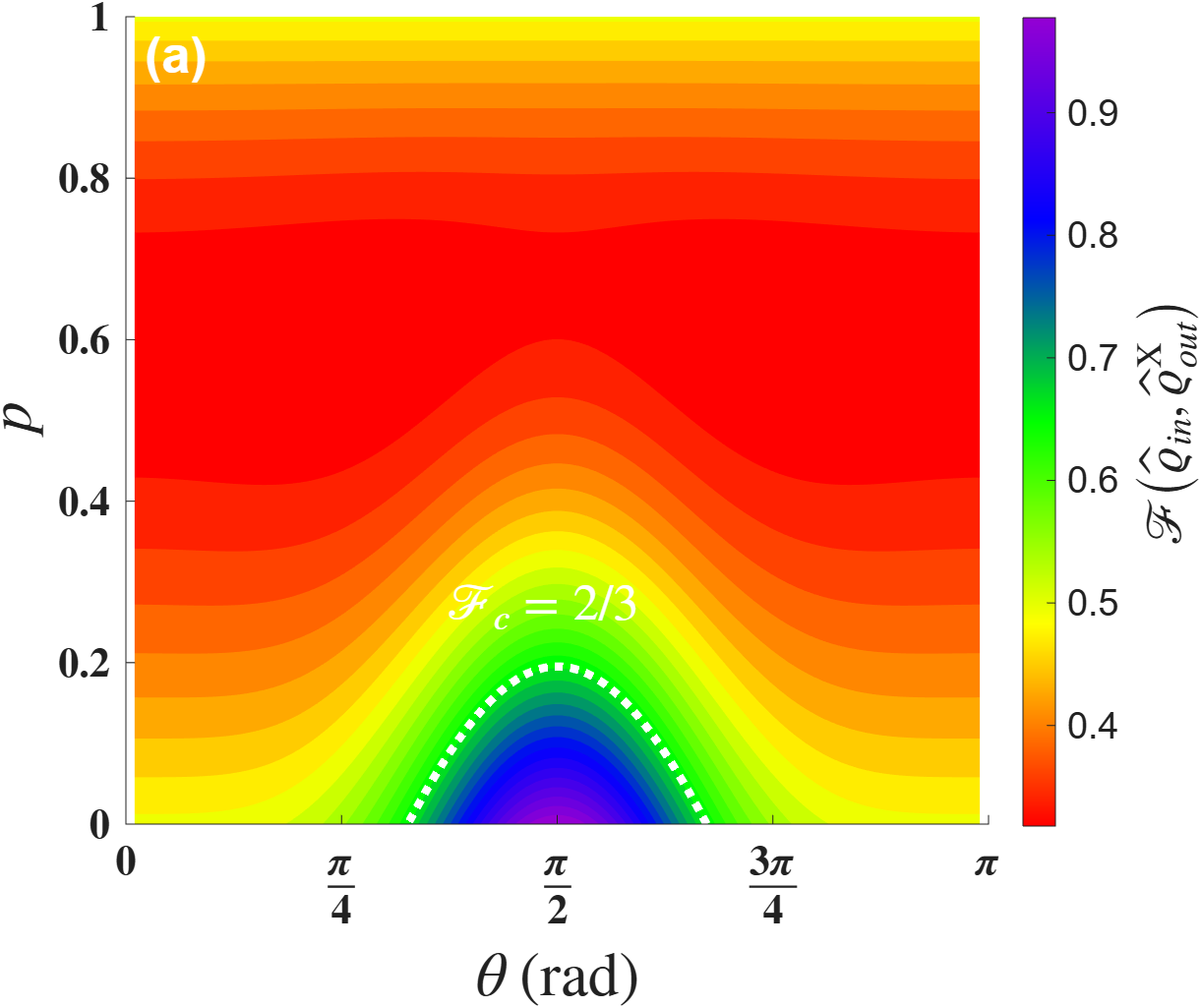}
\includegraphics[scale=0.28]{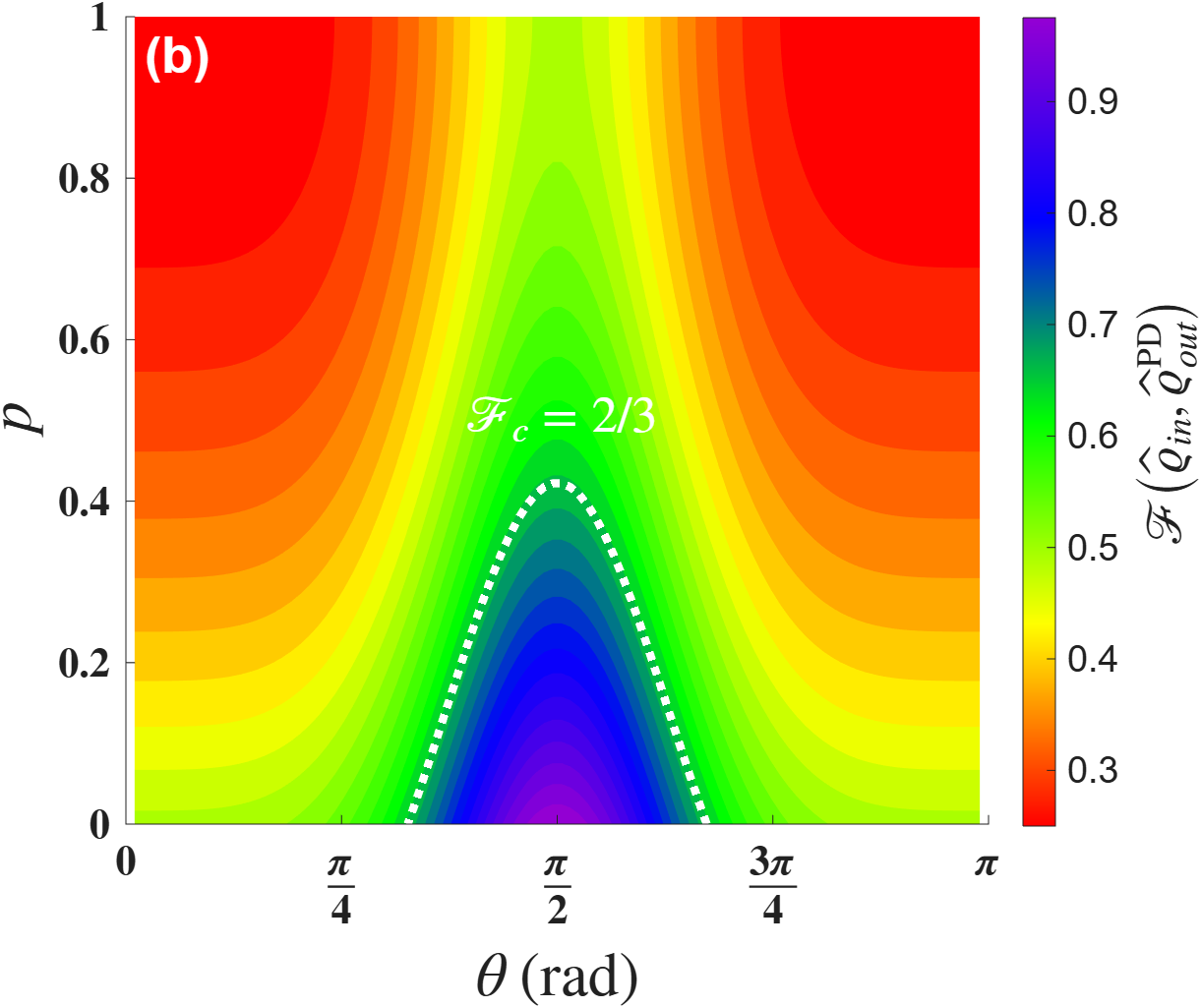}
\includegraphics[scale=0.28]{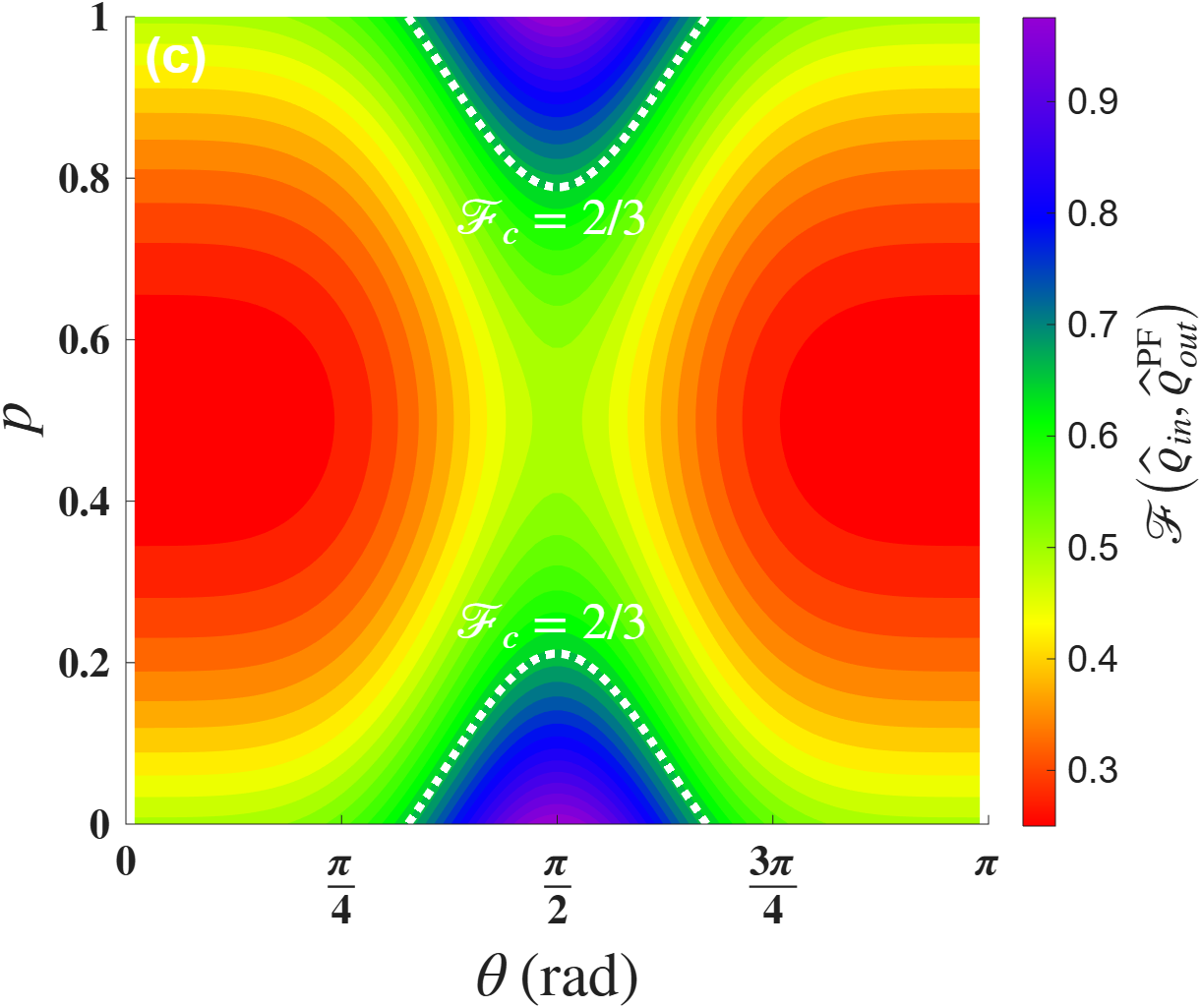}
\caption{Plots of the fidelity $\mathcal{F}$ versus the production angle $\theta$ and the decoherence parameter $p$, with $\varphi = \pi/2$, $\phi = 0$, in the ultrarelativistic regime ($\beta \to 1$).}
\label{fig:F3}
\end{figure*}

\begin{figure*}[!h]
\includegraphics[scale=0.28]{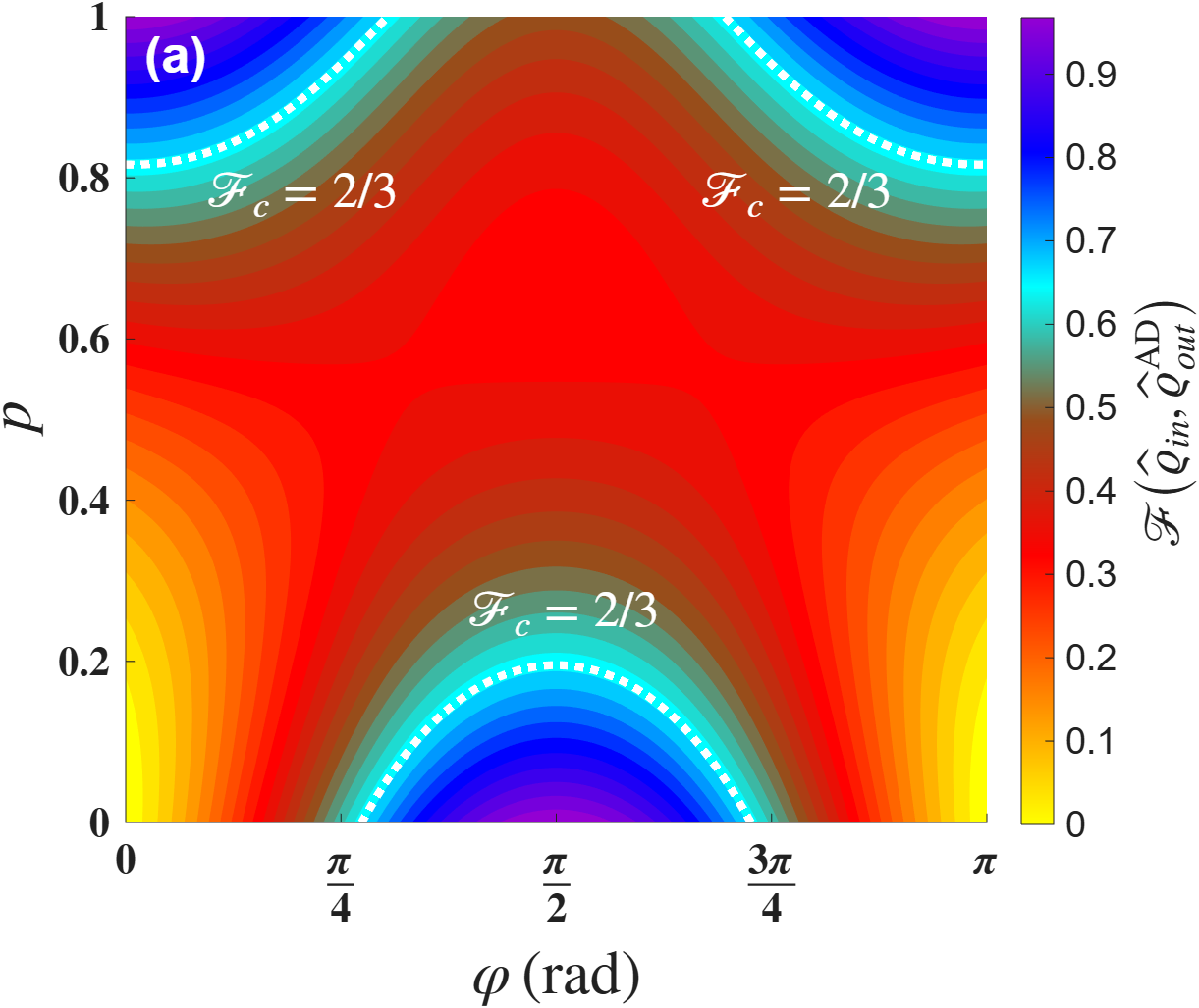}
\includegraphics[scale=0.28]{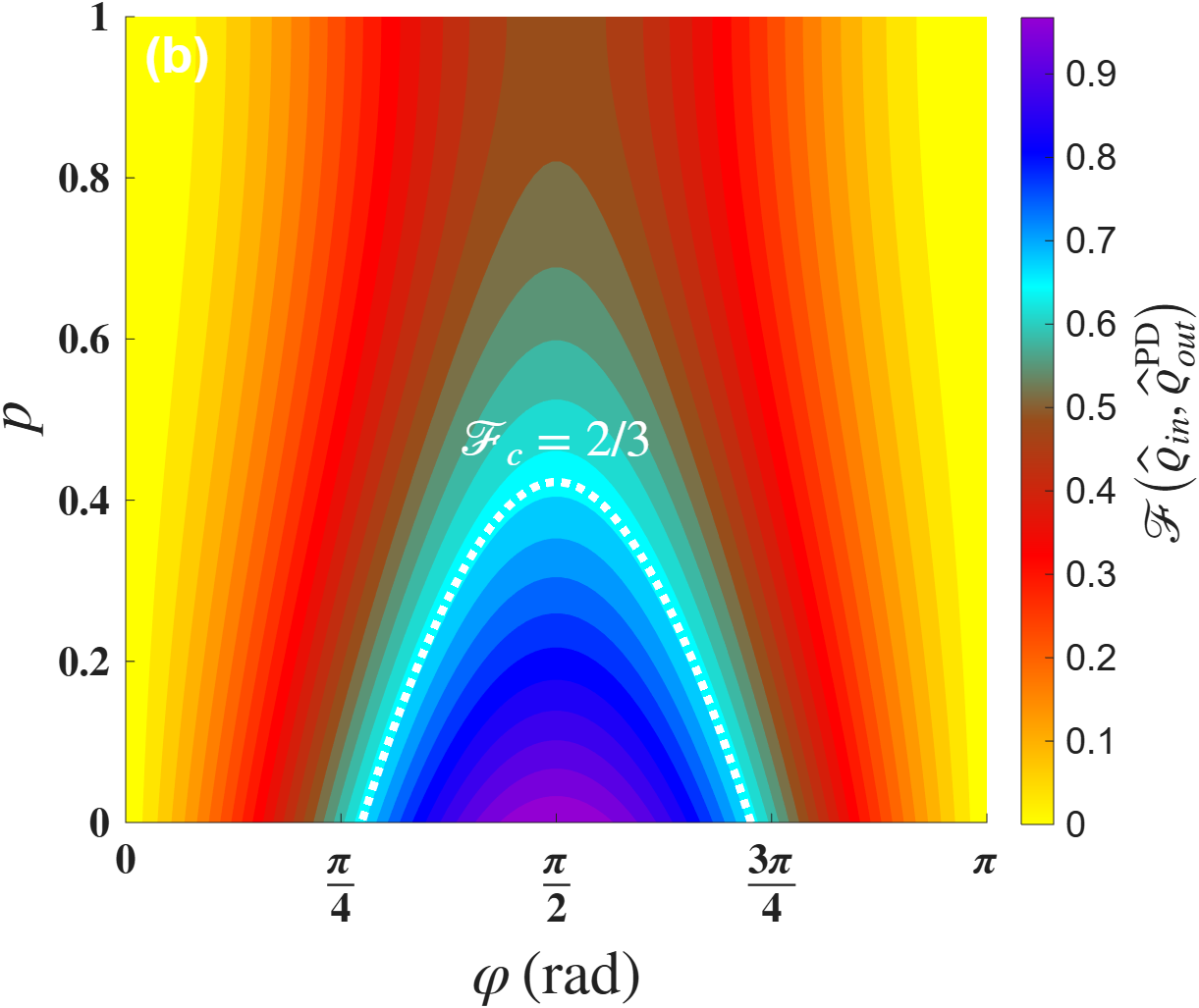}
\includegraphics[scale=0.28]{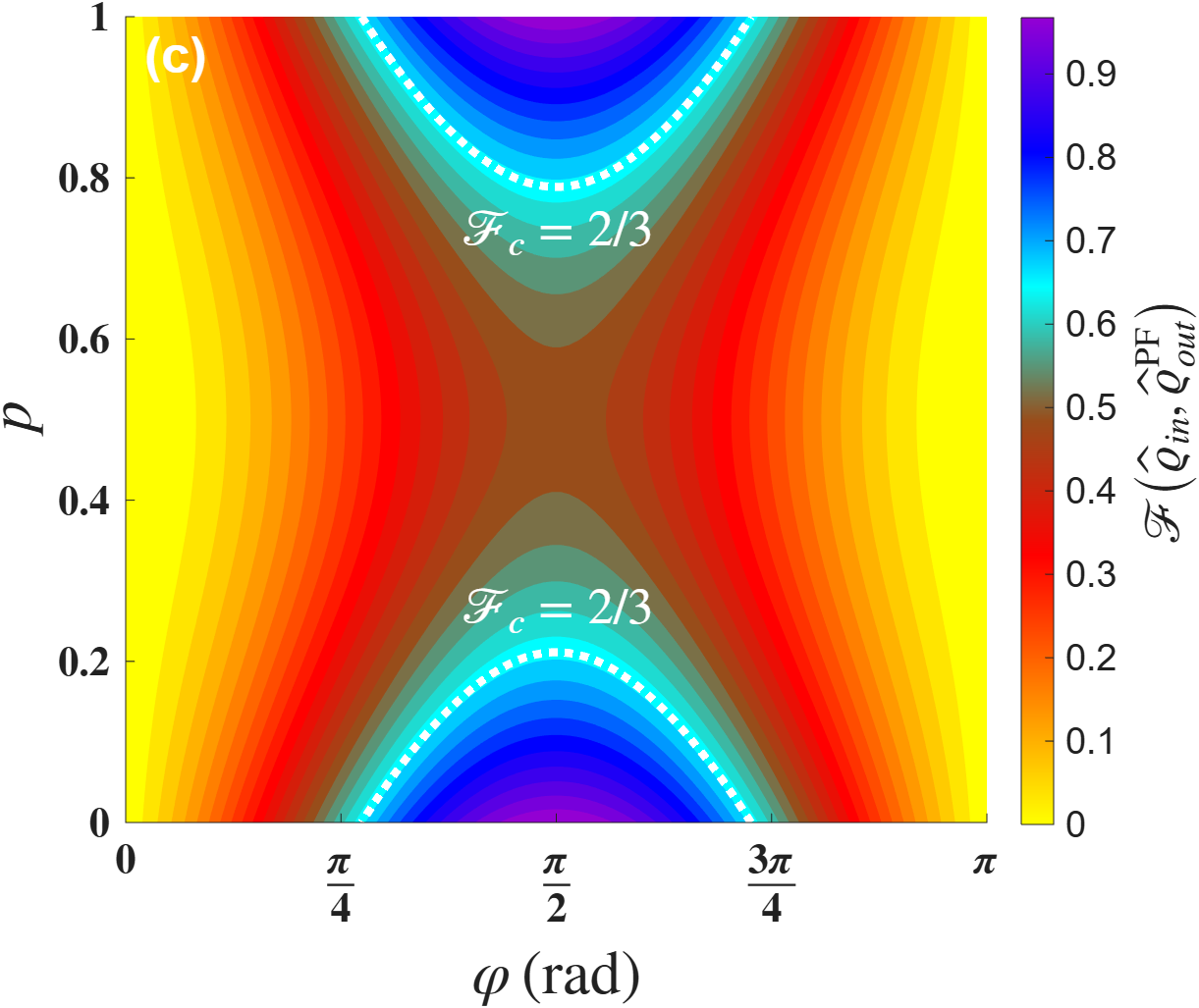}
\caption{Representations of the fidelity $\mathcal{F}$ as a function of the amplitude angle $\varphi$ and the decoherence parameter $p$, for $\theta = \pi/2$, $\phi = \pi/2$ and $\phi = 0$, in the ultrarelativistic limit ($\beta \to 1$).}
\label{fig:F4}
\end{figure*}
Hereafter, in this subsection, we present a three-dimensional analysis of the quantum teleportation fidelity obtained from the reconstructed spin density matrix of the $t\bar{t}  $ system.
We investigate the dependence of this fidelity on the effective noise parameter $p$ (where $p =0$ corresponds to the noiseless case and $p = 1$ to full decoherence) and on the production angle $\theta$ in the center-of-mass frame. The shared state is subjected to three standard quantum channels: Amplitude Damping (AD), Phase Damping (PD), and Phase Flip (PF). These channels are employed here as phenomenological tools to probe the robustness and the underlying structure of the spin correlations, rather than as a realistic description of physical dynamical evolution. The analysis is performed in the ultra-relativistic regime corresponding to $\beta = 1$, where the $q\bar{q} \to t\bar{t}$ production channel becomes marginal and the process is dominated by the gluonic channel $gg \to t\bar{t}$.
Figure~\ref{fig:F3}(a) illustrates the three-dimensional behavior of the teleportation fidelity in the presence of the amplitude-damping (AD) channel, revealing the effect of decoherence on the stability of quantum correlations. The fidelity reaches its highest value at $p=0$ and $\theta=\pi/2$, corresponding to the noiseless regime with maximal quantum correlations. As the decoherence parameter $p$ increases, the fidelity decreases progressively, followed by a partial revival before approaching its minimum value in the ultra-relativistic regime ($\beta \to 1$). For sufficiently strong decoherence, the fidelity eventually vanishes. This nontrivial behavior demonstrates the significant impact of amplitude damping on the efficiency of quantum information transfer.

A similar trend is observed for the phase-damping (PD) channel. As displayed in Fig.~\ref{fig:F3}(b), the teleportation fidelity achieves its maximum for $p=0$ and $\varphi=\pi/2$, corresponding to the absence of environmental noise and to the phase configuration that optimizes the quantum correlations of the system. With increasing decoherence strength $p$, the fidelity decreases approximately linearly due to the gradual suppression of phase coherence. This behavior illustrates the destructive influence of dephasing noise on quantum correlations and highlights the importance of coherence-preservation techniques, such as quantum control protocols, feedback mechanisms, and decoherence-resistant quantum states, for reliable quantum information processing.

The evolution of the teleportation fidelity, depicted in Figure~\ref{fig:F3}(c), provides key insights into the behavior of the teleported state as a function of the production angle  $\theta$ and the decoherence parameter $p$ under the phase-flip (PF) channel. The fidelity reaches its maximum value at a diffusion angle of $\theta=\pi/2$ and at the extreme values of the decoherence parameter, namely $p=0$ (ideal channel) and $  p=1  $ (full decoherence). A detailed analysis of fidelity as a function of $p$ reveals distinct dynamical regimes. As shown in Figure~\ref{fig:F3}(a), the fidelity initially decreases monotonically with increasing $p$. This decline continues up to $p=0.5$, where the fidelity attains its minimum. Beyond this point, the fidelity begins to increase monotonically, recovering its maximum value as $p$ approaches $1$. This non-monotonic and counter-intuitive behavior highlights the complex, nonlinear impact of phase damping noise on quantum teleportation. It demonstrates that, under certain conditions, partial restoration of fidelity can occur despite growing environmental decoherence. Such a revival is characteristic of the phase-flip channel, particularly for input states where the production angle $\theta$ favors components that are more robust against phase errors.

We investigate the quantum teleportation fidelity as a function of the production angle $\theta$ and the decoherence parameter $p$ for three different noise channels. Under the Amplitude Damping (AD) channel, the fidelity reaches its maximum at $p=0$ and $\theta=\pi/2$, then decreases monotonically, followed by a partial recovery before attaining a minimum as $\beta \to 1$. The Phase Damping (PD) channel exhibits a different behavior, characterized by an almost linear decay from its maximum value ($p=0$, $\varphi=\pi/2$), highlighting the consistently destructive effect of phase decoherence on quantum correlations. Finally, the Phase Flip (PF) channel shows the richest dynamics: the fidelity is maximized at $\theta=\pi/2$ for the extreme values $p=0$ and $p=1$, decreases to a minimum around $p=0.5$, and then increases back to its initial maximum. This counter-intuitive behavior reveals a partial restoration of correlations, linked to input states whose production angle $\theta$ enhances components more robust against phase errors.

\subsection{Quantum correlations}

In this section, we present a detailed investigation of the dynamics of quantum correlations in the teleported $  t\bar{t}$ state produced via the partonic channels $gg \to t\bar{t}$ and $q\bar{q} \to t\bar{t}$. We quantify several measures of quantumness, namely Bell nonlocality, quantum steering, concurrence, and geometric quantum discord, in the presence of three paradigmatic decoherence channels: amplitude damping (AD), phase damping (PD), and phase flip (PF).
Our analysis systematically examines the influence of the decoherence parameter, the physical parameters of the system (such as top-quark mass, width, and production kinematics), and the characteristics of the input state on the evolution of the $  t\bar{t}  $ density matrix. Particular emphasis is placed on how these parameters govern the decay, preservation, or revival of quantum correlations under realistic noisy environments.

\begin{figure*}[!h]
\includegraphics[scale=0.4]{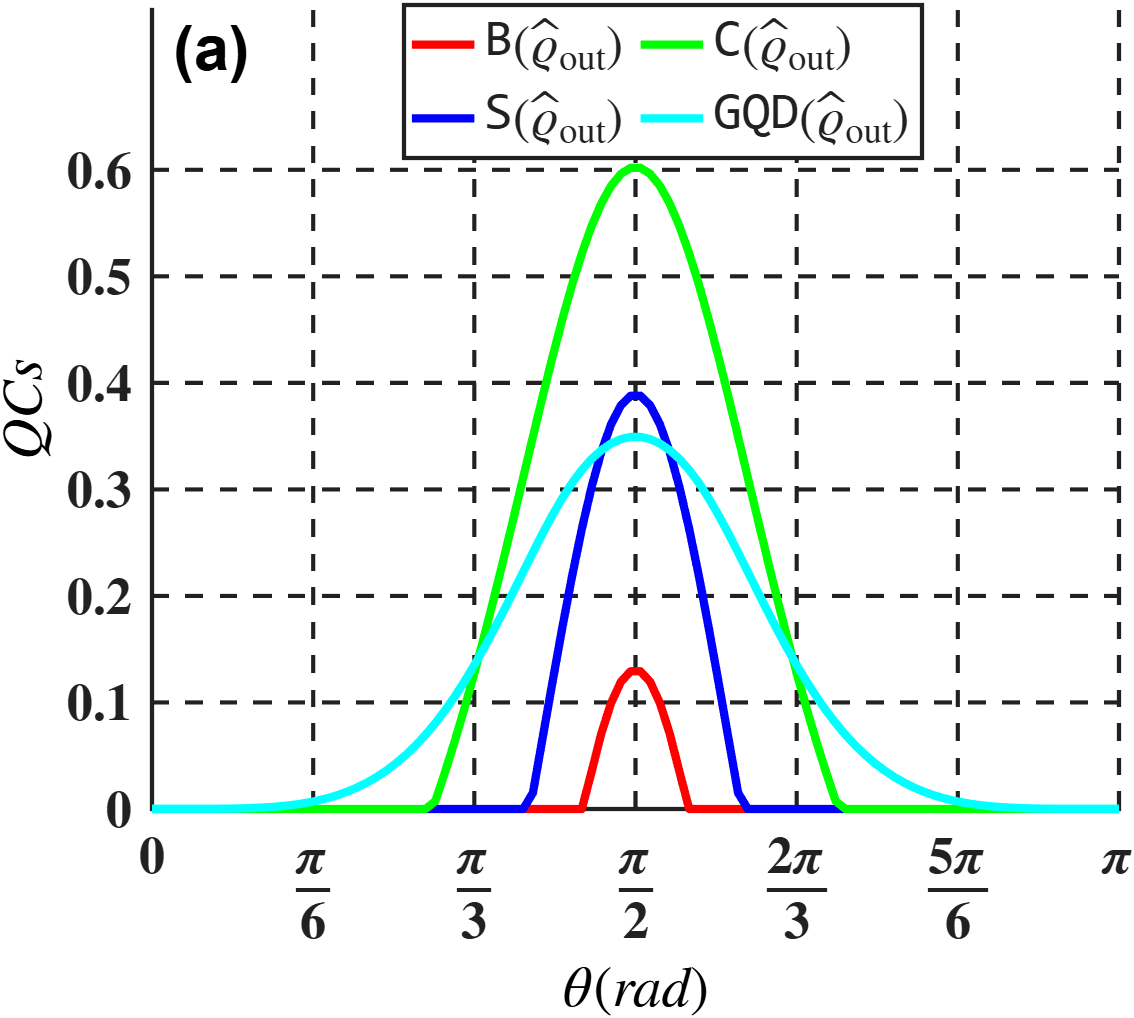}
\includegraphics[scale=0.4]{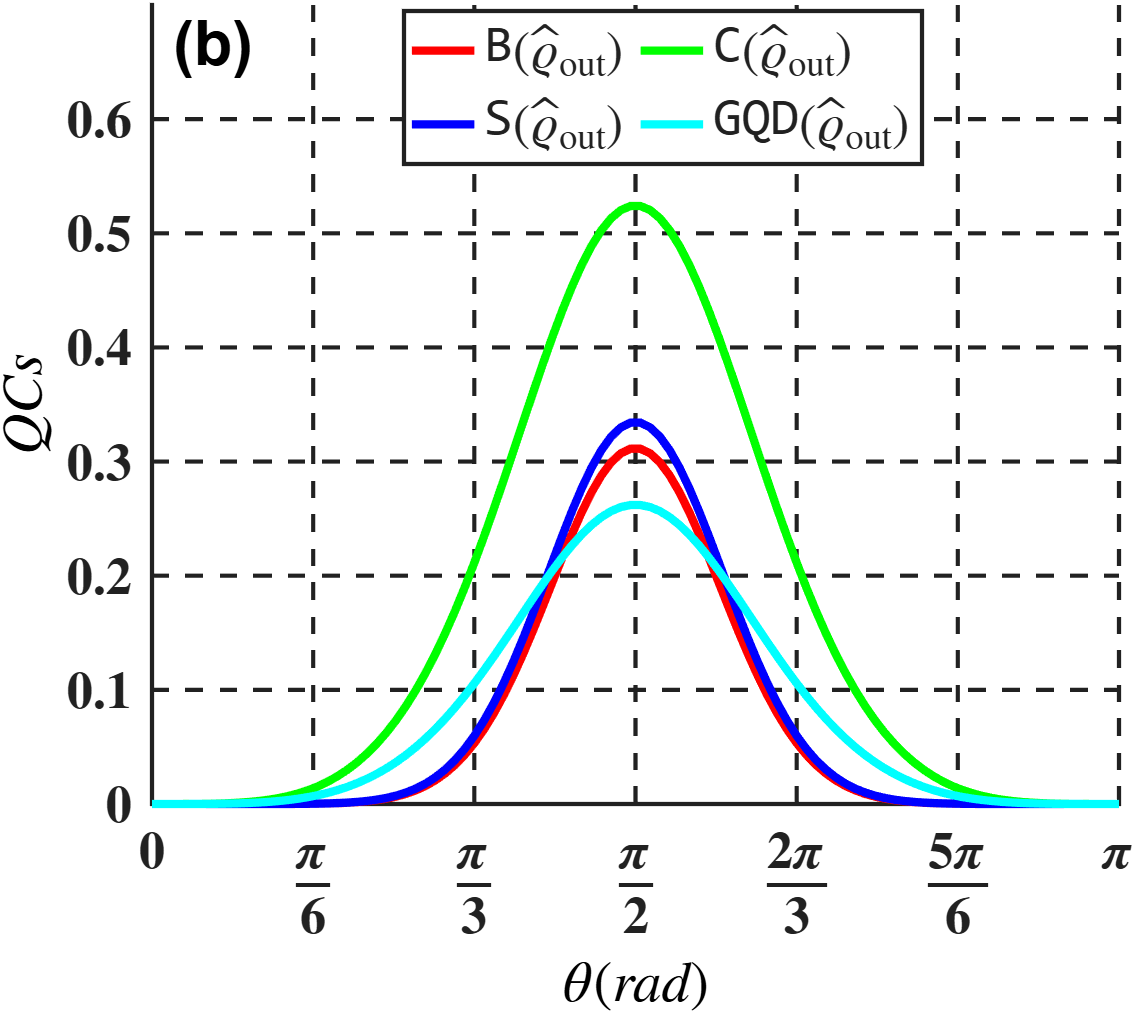}
\caption{Quantum correlations as functions of the production angle $\theta$ in the absence of decoherence ($p=0$), for $\varphi=\pi/2$ and $\phi=0$. Panel (a) illustrates the gluon-fusion process $gg \to t\bar{t}$ with $2m_{t\bar{t}}=346,\mathrm{GeV}$ and $M_{t\bar{t}}=10,m_{t\bar{t}}$, whereas panel (b) corresponds to the quark--antiquark annihilation process $q\bar{q}\to t\bar{t}$ with $2m_{t\bar{t}}=346\mathrm{GeV}$ and $M_{t\bar{t}}=5m_{t\bar{t}}$.}
\label{fig:QCs1}
\end{figure*}

\begin{figure*}[!h]
\includegraphics[scale=0.4]{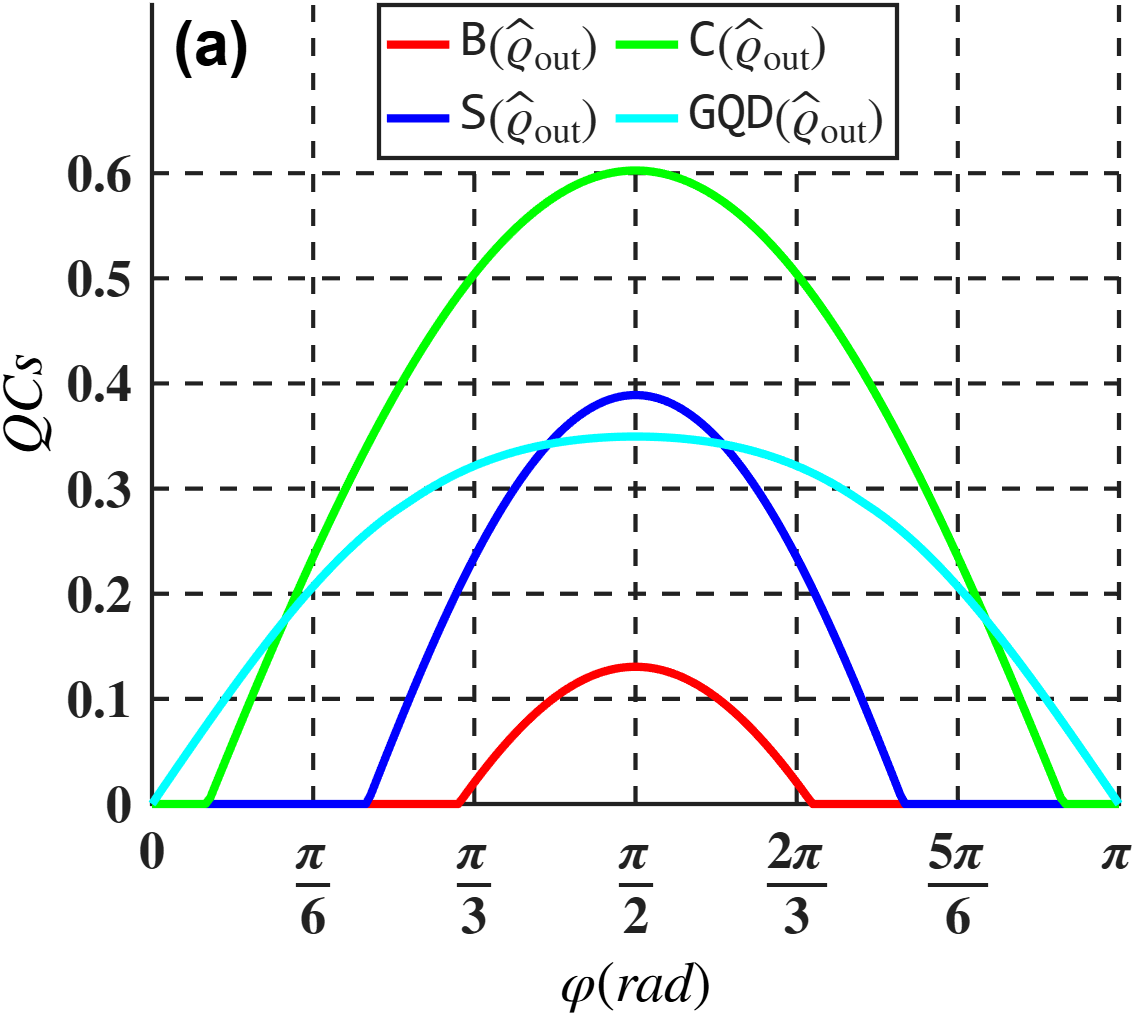}
\includegraphics[scale=0.4]{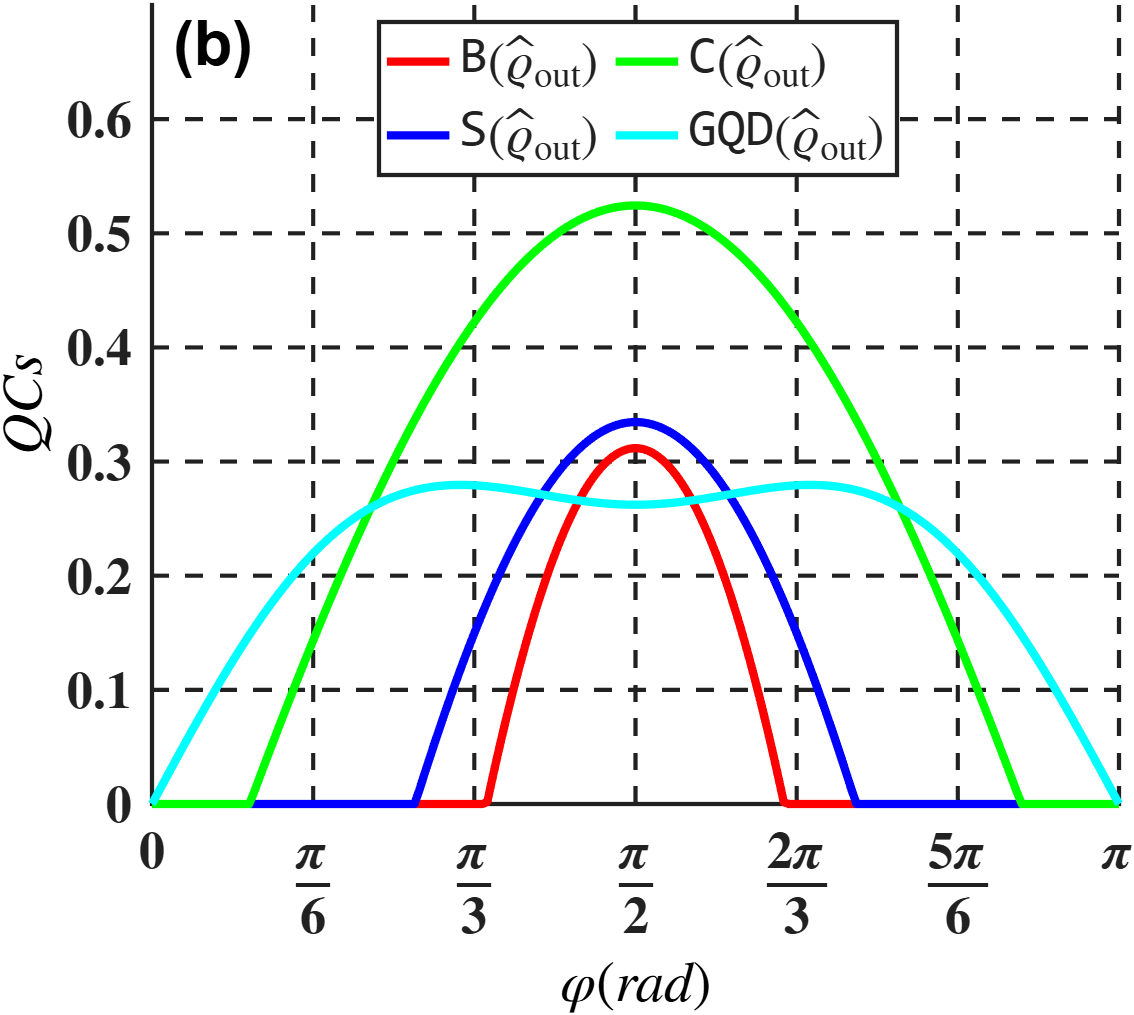}
\caption{Quantum correlations as functions of the amplitude angle $\varphi$ in the absence of decoherence ($p=0$), for $\theta=\pi/2$ and $\phi=0$. Panel (a) corresponds to the gluon-fusion process $gg \to t\bar{t}$ with $2m_{t\bar{t}}=346\mathrm{GeV}$ and $M_{t\bar{t}}=10m_{t\bar{t}}$, while panel (b) describes the quark--antiquark annihilation process $q\bar{q}\to t\bar{t}$ with $2m_{t\bar{t}}=346,\mathrm{GeV}$ and $M_{t\bar{t}}=5m_{t\bar{t}}$.}
\label{fig:QCs2}
\end{figure*}

Figures~\ref{fig:QCs1} and \ref{fig:QCs2} present a comparative analysis of four measures of quantum correlations---Bell nonlocality ($\mathtt{B}$), quantum steering ($\mathtt{S}$), concurrence ($\mathtt{C}$), and geometric quantum discord ($\mathtt{QGD}$)---displayed as functions of the production angle $\theta$ and the amplitude parameter $\varphi$. The analysis is performed for both the gluon-fusion channel $gg \to t\bar{t}$ and the quark--antiquark annihilation channel $q\bar{q} \to t\bar{t}$.

In Fig.~\ref{fig:QCs1}(a-b), we display the dependence of the four quantum quantifiers on the production angle $\theta$ for both the gluon fusion process $gg\to t\bar{t}$ and the quark-antiquark annihilation process $  q\bar{q}\to t\bar{t}$. The results show that all four measures are symmetric about $\theta=\pi/2$ across the full range $\theta\in [0, \pi]$. They start at zero for $  \theta=0  $ and grow monotonically as $\theta$ increases toward $\pi/2$, where they simultaneously attain their peak values. Beyond this point, the quantifiers decrease symmetrically and return to zero at $\theta=\pi$. This angular dependence emphasizes that quantum correlations in the $t\bar{t} $ pair reach their maximum strength at $\theta=\pi/2$, demonstrating the key role played by the production angle in controlling the degree of entanglement generated in these high-energy processes.

In Figs.~\ref{fig:QCs1}(a) and (b), we present the variation of the four quantum quantifiers as a function of the amplitude angle $\varphi$ for gluon fusion $gg\to t\bar{t}$ [Fig.~\ref{fig:QCs1}(a)] and quark-antiquark annihilation $q\bar{q}\to t\bar{t}$ [Fig.~\ref{fig:QCs1}(b)], respectively. The results reveal that all four quantifiers exhibit symmetry with respect to $\varphi=\pi/2$ over the angular range $\varphi \in [0, \pi]$. As $\varphi$ increases from $0$ to $\pi/2$, the Bell non-locality, quantum steering, and concurrence increase monotonically from zero, reaching their maximum values at $\varphi=\pi/2$. They then decrease symmetrically and vanish at $\varphi=\pi$. Although the quantum geometric discord displays the same symmetry, it remains finite over nearly the entire interval $[0,\pi]  $, vanishing only in the collinear configurations ($  \varphi=0$ and $\varphi=\pi$). This behavior highlights the central role of the production angle $\varphi=\pi/2$, at which the quantum correlations in the $t\bar{t}$ system are strongest. It underscores the crucial influence of the production angle on the degree of entanglement in high-energy processes.

These observations are consistent with the general hierarchy of quantum correlations for any bipartite system \cite{Hr1,Hr2,Hr3,BachainEPJC}:
\begin{equation}
\text{Bell non-locality} \subseteq \text{Steering} \subseteq \text{Entanglement} \subseteq \text{Discord}.
\label{eq:Hr}
\end{equation}

Fig.~\ref{fig:QCs1} and Fig.~\ref{fig:QCs2} clearly illustrates this hierarchy
\[
\mathtt{B}(\rho_{t\bar{t}}) \subset \mathtt{S}(\rho_{t\bar{t}}) \subset \mathtt{C}(\rho_{t\bar{t}}) \subset \mathtt{QGD}(\rho_{t\bar{t}}).
\]
While quantum geometric discord remains finite across almost the full angular range, concurrence is restricted to a narrower region centered around $\theta = \pi/2$, and both Bell non-locality and steering occupy an even smaller domain.

\begin{figure*}[t]
\includegraphics[scale=0.28]{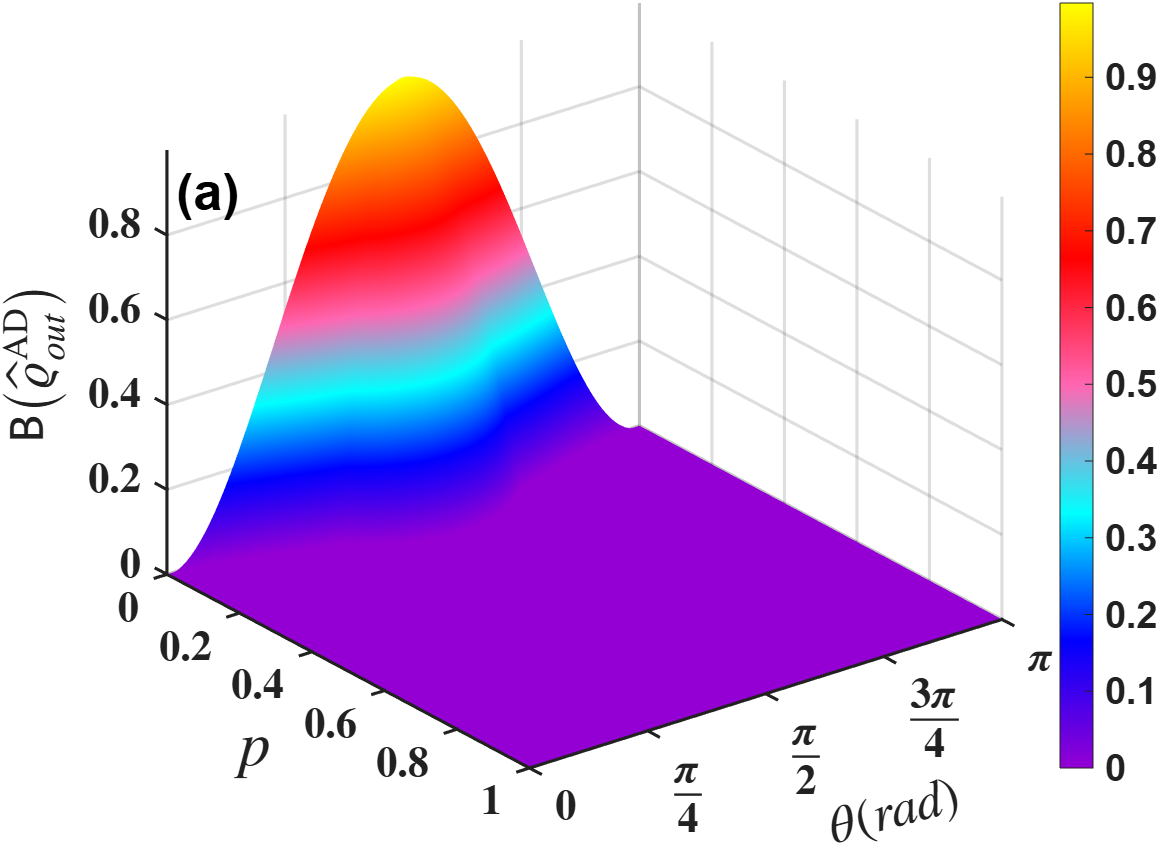}
\includegraphics[scale=0.28]{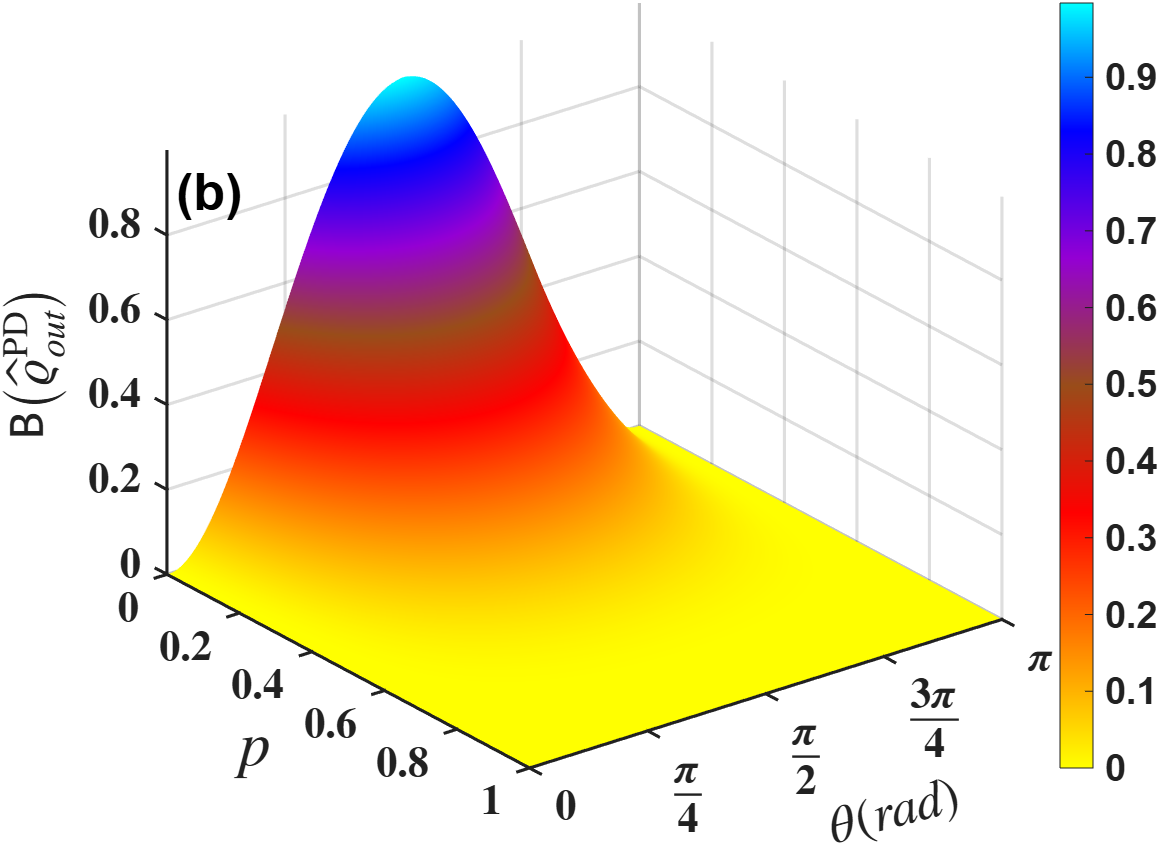}
\includegraphics[scale=0.28]{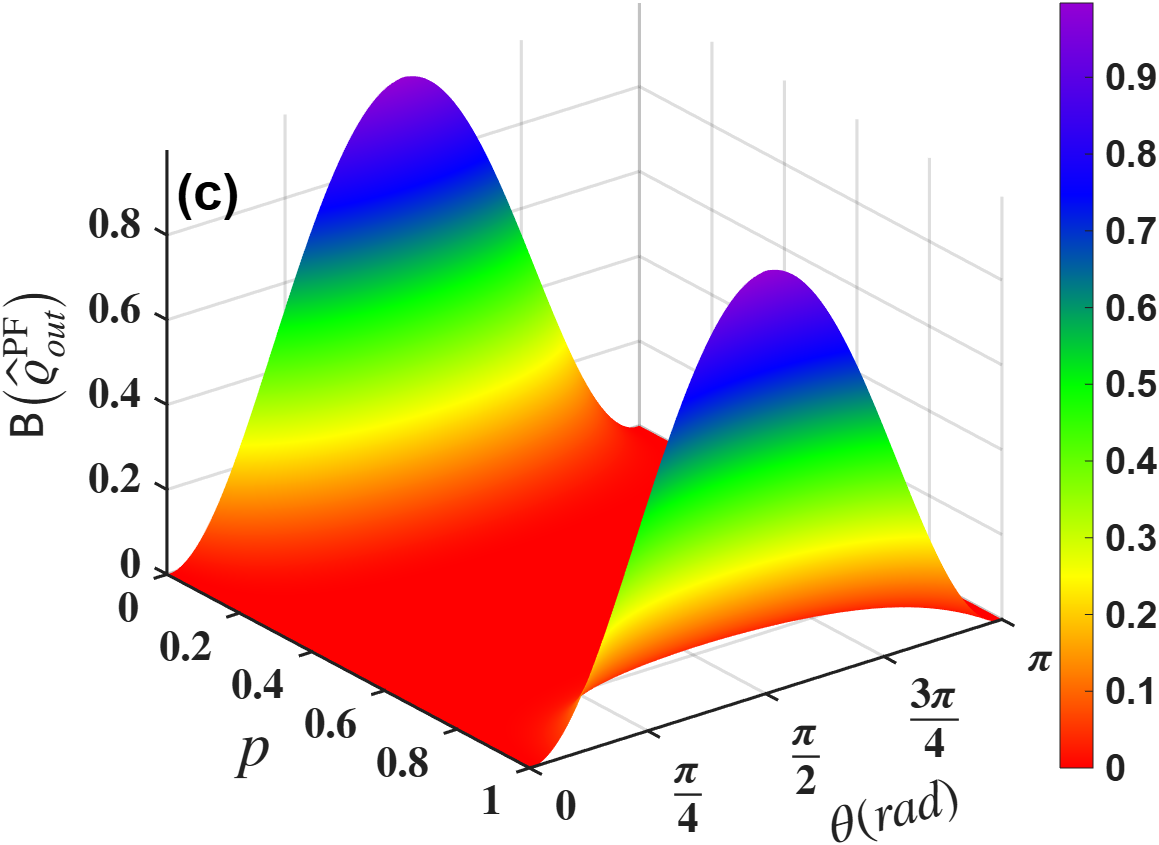}\\
\includegraphics[scale=0.28]{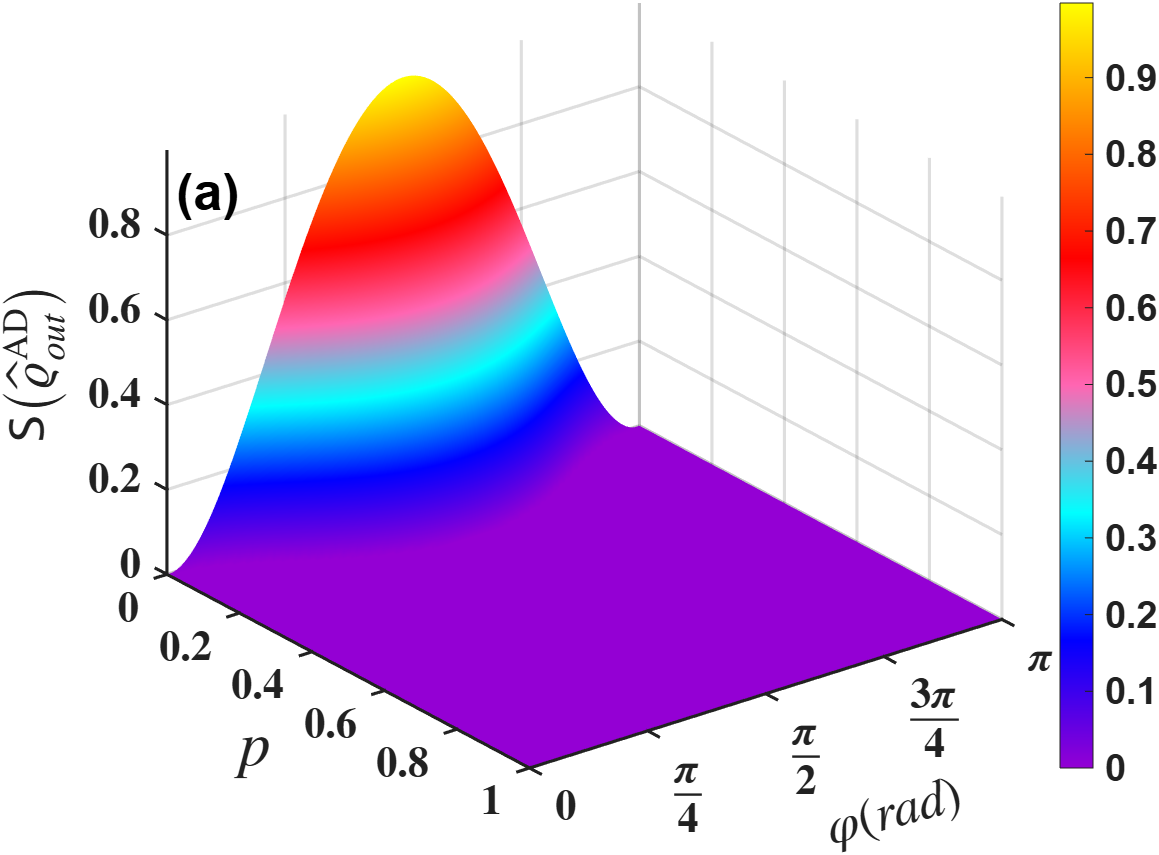}
\includegraphics[scale=0.28]{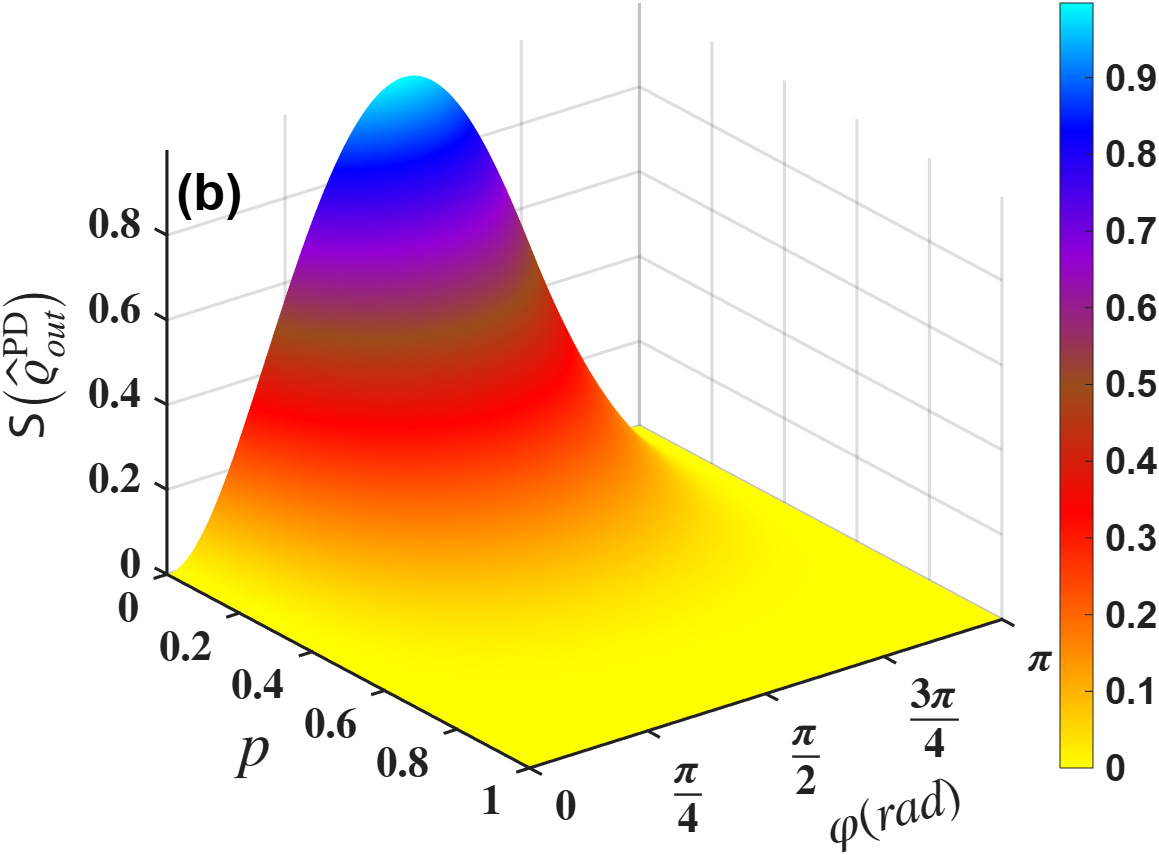}
\includegraphics[scale=0.28]{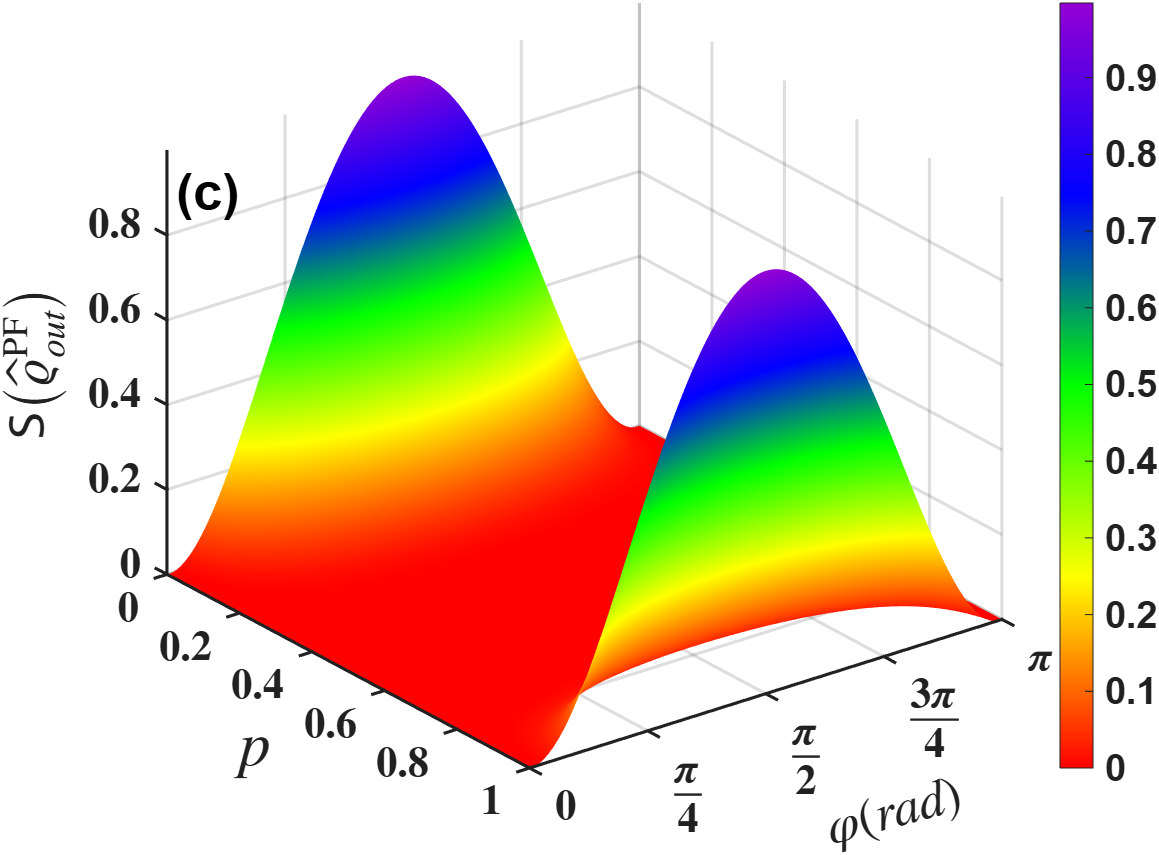}\\
\includegraphics[scale=0.28]{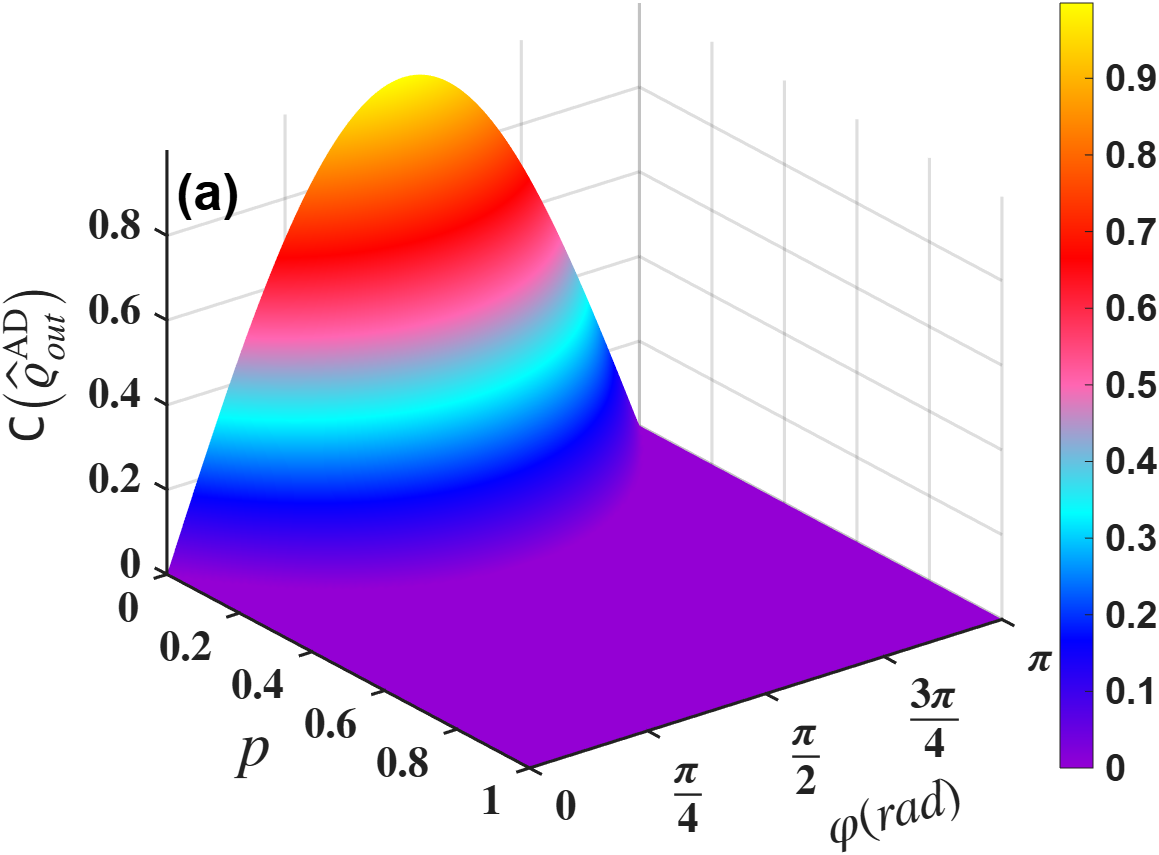}
\includegraphics[scale=0.28]{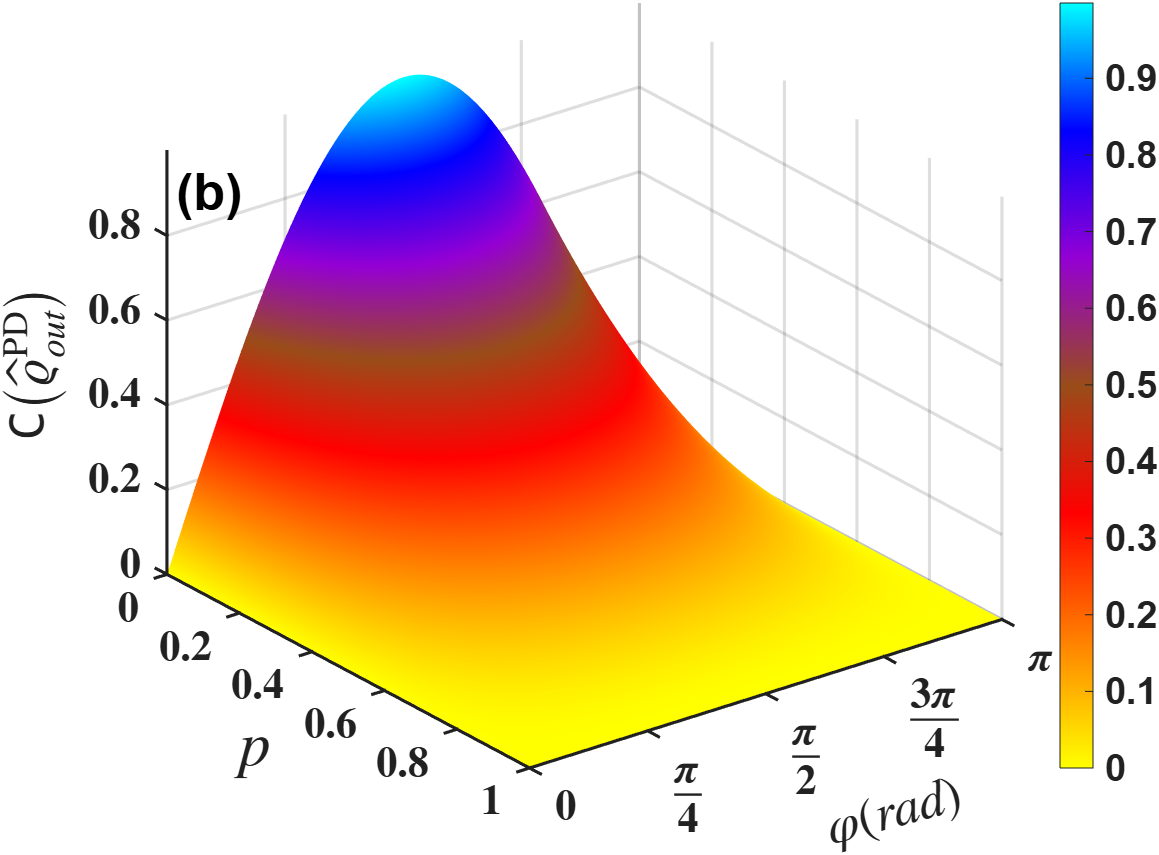}
\includegraphics[scale=0.28]{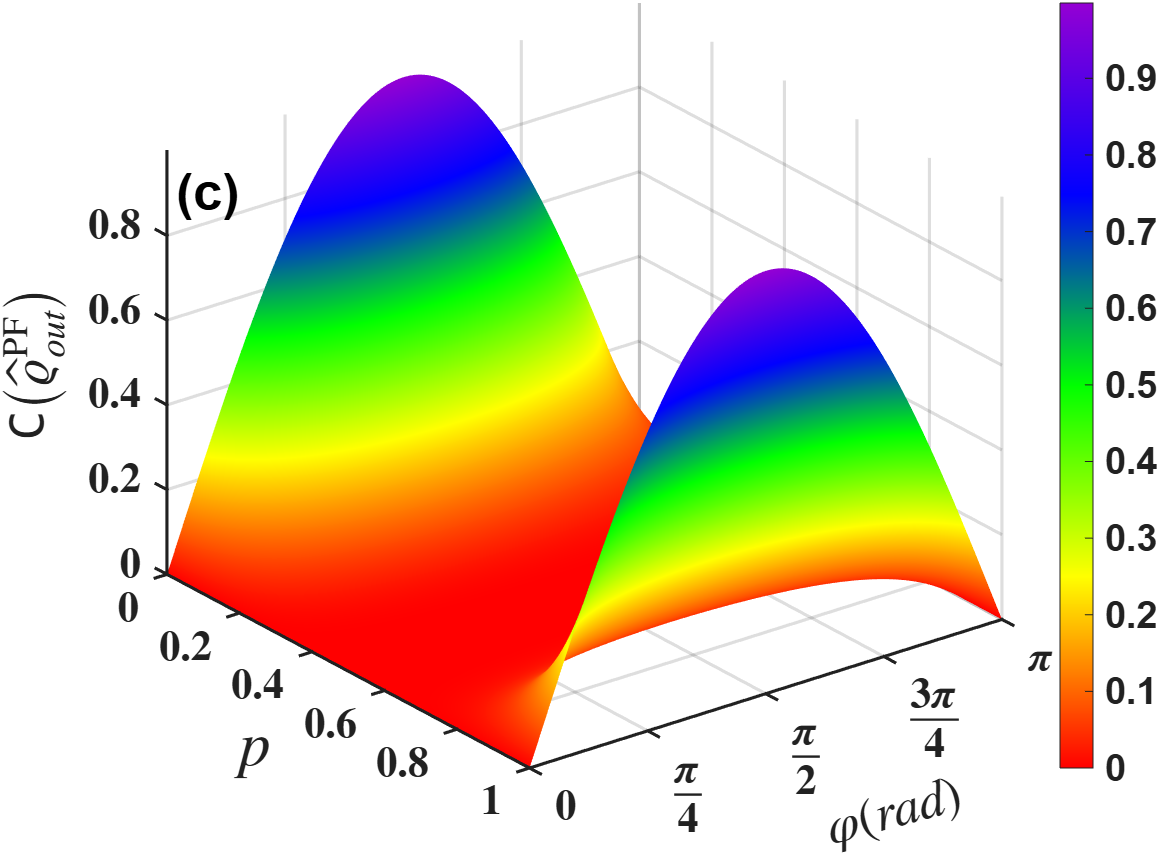}\\
\includegraphics[scale=0.28]{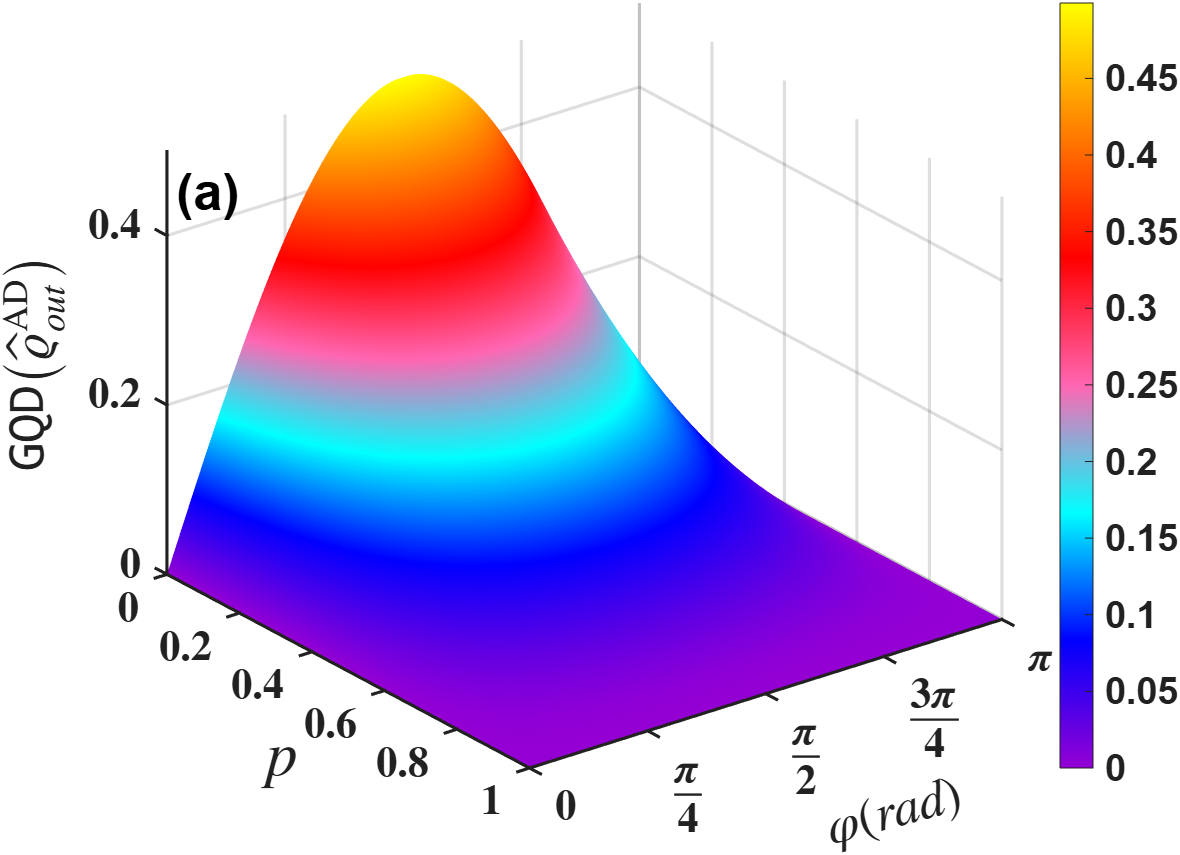}
\includegraphics[scale=0.28]{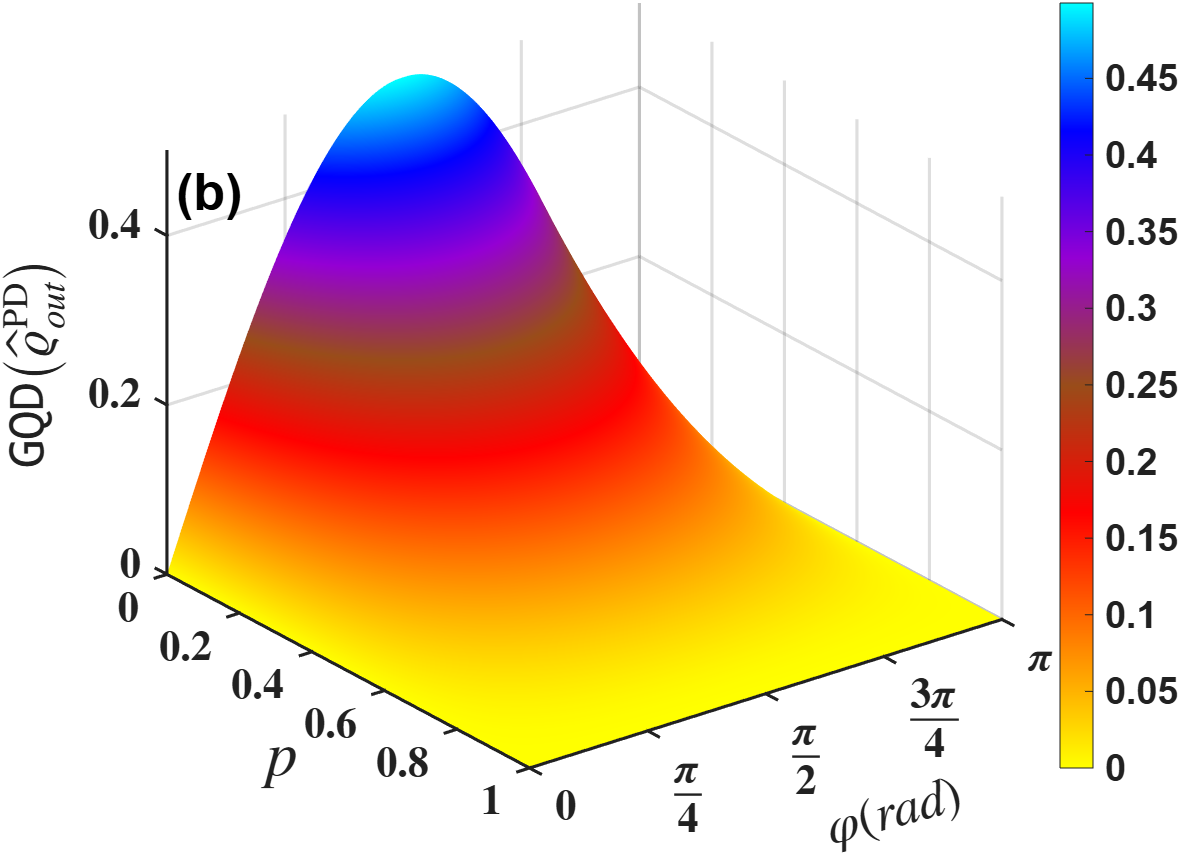}
\includegraphics[scale=0.28]{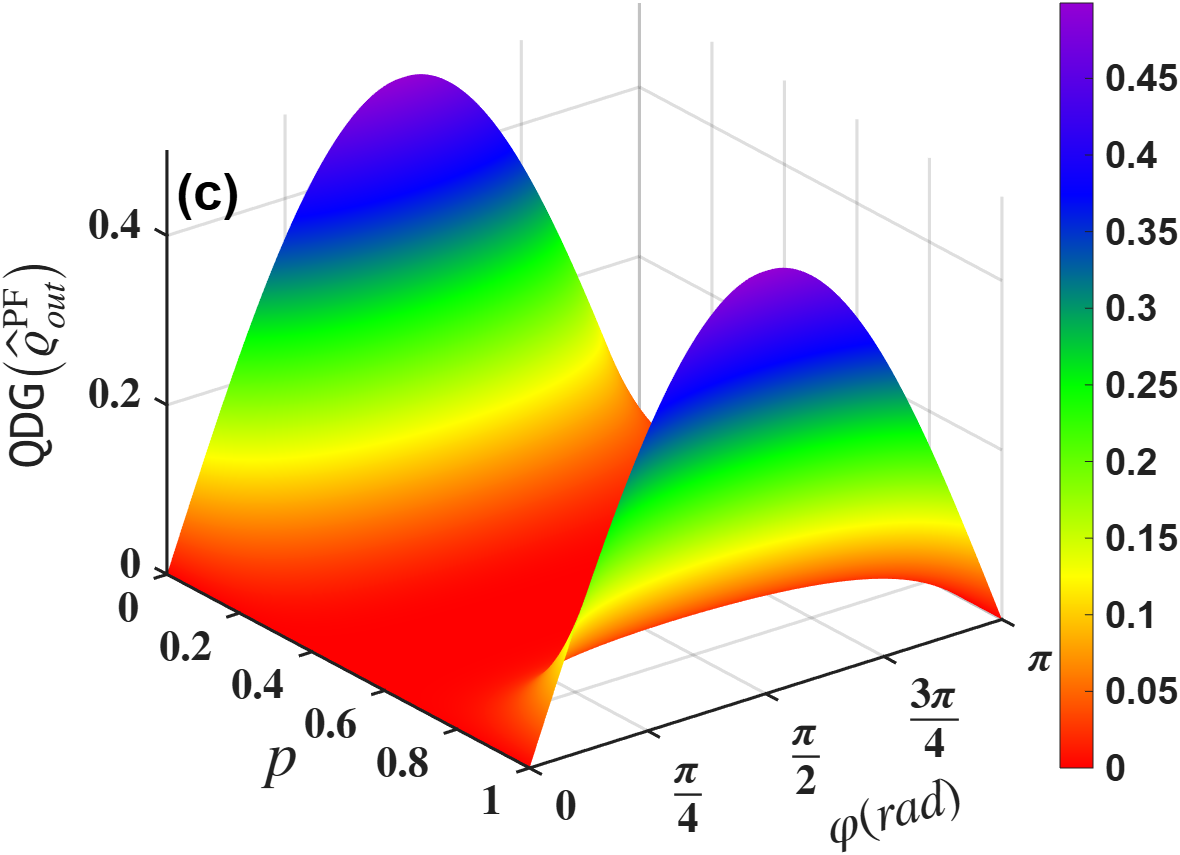}\\
\caption{Quantum correlations in the presence of decoherence: Bell nonlocality, quantum steering, concurrence, and geometric quantum discord as functions of the production angle $\theta$ and the decoherence parameter $p$. Results are shown for $\beta=1$ in the (a) amplitude-damping (AD), (b) phase-flip (PF), and (c) phase-damping (PD) channels. The limit $\beta=1$ corresponds to the regime where the $q\bar{q}\to t\bar{t}$ process approaches the gluon-initiated channel $gg\to t\bar{t}$.}
\label{fig:Te}
\end{figure*}

Figure~\ref{fig:Te}(a-c) illustrates the evolution of the quantum correlation quantifiers as a function of the amplitude angle $\varphi$ and the decoherence parameter $p$ in the relativistic limit $\beta \to 1$ (corresponding to $M_{t\bar{t}} \gg m_t$). In this high-energy regime, the quark-antiquark annihilation channel $q\bar{q} \to t\bar{t}$ becomes subdominant and the production mechanism converges toward the gluon-gluon fusion process $gg \to t\bar{t}$.

We investigate the effect of the decoherence parameter $p$ on the dynamics of quantum correlations encoded in the teleported $t\bar{t}$ state. The state is propagated through three different noisy quantum channels -- amplitude-damping (AD), phase-flip (PF), and phase-damping (PD) -- with the production angle fixed at $\theta = \pi/2$.

A pronounced symmetry in the quantum correlations is observed around $\varphi = \pi/2$. At this specific angle, all quantifiers reach their maximum values, which corresponds to a regime of maximal quantum entanglement for the teleported $t\bar{t}$ state. The quantifiers under study---namely, Bell non-locality, quantum steering, quantum concurrence, and geometric quantum discord---demonstrate notably different robustness depending on the noise channel. Specifically, they remain significantly more robust under the phase-flip (PF) channel compared to the amplitude-damping (AD) and phase-damping (PD) channels in the strong decoherence regime ($p \to 1$), as clearly shown in Fig.~\ref{fig:Te}(c). In contrast, these quantifiers completely vanish under the AD and PD channels in the same limit, as illustrated in Fig.~\ref{fig:Te}(a-b). Furthermore, the quantifiers exhibit a perfectly symmetric behavior with respect to $\varphi$ under the phase-flip channel. Overall, these results emphasize the high sensitivity of the teleported quantum correlations to decoherence effects and highlight the importance of the noise model in quantum information protocols involving top quark pairs.

\section*{Acknowledgments}

The research work was supported by Princess Nourah bint Abdulrahman University Researchers Supporting Project number (PNURSP2026R59), Princess Nourah bint Abdulrahman University, Riyadh, Saudi Arabia. The authors are thankful to the Deanship of Graduate Studies and Scientific Research at University of Bisha for supporting this work through the Fast-Track Research Support Program.

\section{Conclusion}\label{sec:6}

This work conducts a comprehensive analysis of quantum correlations in top–antitop quark pair ($t\bar{t}$) production, reinforcing the growing connection between quantum information science and high-energy physics. We employed key quantum-information quantifiers — Bell nonlocality, quantum steering, concurrence, and geometric quantum discord — to characterize the correlations in both the dominant gluon-fusion ($gg \to t\bar{t}$) and quark–antiquark annihilation ($q\bar{q} \to t\bar{t}$) channels.
For gluon fusion, strong quantum correlations emerge in the low-velocity regime across a broad range of production angles $\theta$. These correlations persist and are significantly enhanced in the ultra-relativistic limit ($\beta=1$), reaching a maximum near $\theta=\pi/2  $. Conversely, in the quark–antiquark annihilation channel, the correlations strengthen progressively with increasing top-quark velocity, also peaking in the ultra-relativistic regime near $\theta=\pi/2$.
Additionally, we examined the robustness of these quantum resources under realistic noisy conditions using three major decoherence channels: amplitude damping (AD), phase damping (PD), and phase flip (PF). For non-teleported states, AD and PD cause gradual degradation of all quantum measures with rising decoherence parameter $p$, while PF displays characteristic symmetry around $p=0.5$. In teleported states, all quantum quantifiers and the teleportation fidelity achieve their highest values at $\varphi=\pi/2$. Although this angle optimizes resource preservation, robustness varies markedly across channels. The PD channel strongly suppresses fidelity as $p\to 1$, while the PF channel remarkably maintains fidelity above the classical limit of $2/3$ near $p=0$ and $p=1$, revealing that noise can, in specific cases, partially mitigate decoherence effects.
Overall, these findings illuminate the complex interplay between decoherence and quantum correlations in top-quark systems, opening new avenues for controlling quantum resources in noisy high-energy environments and establishing collider-based processes as promising platforms for quantum information research.

\bibliography{sample}
\end{document}